\documentclass[a4paper,11pt]{article}
\usepackage[utf8]{inputenc}
\usepackage[T1]{fontenc}
\usepackage{comment}
\usepackage{tensor}
\usepackage{bbm}
\usepackage{amssymb}
\usepackage{authblk}
\usepackage[dvipsnames]{xcolor}
\usepackage{graphicx}
\usepackage[new]{old-arrows}
\usepackage{jheppub}
\usepackage{bm}

\usepackage[multiple]{footmisc}
\usepackage{subcaption}
\usepackage{multicol}
\usepackage{multirow}
\usepackage{rotating}
\usepackage{array}
\usepackage{mathtools}
\usepackage{longtable}
\usepackage{booktabs}
\usepackage[mathscr]{euscript}
\usepackage{tikz}
\usepackage{tikz-cd}
\usepackage{braket}
\usepackage{dsfont}
\usepackage{slashed}
\usepackage{marginnote}
\usepackage[dvipsnames]{xcolor}

\newcommand{\RNum}[1]{\uppercase\expandafter{\romannumeral #1\relax}}

\tikzset{
	partial ellipse/.style args={#1:#2:#3}{
		insert path={+ (#1:#3) arc (#1:#2:#3)}
	}
}

\tikzset{snake it/.style={decorate, decoration=snake}}

\usetikzlibrary{calc,decorations.markings}
\usetikzlibrary{decorations.pathmorphing}
\usetikzlibrary{decorations.pathreplacing}
\usetikzlibrary{shapes,backgrounds}
\usetikzlibrary{fadings}
\usetikzlibrary{cd}
\usetikzlibrary{hobby}
\usetikzlibrary{decorations.pathreplacing}
\usetikzlibrary{knots}
\usetikzlibrary{3d}
\tikzfading
[
	name=fade out,
	inner color=transparent!0,
	outer color=transparent!100
]

\newif\ifdraft
\draftfalse

\newcommand{\be}{ \begin{equation}}
\newcommand{\ee}{\end{equation}}
\newcommand{\bi}{ \begin{itemize}}
\newcommand{\ei}{\end{itemize}}

\def\pd{\partial}
\def\ol{\overline}
\def\pd{\partial}
\newcommand{\g}{{\gamma}}

\newcommand{\lr}[1]{\left( #1 \right)}

\makeatletter
\newcommand*\bigcdot{\mathpalette\bigcdot@{.65}}
\newcommand*\bigcdot@[2]{\mathbin{\vcenter{\hbox{\scalebox{#2}{$\m@th#1\bullet$}}}}}
\makeatother

\title{\boldmath The Lion, the Witch, and the Wormhole:\\Ensemble averaging the symmetric product orbifold}
\author[a]{Joshua Kames-King,}
\author[b]{Alexandros Kanargias,}
\author[c]{Bob Knighton,}
\author[d]{Mykhaylo Usatyuk}
\affiliation[a]{Laboratory for Theoretical Fundamental Physics, Institute of Physics,\\
\'Ecole Polytechnique F\'ed\'erale de Lausanne, Switzerland}
\affiliation[b]{PRISMA+ Cluster of Excellence \& Mainz Institute for Theoretical Physics,\\
Johannes Guttenberg-Universit\"at Mainz,
55099 Mainz, Germany}
\affiliation[c]{Institut f\"{u}r Theoretische Physik, ETH Z\"{u}rich\\
Wolfgang-Pauli-Strasse 27, 8093 Z\"{u}rich, Switzerland}
\affiliation[d]{Center for Theoretical Physics and Department of Physics\\
Berkeley, CA, 94720, USA}
\emailAdd{jvakk@yahoo.com}
\emailAdd{kanargias@uni-mainz.de}
\emailAdd{robejr@ethz.ch}
\emailAdd{musatyuk@berkeley.edu}
\abstract{We consider the ensemble average of two dimensional symmetric product orbifold CFTs $\text{Sym}^N(\mathbb{T}^D)$ over the Narain moduli space. We argue for a bulk dual given by $N$ copies of an abelian Chern-Simons theory coupled to topological gravity, endowed with a discrete gauge symmetry exchanging the $N$ copies. As a check of this proposal, we calculate the ensemble average of various partition and correlation functions of the symmetric product orbifold theory and compare the resulting expressions to gauge theory quantities in the bulk. We comment on the ensemble average of the tensionless string partition function on $\text{AdS}_3 \times \text{S}^3 \times \mathbb T^4$ by considering the specific case of $D=4$ with the addition of supersymmetry.
}

\arxivnumber{}

\begin{document}

\maketitle
\flushbottom
\begingroup\allowdisplaybreaks

\section{Introduction}
The AdS/CFT correspondence states that a given conformal field theory is dual to a theory of quantum gravity. However, it has recently been appreciated that some simple theories of quantum gravity, defined by a sum over geometries and  weighed by a semiclassical action, are dual to an average over a suitable ensemble of boundary theories. The clearest example of this is JT gravity which is a two-dimensional gravity theory dual to an ensemble of one dimensional quantum mechanical theories \cite{Saad:2019lba}. 

This idea has been extended to a wide class of two-dimensional gravitational theories \cite{Stanford:2019vob,Witten:2020wvy,Maxfield:2020ale,Turiaci:2020fjj,Forste:2021roo}, and progress has also been made in extending these concepts to higher-dimensional theories of gravity \cite{Maloney:2020nni,Afkhami-Jeddi:2020ezh,Cotler:2020ugk,Chandra:2022bqq,Collier:2022emf}. In more than two bulk dimensions it is a priori unclear how to construct ensemble averages over dual microscopic theories, so many approaches have focused on two-dimensional CFTs with a large number of symmetries such as Narain CFTs \cite{Datta:2021ftn,Perez:2020klz,Raeymaekers:2021ypf,Ashwinkumar:2021kav,Dymarsky:2020qom,Dymarsky:2020pzc}, and WZW models \cite{Dong:2021wot,Meruliya:2021utr,Meruliya:2021lul}. We will consider the former. In \cite{Maloney:2020nni,Afkhami-Jeddi:2020ezh} it was demonstrated that averaging over the family of $\mathbb{T}^D$ Narain CFTs is dual to a bulk theory given by Chern-Simons coupled to topological gravity. Narain CFTs can be defined by the action
\begin{equation}
    I=\int \mathrm{d}^2 z \lr{G_{mn}\delta^{\alpha\beta}\partial_\alpha X^m\partial_\beta X^n+i B_{mn}\varepsilon^{\alpha\beta}\partial_\alpha X^m \partial_\beta X^n}\,,
\end{equation}
where the choice of target metric $G_{m n}$ and $B_{m n}$ is a choice of moduli and defines the theory. We denote the Narain CFT partition function by $Z_{\mathbb{T}^D}(m, \Sigma)$ where $\Sigma$ is a two dimensional Riemann surface on which the theory is defined and $m$ is a particular choice of the moduli defining the theory. Averaging over the moduli with an appropriate measure, it was found that the resulting averaged partition function $\langle Z_{\mathbb{T}^D}(m, \Sigma)\rangle$ could be reproduced by a bulk Chern-Simons calculation. The bulk theory takes the form of $2D$ copies of abelian Chern-Simons with total gauge group $G = \text{U(1)}^D \times \text{U(1)}^D$ and action given by
\be \label{eqn:introaction}
S_{\text{CS}} = \sum_{i=1}^D \int_M \left(A_i \wedge d A_i - B_i \wedge d B_i \right),
\ee
where the $2D$ gauge fields $A_i,B_i$ transform under independent copies of $\text{U}(1)$. It was found that summing over a class of bulk three-manifolds, bulk handlebodies $M$ with asymptotic boundary $\pd M = \Sigma$, precisely reproduces the average over Narain CFTs
\be
\langle Z_{\mathbb{T}^D}(m, \Sigma)\rangle = \sum_{\text{handlebodies } M} Z_{G}(M),
\ee
where on the right we evaluate the Chern-Simons path integral with action \eqref{eqn:introaction} on each handlebody. The subscript $G$ on the partition function denotes the gauge group of the Chern-Simons theory. We go into additional details on the proposed duality between the Narain average and a bulk Chern-Simons theory in Section \ref{sec:narain-review}.

In this paper we will extend this duality by ensemble averaging over a related family of two-dimensional CFTs, symmetric product orbifolds of Narain CFTs. The process to construct a symmetric product orbifold is to take $N$ tensor copies of a seed CFT $X$, and gauge the $S_N$ permutation symmetry  exchanging the copies of the theory
\be
\text{Sym}^N(X) = X^{\otimes N}/S_N\, .
\ee
We go into additional details on constructing such theories in Section \ref{subsec:permutation-orbifolds}. Applying this procedure to the Narain theories we can construct a family of CFTs, denoted by Sym$^N(\mathbb{T}^D)$, labelled by a choice of integer $N$ and a point in moduli space $m$. We denote the partition function of such theories by $Z_{\mathbb{T}^D \wr S_N}(m, \Sigma)$. Since this family of theories has the same moduli space as the Narain theories we can again perform the ensemble average over such theories. We now summarize our main results.

\subsection{Summary of main results}

\textbf{Ensemble averaging $\text{Sym}^N(\mathbb{T}^D)$:} The goal of this paper is to provide a bulk dual for the ensemble average of the symmetric product orbifold of Narain CFTs. Following the standard holographic prescription, the bulk dual should be given by a sum over a suitable set of bulk geometries. The philosophy we will take in this work is that the rules for the bulk path integral should be dictated by the boundary ensemble average. In particular, the choice of which bulk geometries to include is determined by consistency with the boundary answer \cite{Maloney:2020nni}.

We now restrict our attention to Sym$^N(\mathbb{T}^D)$ CFTs defined on a boundary torus with modular parameter $\tau$. In Section \ref{sec:averaging-orbifold} we explain how to ensemble average the partition function of this class of theories $\langle Z_{\mathbb{T}^D \wr S_N}(m, \tau) \rangle$. The final result is given in equation \eqref{eqn:sec3AverageZFinal}, and is a formal expression in terms of the Siegel-Weil Formula \eqref{eq:generic-siegel-weil-disconnected}, which we introduce and explain in Section \ref{sec:narain-review}. We expect this average to be holographically dual to a sum over bulk geometries with an asymptotic boundary torus. We find this is partially realized. The averaged partition function can be schematically expressed as follows\footnote{As we will discuss later in Section \ref{sec:preliminaries}, both the average over the Narain moduli space and sum over handlebodies can diverge, assuming the degree $N$ of the orbifold group and the boundary genus $g$ are sufficiently large (in a way that will be made precise below), see also \cite{Maloney:2020nni}. We will largely ignore these divergence issues since, even when the sum over handlebodies diverges, the individual summands still make sense as on-shell bulk partition functions. Of course, if one considers the full microscopic theory without averaging, no such divergence should appear.}
\be \label{eqn:introbdyaverage}
\langle Z_{\mathbb{T}^D \wr S_N} (m,\tau) \rangle = \sum_{\substack{\text{handlebodies $M$,} \\ \text{vortices}}} Z_{\text{Bulk}}(M) + \text{non-semiclassical geometries}.
\ee
In the above we have split the boundary average into two terms. We will refer to the first term as a ``semiclassical'' contribution while the second term is a ``non-semiclassical'' contribution. We define contributions as semiclassical or not based on whether they can be reproduced by the standard rules of the gravitational path integral. Let us first explain the bulk origin of the semiclassical contribution.

\vspace{0.25cm}

\noindent\textbf{Semiclassical Contributions:} The semiclassical contribution is reproduced by a standard gravitational path integral where we sum over handlebody geometries bounding the asymptotic torus with the inclusion of ``vortices'' (analogous to `t Hooft loops for discrete gauge groups) running along the non-contractible cycle of the geometry. We have schematically represented the contribution of each such geometry by $Z_{\text{Bulk}}(M)$, which is given by a one-loop exact Chern-Simons calculation. In the case of a boundary torus, handlebody geometries are three manifolds of the form $D^2 \times S^1$. On each handlebody geometry we evaluate the partition function of a Chern-Simons theory with gauge group $\text{U}(1)^D \times \text{U}(1)^D \wr S_N$.\footnote{The notation $\wr$ indicates that the gauge group takes the form of a wreath product group, which is defined by taking $N$ copies of $\text{U}(1)^D \times \text{U}(1)^D$ and gauging the symmetry permuting them. We explain this structure in Section \ref{sec:bulk-theory}.} The bulk action of the Chern-Simons theory with this group is given by $N$ copies of the previous action in equation \eqref{eqn:introaction}
\begin{equation} \label{intro:CSAction}
S_{\text{CS}}=\sum_{i=1}^{N}\int_{M}\left(A_{(i)}\wedge\mathrm{d}A_{(i)}-B_{(i)}\wedge\mathrm{d}B_{(i)}\right)\, ,
\end{equation}
where we have suppressed the summation over the $D$ indices present in \eqref{eqn:introaction} for simplicity. In total the theory has $2DN$ gauge fields. Evaluating the partition function of this theory is slightly non-trivial since the structure of the gauge group is a wreath product, and in Section \ref{sec:4.1} we explain how to accomplish this for bulk handlebody geometries.

The final aspect of equation \eqref{eqn:introbdyaverage} that we must explain is the summation over vortices. A vortex is a gauge theory line operator that we choose to place along the non-contractible cycle of the handlebody. This operator implements twisted boundary conditions on the gauge fields $A_{(i)}, B_{(i)}$ as they travel around the vortex
\be\label{eq:CSmonodromy}
A_{(i)} \to A_{\pi(i)}, \qquad B_{(i)} \to B_{\pi(i)},
\ee
where $\pi \in S_N$ are permutations. The inclusion of vortices amounts to including gauge field configurations that are singular in the interior of the handlebody, and we explain how to evaluate the path integral on a handlebody with a vortex insertion in Section \ref{sec:4.1}. The summation over vortices amounts to a summation over all twisted boundary conditions implemented by permutations $\pi$ on the gauge fields. 

Putting everything together, in sections \ref{sec:cs-calculations-n=2} and \ref{sec:4.3} we evaluate the Chern-Simons partition function on handlebody geometries with vortex operator insertions and show that we precisely reproduce what we denote as the semiclassical contribution to the averaged partition function
\be
\langle Z_{\mathbb{T}^D \wr S_N} (m,\tau) \rangle \supset \sum_{\substack{\text{handlebodies $M$,} \\ \text{vortices}}} Z_{G \wr S_N} (M) = ~\raisebox{-.85cm}{\includegraphics[width=.2\textwidth]{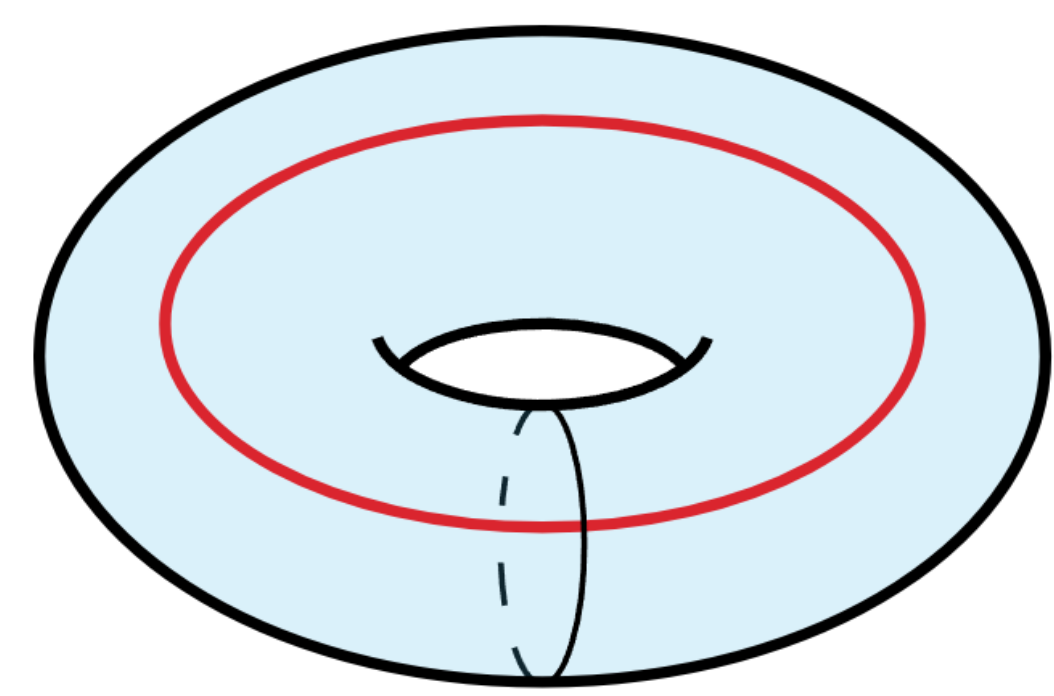}}\, .
\ee
In the above figure the vortex is the red line running in the interior of the handlebody. The Chern-Simons partition function $Z_{G \wr S_N}$ implicitly depends on the twisted boundary conditions the vortex implements. Thus, at least a portion of the ensemble averaged partition function can be reproduced by a standard bulk theory given by $U(1)^D \times U(1)^D \wr S_N$ Chern-Simons with the inclusion of bulk vortices. 

\vspace{0.25cm}

\noindent\textbf{Non-semiclassical Contributions:} Let us now explain the ``non-semiclassical'' contribution to equation \eqref{eqn:introbdyaverage}. The ensemble average of the partition function $\langle Z_{\mathbb{T}^D \wr S_N}\rangle$ contains averages over multiple disconnected products of partition functions of the seed theory. A useful example is to consider the case of $N=2$, where the boundary average contains a contribution
\be \label{eqn:introN2disconnected}
\langle Z_{\mathbb{T}^D \wr S_2} (m,\tau) \rangle \supset \frac{1}{2} \langle Z_{\mathbb{T}^D}(m,\tau)Z_{\mathbb{T}^D}(m,\tau)\rangle  = \sum_{\text{geometries}} ~\raisebox{-1cm}{\includegraphics[width=.3\textwidth]{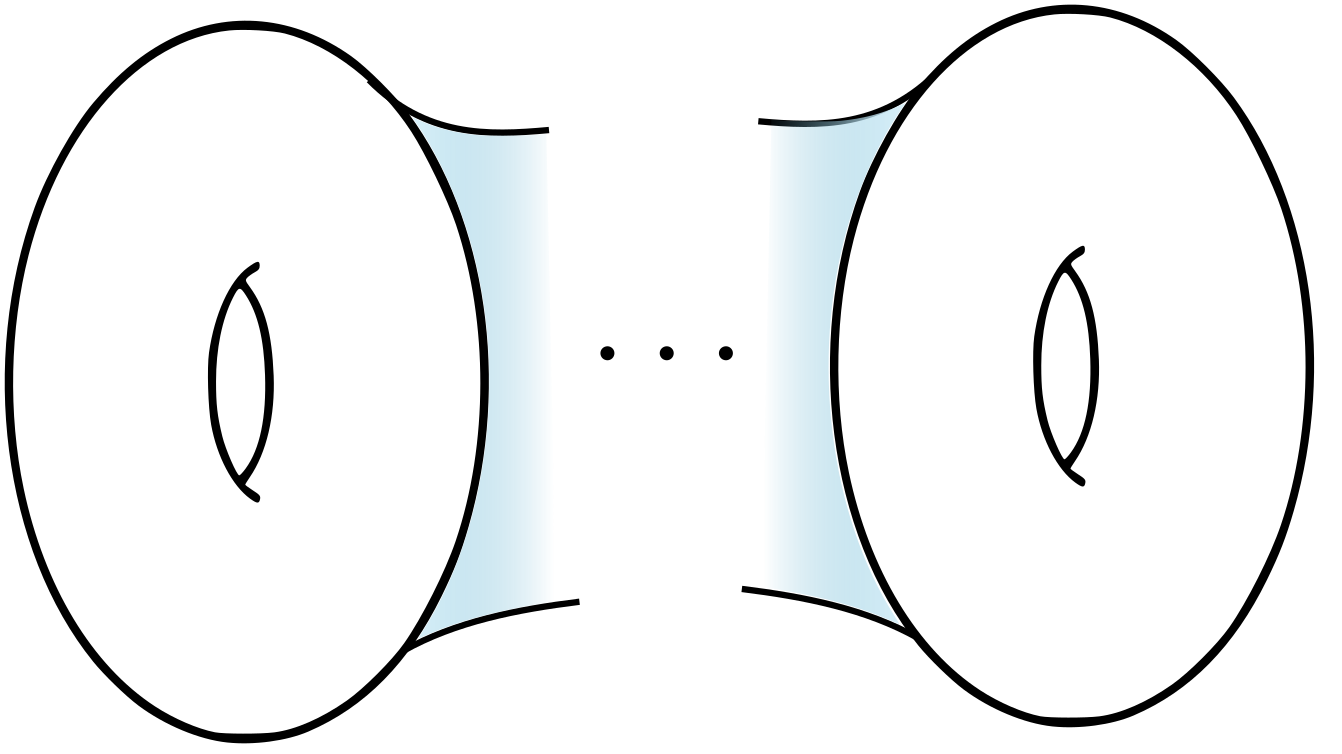}} \, .
\ee
Where in the above figure we have used the results summarized in Section \ref{sec:narain-review} to represent the average $\langle Z_{\mathbb{T}^D}^2 (m, \tau)\rangle$ as a gravitational path integral with two asymptotic boundary tori \cite{Maloney:2020nni}. Generic non-semiclassical contributions arise from geometries of a similar nature, where for general $N$ the boundary average instructs us to include terms  with up to $N$ asymptotic boundary tori. 

While such terms have a geometric interpretation as bulk configurations with multiple asymptotic boundaries, typically, they do not have a holographic interpretation as a single geometry with a single asymptotic boundary. To assign a holographic interpretation to a bulk wormhole geometry $M$ with $n$ asymptotic boundaries we require that $M$ is suitably ``symmetric''. The precise notion of this symmetry is subtle and we elaborate on it in Section \ref{sec:4.3}, but it is reminiscent of the $\mathbb{Z}_n$ replica symmetry in the context of the replica trick \cite{Lewkowycz:2013nqa}. In the case that $M$ is completely disconnected, the requirement is that the same boundary cycle is contractible in the interior of each disconnected bulk geometry. Such a symmetry is highly non-generic, and most wormhole configurations contributing to the average do not have a simple interpretation. However, a subset of such geometries do have such a symmetry, and have already been implicitly included in the ``semi-classical'' Chern-Simons computation, see Section \ref{sec:4.3}.

It's useful to give a simple example of a geometry that does not have a semiclassical interpretation. Consider a contribution to equation \eqref{eqn:introN2disconnected} where one torus is filled in with a handlebody with contractible spatial cycle, while the other is filled in with a handlebody with contractible time cycle
\be
\langle Z_{\mathbb{T}^D \wr S_2} (m,\tau) \rangle \supset ~\raisebox{-1.2cm}{\includegraphics[width=.5\textwidth]{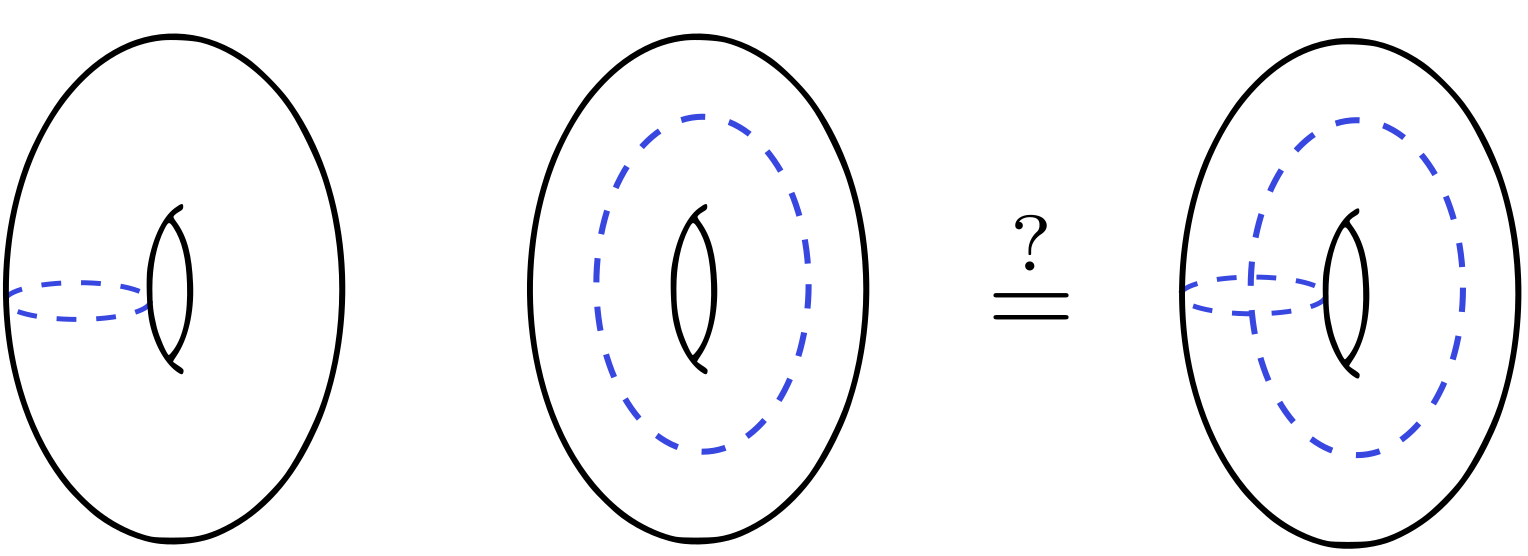}}\, ,
\ee
where the contractible cycles are identified by the dashed lines in the figure. A holographic interpretation would require a single bulk geometry with both spatial and time cycles contractible in the interior. A bulk manifold $M$ cannot have both of these cycles contract, and so such a contribution has no hope of being reproduced by a standard sum over geometries. To include such contributions we would need to seriously modify the standard rules for the bulk path integral, and allow different gauge fields $A_I$ to live on ``independent'' manifolds with different contractible cycles. See Section \ref{sec:cs-calculations-n=2} and the discussion in Section \ref{sec:7}.

\vspace{0.25cm}

\noindent\textbf{Averaging correlators:} In Section \ref{sec:correlators} we consider ensemble averaging correlation functions of twist operators in the symmetric orbifold, focusing specifically on the case of the $\text{Sym}^2(\mathbb{T}^D)$ orbifold. These are non-local operators implementing twisted boundary conditions for the fundamental fields in the orbifolded theory. Hence, these objects are naturally identified as being the dual to the vortices of the the bulk Chern-Simons theory mentioned around \eqref{eq:CSmonodromy}. Following the elegant approach of references \cite{Lunin:2000yv,Lunin:2001pw}, the monodromy implemented by twist fields trivialises on the covering space and it can be shown that the correlation functions reduce to a product of covering map data and the seed partition function on the (branched) covering space. For example for the case of the sphere we get \eqref{eq:correlationfunctiontwistfieldsgeneral}. Therefore the Siegel-Weil formula may be used in performing the average of the latter resulting in a modular sum. How is this interpreted from the bulk perspective? Using the identification of vortices and twist operators we will show that summing over all inequivalent configurations of vortices ending on pairs of equivalent twist operators reduces to the aforementioned modular sum. More specifically, as in \cite{Benjamin:2021wzr} we consider vortex configurations, which are rational tangles in the language of knot theory. Such tangles exhibit handlebodies as branched covering spaces such that the modular sum can be understood as a sum over hyperbolic metrics.

\vspace{0.25cm}

\noindent\textbf{Averaging Supersymmetric Narain CFTs:} In Section \ref{sec:6}, we consider the ensemble average over the supersymmetric version of Narain CFTs. We propose that the ensemble average is holographically dual to a supersymmetric version of $\text{U}(1)^D \times \text{U}(1)^D$ Chern-Simons, and we reproduce the boundary ensemble averaged torus partition function from a bulk supersymmetric Chern-Simons sum over handlebody geometries. Furthermore, we consider the symmetric product orbifold $\text{Sym}^N(\mathbb{T}^D)$ of supersymmetric Narain theories and show that, similar to the non-supersymmetric case, a supersymmetric Chern-Simons theory with gauge group $\text{U}(1)^D \times \text{U}(1)^D \wr S_N$ reproduces many ``semiclassical'' contributions to the averaged partition function.

\vspace{0.25cm}

\noindent\textbf{Averaging the Tensionless String:} The symmetric orbifold $\text{Sym}^N(\mathbb{T}^4)$ at large $N$ is dual to type IIB string theory on $\text{AdS}_3\times\text{S}^3\times\mathbb{T}^4$ with one unit of pure NS-NS flux (the so-called `tensionless string') \cite{Eberhardt:2018ouy,Eberhardt:2019ywk,Eberhardt:2020akk,Eberhardt:2020bgq,Eberhardt:2021jvj,Dei:2020zui,Knighton:2020kuh,Bertle:2020sgd,Gaberdiel:2021njm,Gaberdiel:2022oeu,Naderi:2022bus}. Part of our motivation for considering the average of $\text{Sym}^N(\mathbb{T}^D)$ theories is to understand whether the tensionless string can be ensemble averaged to produce a semiclassical sum over geometries. We are partially successful, averaging a single string propagating on an AdS$_3$ background gives a ``semiclassical'' geometry as defined above. Averaging over multiple strings on a single background gives rise to the ``non-semiclassical'' geometries. We leave a more complete discussion of this to Section \ref{sec:7}.

\vspace{0.25cm}

In Section \ref{sec:preliminaries} we review some preliminary results necessary for the rest of the work. We explain the Narain average/Chern-Simons duality of \cite{Maloney:2020nni,Afkhami-Jeddi:2020ezh}, and we review the basic construction of symmetric product orbifold CFTs. In Section \ref{sec:averaging-orbifold} we begin by explaining how to ensemble average over a symmetric product CFT. We then apply this to the class of Narain CFTs $\text{Sym}^N(\mathbb{T}^D)$ to obtain a boundary answer for the average. In Section \ref{sec:bulk-theory} we interpret the boundary average as a holographic Chern-Simons theory with the inclusion of bulk vortices. In Section \ref{sec:correlators} we consider correlation functions of twist operators in the CFT, and we show that a bulk dual is given by a sum over vortex configurations equivalent to a specific sum over hyperbolic three-manifolds. In Section \ref{sec:6} we consider the supersymmetric extension of the Narain theories, and we provide a bulk dual given by supersymmetric Chern-Simons theory, with additional details left to appendix \ref{sec:susyappendix}. In Section \ref{sec:7} we end with a discussion of our results.

\subsection*{Note:} 

The title of this paper was chosen keeping in tradition with \cite{Benjamin:2021wzr}, which in turn was named in homage to a video introduction to branched coverings over knots by W. Thurston \cite{Thurston:video}.

\section{Preliminaries}\label{sec:preliminaries}

In this section we review the necessary technology used throughout the main parts of the paper. In particular, we review the Narain-ensemble/$\text{U}(1)$ gravity duality proposed by \cite{Maloney:2020nni, Afkhami-Jeddi:2020ezh}, as well as the basics of permutation orbifolds. Readers familiar with Narain averaging and permutation orbifolds should feel free to skip this section.

\subsection{Narain averaging and the sum over geometries}\label{sec:narain-review}
Consider a sigma-model with target space a $D$-dimensional torus $\mathbb{T}^D$. The action reads
\begin{equation}
    I=\int \mathrm{d}^2 z \lr{G_{mn}\delta^{\alpha\beta}\partial_\alpha X^m\partial_\beta X^n+i B_{mn}\varepsilon^{\alpha\beta}\partial_\alpha X^m \partial_\beta X^n}\ .
\end{equation}
where $G_{mn}$ is the metric on the $\mathbb{T}^D$ target space and $B_{mn}$ is a two-form field, and the target coordinates are compact $X^m\sim X^m+2\pi$. We take the theory to be defined on a Riemann surface with locally flat metric $\delta^{\alpha\beta}$, with $\varepsilon^{\alpha\beta}$ being the Levi-Civita symbol. This CFT belongs to a family of two-dimensional CFT's with left and right-moving current algebras of type $\text{U}(1)^D \times \text{U}(1)^D$ with central charges $(c_L,c_R)=(D,D)$, namely the \textit{Narain} family of CFTs. The moduli space of Narain CFTs is parameterised by $D^2$ parameters encoded in choosing a target metric $G_{m n}$ and the two form field $B_{m n}$. Equivalently the moduli space is given by the double quotient space (for more details, see e.g. \cite{Blumenhagen:2013fgp}):
\begin{equation}
    \mathcal{M}_D=\text{O}\lr{D,D;\mathbb{Z}}\backslash \text{O}\lr{D,D} / \text{O}\lr{D}\times  \text{O}\lr{D}\,.
\end{equation}
The partition function on a torus with modular parameter $\tau$ is given by
\begin{equation}
    Z_{\mathbb{T}^D}(m,\tau)=\frac{\Theta(m,\tau)}{\left|\eta(\tau)\right|^{2D}}\,,
\end{equation}
where $m \in \mathcal{M}_D$ labels a particular point in the moduli space of Narain CFTs, and $\eta(\tau)$ is the Dedekind eta function. The only moduli dependence enters through $\Theta(m,\tau)$ which is the Siegel-Narain theta function.\footnote{The partition function $Z(m,\tau)$, the theta function $\Theta(m,\tau)$, and the Eisenstein series $E_{s}(\tau)$ which we introduce later, also depend on $\overline{\tau}$. To avoid clutter, we omit this dependence in the notation and keep it implicit.} As this moduli space carries a natural measure via the Zamolodchikov metric,\footnote{This metric is calculated by computing the two-point function of the exactly marginal operator\linebreak $\mathcal{O}\sim \delta G_{mn}\delta^{\alpha\beta}\partial_\alpha X^m\partial_\beta X^n+i \delta B_{mn}\varepsilon^{\alpha\beta}\partial_\alpha X^m \partial_\beta X^n $. This metric is equivalently the Haar measure of $\text{O}(D,D;\mathbb{R})$ descendend to the quotient.} one can ensemble-average over the space of Narain CFTs. Since the only dependence on the moduli enters through $\Theta(m,\tau)$ we must consider the formal expression:
\begin{equation} \label{eqn:SiegelWeilTorus}
    \langle \Theta(m,\tau)\rangle := \int_{\mathcal{M}_D} \mathrm{d}\mu(m) \Theta(m,\tau)\, = \frac{E_{\frac{D}{2}}(\tau)}{\left(\operatorname{Im} \tau\right)^{\frac{D}{2}}} \, ,
\end{equation}
with $\mathrm{d}\mu(m)$ being the normalized Zamolodchikov measure. The expression for $\langle \Theta(m,\tau)\rangle$ is found by the use of the \textit{Siegel-Weil} \cite{Siegel1951,Weil1,Weil2,Maas} formula, and it is given by the real analytic Eisenstein series $E_s(\tau)$ which is defined as
\begin{align}\label{eq:Eisensteinseries1}
   & \qquad E_s(\tau) = \sum_{\gamma\in\Gamma_{\infty}\backslash\text{SL}(2,\mathbb{Z})}\lr{ \operatorname{Im} \gamma \cdot \tau}^s\, .
\end{align}
Here $\Gamma_{\infty}$ is the subset of the modular group which leaves invariant the imaginary part $\operatorname{Im}\tau$, and $\Gamma_{\infty}\backslash\text{SL}(2,\mathbb{Z})$ is the left quotient \cite{Maloney:2020nni}.\footnote{The subgroup $\Gamma_{\infty}$ consists of all $\text{SL}(2,\mathbb{Z})$ matrices of the form
$\begin{pmatrix} \pm 1 & n\\ 0 & \pm 1 \end{pmatrix}$. Two matrices $\gamma,\gamma'\in\text{SL}(2,\mathbb{Z})$ are considered equivalent in $\Gamma_{\infty}\backslash\text{SL}(2,\mathbb{Z})$ if $\gamma=h\cdot\gamma'$ for some $h\in\Gamma_{\infty}$. The sum in \eqref{eq:Eisensteinseries1} includes one matrix from each equivalence class in $\Gamma_{\infty}\backslash\text{SL}(2,\mathbb{Z})$.} The sum over modular images can be represented by matrices $\gamma= \begin{pmatrix}
a & b \\
c & d 
\end{pmatrix} \in \text{SL}(2,\mathbb{Z})$ with coprime $(c,d)=1$. Using the above we can write the final expression for the ensemble-averaged torus partition function as a sum over modular images
\begin{equation} \label{eqn:TorusAverage}
    \langle Z_{\mathbb{T}^D}(m,\tau)\rangle=\frac{E_{\frac D2}(\tau)}{\left(\operatorname{Im}\tau\right)^{\frac{D}{2}}\left|\eta(\tau)\right|^{2D}}\,=\sum_{\gamma\in\Gamma_{\infty}\backslash\text{SL}(2,\mathbb{Z})}\frac{1}{|\eta(\gamma\cdot\tau)|^{2D}}\, .
\end{equation} 
The above averaging procedure can be generalized to partition functions on arbitrary Riemann surfaces by using the higher genus analogue of the Siegel-Weil formula. The averaged partition function of the sigma-model on a genus $g$ surface $\Sigma_g$ with period matrix $\Omega$ takes on the form
\begin{equation}\label{eq:generic-siegel-weil}
    \langle Z_{\mathbb{T}^D}(m,\Omega)\rangle=\frac{E_{\frac D2}(\Omega)}{\lr{\text{det}\text{ Im}\,\Omega}^\frac D2 \left|\text{det}' \ \overline{\partial}\right|^D}\,.
\end{equation}
The determinant $\text{det}' \ \overline{\partial}$ appearing in \eqref{eq:generic-siegel-weil} is of the operator $\overline{\partial}$ on $\Sigma_g$ omitting the zero-modes. We have introduced the higher genus generalization of the Eisenstein series
\begin{equation}\label{eq:Eisensteinseriesarbitrarygenus}    E_s(\Omega)=\sum_{\Gamma_0}\lr{\text{det}\text{ Im}\,\Omega_{\Gamma_0}}^s\,,
\end{equation}
with $\Omega_{\Gamma_0}$ being the period matrix defined with respect to what is known as a Lagrangian sublattice $\Gamma_0$. We defer the discussion of Lagrangian sublattices to slightly later in this section, but for now a particular $\Gamma_0$ should be thought of as specifying a distinguished set of asymptotic boundary cycles which will be contractible in the bulk manifold when we interpret the ensemble average holographically, see Figure \ref{fig:LagrangianSublattice}.

\begin{figure}
    \centering
    \includegraphics[width=0.4\textwidth]{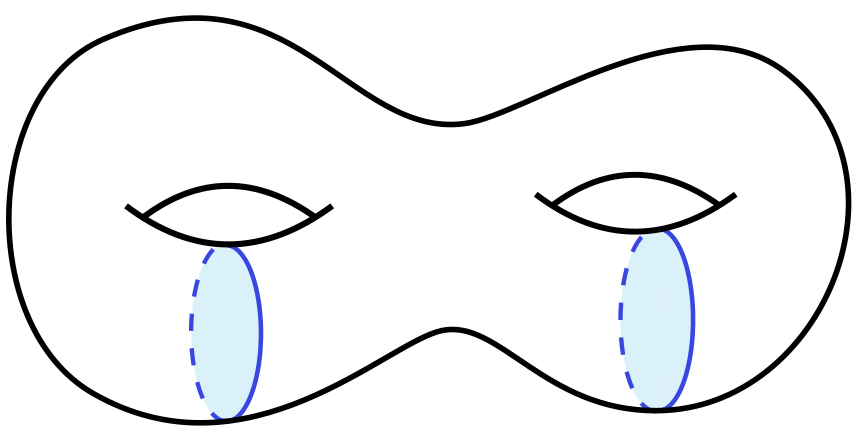}
    \caption{A given Lagrangian sublattice $\Gamma_0$ appearing in the sums in \eqref{eqn:TorusAverage}, \eqref{eq:generic-siegel-weil}, and \eqref{eq:generic-siegel-weil-disconnected} will be holographically associated with a choice of asymptotic boundary cycles that becomes contractible in the interior of the geometry. In this figure we have genus two handlebody with the drawn cycles contractible in the bulk.}
    \label{fig:LagrangianSublattice}
\end{figure}

The average can also be defined over products of partition functions on disconnected Riemann surfaces. Suppose we have a product of $n$ partition functions on associated Riemann surfaces of genus $g_i$ with period matrices $\Omega_i$. We can form a matrix $\Omega$ which is the direct sum of the period matrices of the respective Riemann surfaces
\be
\Omega = \bigoplus_{i=1}^n \Omega_i.
\ee
The ensemble average over disconnected Riemann surfaces is then given by the following generalization of the Siegel-Weil formula
\begin{equation}\label{eq:generic-siegel-weil-disconnected}
    \langle Z_{\mathbb{T}^D}(m,{\Omega_1})\ldots Z_{\mathbb{T}^D}(m,\Omega_n)\rangle=\frac{E_{\frac D2}(\Omega)}{{\displaystyle \prod_{i=1}^n} \lr{\text{det}\text{ Im}\,\Omega_i}^\frac D2 \left|\text{det}' \ \overline{\partial}_{\Sigma_{g_i}}\right|^D}\,,
\end{equation}
where in the above $\Omega$ is no longer the period matrix of a single Riemann surface, but a direct sum of period matrices of disconnected Riemann surfaces. 

\subsection*{Double Torus Average}
We will primarily be interested in the Narain average over products of partition functions on disconnected tori boundaries. In this case the period matrices are just the modular parameters of the tori $\Omega_i =\tau_i$, and $\Omega = \operatorname{diag}(\tau_1,\ldots, \tau_n)$ is a diagonal square matrix. The averaged partition function for products of disconnected torus boundaries is then given by \eqref{eq:generic-siegel-weil-disconnected} with the appropriate diagonal matrix $\Omega$. We explicitly work out the case of two tori with identical modular parameters $\tau$ since it will be used later. Applying \eqref{eq:Eisensteinseriesarbitrarygenus} and \eqref{eq:generic-siegel-weil-disconnected} to the case $\Omega = \operatorname{diag}(\tau, \tau)$ we obtain
\be
\braket{Z_{\mathbb{T}^D}(m,\tau)Z_{\mathbb{T}^D}(m,\tau)}=\frac{1}{\operatorname{Im} (\tau)^{D}|\eta(\tau)|^{4D}}\sum_{\Gamma_0\subset H_1(\Sigma\sqcup\Sigma,\mathbb{Z})}(\det\text{Im}(\Omega_{\Gamma_0}))^{D/2}\,.
\ee
Where we have used that $\operatorname{det}' \ol{\partial} = |\eta(\tau)|^2$ on the torus. In the above $\Gamma_0\subset H_1(\Sigma\sqcup\Sigma,\mathbb{Z})$ is a sum over possible contractible cycles on the two tori. This sum contains a set of contribution that give the disconnected average $\braket{Z(\tau)}^2$, in addition to wormhole contributions.

We now explain how to see the contribution of the disconnected average $\braket{Z(\tau)}^2$ in the sum. Let $\mathcal{A}^{(1)}, \mathcal{A}^{(2)}$ be the A-cycles of the two tori, while $\mathcal{B}^{(1)}, \mathcal{B}^{(2)}$ are their B-cycles. Take the contractible cycles specified by $\Gamma_0$ to be given by independent modular transformations $\gamma_i$ of the $\mathcal{A}^{(i)}$ cycles on the two respective torii. This corresponds to a choice of $\Gamma_0$ and $\Omega_{\Gamma_0}$ given by
\be
\Gamma_0=\text{Span}_{\mathbb{Z}}\left(\gamma_1(\mathcal{A}^{(1)}),\gamma_2(\mathcal{A}^{(2)})\right)\, , \qquad \Omega_{\Gamma_0} = \begin{pmatrix}
\gamma_1 \cdot \tau & 0 \\
0 & \gamma_2 \cdot \tau 
\end{pmatrix}.
\ee
The above choice of $\Gamma_0$ is \textit{decomposable}\footnote{Intuitively, a decomposable $\Gamma_0$ amounts to picking independent contractible cycles on all surfaces\cite{Maloney:2020nni}.} and amounts to picking all possible choices of contractible cycles on the two tori independently, and we postpone an explanation of how to obtain $\Omega_{\Gamma_0}$ to the next subsection. This choice immediately gives the following contribution to the average
\begin{align}
 \braket{Z_{\mathbb{T}^D}(m,\tau)Z_{\mathbb{T}^D}(m,\tau)} &\supset \frac{1}{\operatorname{Im} (\tau)^{D}|\eta(\tau)|^{4D}} \sum_{\gamma_1, \gamma_2 \in \Gamma_{\infty}\backslash\text{SL}(2,\mathbb{Z})} \operatorname{Im}(\gamma_1 \cdot \tau)^{\frac{D}{2}} \operatorname{Im}(\gamma_2 \cdot \tau)^{\frac{D}{2}}\, .
\end{align}
By comparing to equation \eqref{eqn:TorusAverage} we notice that this is the disconnected contribution squared $\braket{Z_{\mathbb{T}^D}(m,\tau)}^2$. Wormhole contributions arise from other choices for $\Gamma_0$, an example of which is given by
\begin{equation}
\Gamma_0=\text{Span}_{\mathbb{Z}}\left(\mathcal{A}^{(1)}+\mathcal{A}^{(2)},\mathcal{B}^{(1)}-\mathcal{B}^{(2)}\right)\,.
\end{equation}
The above choice corresponds to a bulk wormhole geometry of the form $\Sigma \times [0,1]$, where $\Sigma$ is a torus. We examine this case in greater detail in Section \ref{sec:bulk-theory}. To summarize, the average over products of partition functions contains disconnected contributions which can be identified with special choices of $\Gamma_0$, alongside wormhole contributions which correspond to more non-trivial choices of contractible cycles.

\subsubsection*{Lagrangian Sublattices}
We now briefly explain Lagrangian sublattices since they appear in the Eisenstein series \eqref{eq:Eisensteinseriesarbitrarygenus}. Consider a Riemann surface $\Sigma_g$ of genus $g$. The surface has $2 g$ canonical cycles which are labelled by $\mathcal{A}_i, \mathcal{B}_i$ with $i= 1,\ldots, g$. The first homology group of the surface $H_1(\Sigma_g)$ is generated by these $2 g$ cycles, and we have
\be
H_1(\Sigma_g) \cong \underbrace{\mathbb{Z} \oplus \ldots \oplus \mathbb{Z}}_{2 g}. 
\ee
Once we have made a choice of cycles $\mathcal{A}_i, \mathcal{B}_i$ we can choose a basis of $g$ holomorphic one-forms $\omega_j$ by imposing the condition that
\be
\oint_{\mathcal{A}_i} \omega_j = \delta_{i j}. 
\ee
The period matrix of $\Sigma_g$ is then defined to be given by
\be
\oint_{\mathcal{B}_i} \omega_j = \Omega_{i j}.
\ee
To all cycles $\gamma, \gamma' \in H_1(\Sigma_g)$ we can associate an intersection number $ \langle \gamma, \gamma' \rangle $ that counts the number of times that $\gamma$ and $\gamma'$ cross. A Lagrangian sublattice is defined to be a primitive\footnote{A primitive sublattice/subgroup $\Gamma_0$ of $H_1(\Sigma)$ is defined such that given $v \in \Gamma_0$ there does not exist an integer $n$ such that $v=n u$ for some $u\in H_1(\Sigma)$. This is to exclude situations where the lattice $\Gamma_0$ is generated by $2[\mathcal{A}]$ in the case of the torus. Such a lattice would be generated by the cycle that winds twice around the $[a]$ cycle of the torus. Holographically, such a choice would require demanding the twice wound $[a]$ cycle be contractible in the bulk.} subgroup $\Gamma_0 \subset H_1(\Sigma_g)$ generated by $g$ cycles $\Tilde{\mathcal{A}}_i$ such that their mutual intersection numbers vanish $\langle \Tilde{\mathcal{A}}_i, \Tilde{\mathcal{A}}_j  \rangle=0$. That is, for a genus $g$ surface a Lagrangian sublattice is a choice of $g$ non-intersecting cycles. Once we have picked $g$ cycles to define the Lagrangian sublattice we must choose a dual pair of cycles $\Tilde{\mathcal{B}}_i$ that do not mutually intersect, but intersect the original cycles once $\langle \Tilde{\mathcal{A}_i}, \Tilde{\mathcal{B}}_j \rangle = \delta_{i j}$. We now define holomorphic differentials $\Tilde{\omega}_j$, constructed out of the original differentials $\omega_j$, such that we have
\be
\oint_{\Tilde{\mathcal{A}}_i} \Tilde{\omega}_j = \delta_{i j}.
\ee
The period matrix $\Omega_{\Gamma_0}$ associated to the Lagrangian sublattice is then defined to be
\be
\oint_{\Tilde{\mathcal{B}}_i} \tilde{\omega}_j = \left(\Omega_{\Gamma_0}\right)_{i j}. 
\ee
In the case of multiple disconnected Riemann surfaces the first homology group is given by a direct sum of the homology groups. For two surfaces of genera $g_1$ and $g_2$ we have $H_1(\Sigma_{g_1}\sqcup\Sigma_{g_2},\mathbb{Z})\cong H_1(\Sigma_{g_1},\mathbb{Z})\oplus H_1(\Sigma_{g_2},\mathbb{Z})$. A Lagrangian sublattice is then a group $\Gamma_0 \subset H_1(\Sigma_{g_1}\sqcup\Sigma_{g_2},\mathbb{Z})$ generated by $g_1+g_2$ cycles that have zero mutual intersection numbers. The period matrix associated to $\Gamma_0$ is defined in an identical way to the case of a single surface, and generalizes to any number of disconnected surfaces. The new ingredient with disconnected surfaces is that the cycles $\tilde{\mathcal{A}}_i$ that define $\Gamma_0$ can now be linear combinations of cycles on disconnected surfaces, as explained for the average over two disconnected tori earlier. To summarize, the sum over Lagrangian sublattices appearing in the Eisenstein series \eqref{eq:Eisensteinseriesarbitrarygenus} is a sum over all possible choices of non-intersecting boundary cycles.
 
\subsubsection*{\boldmath Holographic Dual: $\text{U}(1)$ Gravity}
In references \cite{Maloney:2020nni,Afkhami-Jeddi:2020ezh} a three dimensional bulk dual was proposed for the average over $\mathbb{T}^D$ Narain CFTs. It takes the form of a $\text{U}(1)^D \times \text{U}(1)^D$ Chern-Simons theory with $2D$ independent $\text{U}(1)$ gauge fields $A^i, B^i$ and action
\be \label{eqn:CSAction}
 S_{\text{CS}}=i \sum_{i=1}^{D} \int_M \lr{A^i\wedge \mathrm{d}A^i-B^i\wedge \mathrm{d} B^i}-\frac{1}{2}\int_{\partial M} d^2 z \sqrt{g} g^{a b} \left(A^i_a A^i_b + B^i_a B^i_b \right).
\ee
In the above we have included the proper boundary term with boundary metric $g_{a b}$, which corresponds to a choice of boundary Riemann surface. A choice of boundary conditions that make the variational problem well defined are given by asymptotically fixing $A_{\ol{z}}=0$ and $B_z=0$, see \cite{Datta:2021ftn,Porrati:2021sdc, Kraus:2006nb}.\footnote{It turns out that the only bulk configurations that contribute also have $A_z=B_{\ol{z}}=0$ on the boundary. This can be seen by noticing that the holonomy of $A, B$ around the contractible cycle must vanish since the connection is flat \cite{Kraus:2006nb}.}

In principle to compute the bulk partition function we should specify asymptotic boundary conditions and evaluate the Chern-Simons path integral over all bulk manifolds consistent with those boundary conditions. However, it was shown in reference \cite{Maloney:2020nni} that in the case of a single torus boundary, the bulk partition function defined by summing over only bulk handlebodies exactly reproduced the Narain average.\footnote{The statement that the bulk partition function is given by $\text{U}(1)^{2D}$ Chern-Simons theory is rather subtle. See \cite{Maloney:2020nni} for a discussion on subtleties related to whether the gauge group of the Chern-Simons theory should be U$(1)$ or $\mathbb{R}$.} A torus handlebody is a manifold of the form $M \cong D_2 \times S^1$. There is an entire family of distinct handlebodies labelled by elements of $\Gamma_{\infty}\backslash\text{SL}(2,\mathbb{Z})$, with the distinction being which asymptotic cycle of the boundary torus is contractible in the interior of the handlebody. Summing over the contribution of each handlebody we find \cite{Maloney:2020nni}
\be
\sum_{\text{handlebodies $M$}} Z_{\text{CS}}(M)=\sum_{\gamma\in\Gamma_{\infty}\backslash\text{SL}(2,\mathbb{Z})}\frac{1}{|\eta(\gamma\cdot\tau)|^{2D}},
\ee
where each term in the sum corresponds to a one-loop partition function of Chern-Simons on the handlebody specified by $\gamma$. We note that each choice of $\gamma$ picks out a boundary cycle that is contractible in the interior of the handlebody.\footnote{The sum over $\Gamma_{\infty}\backslash\text{SL}(2,\mathbb{Z})$ also appears in pure AdS$_3$ gravity \cite{Maloney:2007ud}, where it is interpreted as the sum over the family of $\text{SL}(2,\mathbb{Z})$ black holes. As an example, the term with $\gamma \cdot \tau = \tau$ is the contribution from the handlebody with the spatial circle contractible in the interior, while the term with $\gamma \cdot \tau = -1/\tau$ corresponds to the handlebody with contractible time circle. In pure AdS$_3$ gravity these handlebodies would correspond to Thermal AdS$_3$ and the BTZ black hole respectively.} In terms of the representation of $\gamma$ given above equation \eqref{eqn:TorusAverage}, the contractible cycle is given by $c \tau + d$ with $(c,d)=1$. This precisely reproduces the Narain average partition function \eqref{eqn:TorusAverage} on the torus. Similarly, it was shown in \cite{Maloney:2020nni} that summing over handlebodies with a single higher genus asymptotic boundary correctly reproduces the higher genus Narain average \eqref{eq:generic-siegel-weil}, where the contribution of each handlebody is again given by the one-loop Chern-Simons partition function. 

From the above discussion there are two key points:
\begin{itemize}
    \item To reproduce the Narain average, the sum over bulk geometries should not include every manifold with appropriate asymptotic boundary conditions.
    \item The sum over Lagrangian sublattices in the Siegel-Weil formula identifies which asymptotic boundary cycles are contractible in the bulk. 
\end{itemize}
The first point is surprising since naively every bulk geometry should be included, but we take the perspective that the boundary ensemble average will dictate what bulk geometries we ultimately include in the sum. For the second point, there are infinitely many three-manifolds with the same contractible bulk cycles, but $\text{U}(1)$ gravity seems to pick out a distinguished bulk manifold with given contractible cycles. We follow the interpretation of \cite{Maloney:2020nni} where it was proposed that this simple theory of gravity cannot resolve finer topological features of the bulk manifold other than which boundary cycles are contractible in the interior.

In the case of averaging over disconnected partition functions the bulk picture is more complicated. The average given by \eqref{eq:generic-siegel-weil-disconnected}, which we rewrite for convenience, can still be given a bulk interpretation
\begin{equation} \label{eqn:wormholeavg}
    \langle Z_{\mathbb{T}^D}(m,{\Omega_1})\ldots Z_{\mathbb{T}^D}(m,\Omega_n)\rangle=\frac{\sum_{\Gamma_0}\lr{\text{det}\text{ Im}\,\Omega_{\Gamma_0}}^{\frac{D}{2}}}{{\displaystyle \prod_{i=1}^n} \lr{\text{det}\text{ Im}\,\Omega_i}^\frac D2 \left|\text{det}' \ \overline{\partial}_{\Sigma_{g_i}}\right|^D}\,.
\end{equation}
Each Lagrangian sublattice $\Gamma_0$ in the sum again corresponds to a bulk manifold with certain asymptotic cycles contractible in the interior. This sum includes both disconnected handlebody contributions whose each connected component appeared when averaging a single partition function, as well as new wormhole contributions where the bulk manifold connects multiple disconnected boundaries.

For an independent bulk computation of \eqref{eqn:wormholeavg} we should evaluate the Chern-Simons path integral on the bulk manifold specified by $\Gamma_0$. For wormhole geometries there are again infinitely many bulk manifolds with the same contractible cycles specified by a particular $\Gamma_0$, and it's unclear which one is picked out by the Narain average. Furthermore, we are left with the problem of evaluating the Chern-Simons partition function on the given wormhole geometry, for which we know of no general results, but see comments in \cite{Maloney:2020nni}. We will forgo these issues and assume the bulk theory is directly defined by \eqref{eqn:wormholeavg}.

\subsubsection*{Divergence of the Ensemble Average}
The average of the $\mathbb{T}^D$ partition function only converges when $D - 1 > g$, where $g$ is the genus of the boundary. In the case of multiple disconnected boundaries this generalizes to 
\be
\langle Z_{\mathbb{T}^D}(m,{\Omega_1})\ldots Z_{\mathbb{T}^D}(m,\Omega_n)\rangle < \infty, \qquad D - 1 > \sum_{i=1}^n g_i.
\ee
We note that the Eisenstein series \eqref{eq:Eisensteinseriesarbitrarygenus} precisely diverges when $D-1 \leq g$. When the average over moduli space such as \eqref{eqn:SiegelWeilTorus} diverges we cannot, strictly speaking, claim that the average is given by an Eisenstein series since the average is not well defined. Nevertheless, we will define the divergent ensemble average for $D-1 \leq g$ to be given by the standard Eisenstein series \eqref{eq:Eisensteinseriesarbitrarygenus}, where each term in the sum gives a finite contribution but the full sum does not converge. From a bulk perspective each term in the sum may be associated with a finite contribution from a given bulk geometry, but each geometry is not sufficiently suppressed to make the sum convergent.

From a boundary perspective this is slightly puzzling since for every member of the ensemble the partition function is well defined, but when the average is performed a new divergence appears. This can be understood as follows, the Narain CFTs are defined by a target $T^D$ torus. There are points in moduli space where the target torus decompactifies and we get infinitely many light states that give a divergence to the thermal partition function. The simplest example of this is $D=1$ where the target is a circle $S^1$. As the radius of the circle goes to infinity we decompactify to target $\mathbb{R}$, and the momentum modes with zero winding become light. The measure for the ensemble average suppresses these dangerous corners of moduli space, but when averaging sufficiently many products of partition functions the growth of light states near decompactification points eventually wins out over the measure suppression and gives a divergent answer. 

\subsection{Symmetric orbifold CFTs}\label{subsec:permutation-orbifolds}

\subsubsection*{Orbifolds in general}

Before discussing the symmetric orbifold, let us recall some basic facts of orbifold conformal field theories in general. Let $M$ be a smooth manifold with a discrete symmetry group $G$. We can define the quotient space $M/G$ by identifying points
\begin{equation}
p\sim g\cdot p\,,
\end{equation}
where $p$ is a point in $M$ and $g\in G$. If the action of $G$ has fixed points, then a generic point of $M/G$ will not locally look like $\mathbb{R}^n$, but will have conical singularities. Such a space $M/G$ is known as an \textit{orbifold}.

Now, consider a conformal field theory which is a sigma-model with target space $M$, i.e. a CFT whose fundamental fields are maps $\Phi:\Sigma\to M$ from some two-dimensional surface $\Sigma$ into $M$. By abuse of notation, we will also refer to this CFT as $M$. The CFT inherits the symmetry group $G$ of the target space $M$, and we can define a new conformal field theory with target space $M/G$ by starting with the CFT $M$ and `gauging' the discrete symmetry $G$ of the original theory. Specifically, we demand that the field configurations $\Phi$ and $g\cdot\Phi$ are physically equivalent, and in the path integral only integrate over equivalence classes $[\Phi]$ of field configurations under the action of $G$.

\begin{figure}
\centering
\ifdraft
\else
\begin{tikzpicture}[scale = 1.2]
\draw[thick, red] (2.5,0) [partial ellipse = 90:115 : 0.6 and 1.1];
\draw[thick, red, dashed] (2.5,0) [partial ellipse = 115:245 : 0.6 and 1.1];
\draw[thick, red] (2.5,0) [partial ellipse = 245:270: 0.6 and 1.1];
\draw[thick] (0,-1) -- (5,-1);
\draw[thick] (0,1) -- (5,1);
\draw[thick] (0,0) ellipse (0.5 and 1);
\draw[thick] (5,0) [partial ellipse = -90:90:0.5 and 1];
\draw[thick, dashed] (5,0) [partial ellipse = 90:270:0.5 and 1];
\draw[thick, red] (2.5,0) [partial ellipse = 0:90 : 0.6 and 1.1];
\draw[thick, red, -latex] (2.5,0) [partial ellipse = 270:350: 0.6 and 1.1];
\node[right] at (3.1,0.3) {$\Phi(x)$};
\node[right] at (3.1,-0.5) {$g\cdot\Phi(x)$};
\end{tikzpicture}
\fi
\caption{An orbifold CFT $M/G$ allows for twisted boundary conditions of fundamental fields around non-contractible loops.}
\label{fig:twisted-bc-orbifold}
\end{figure}
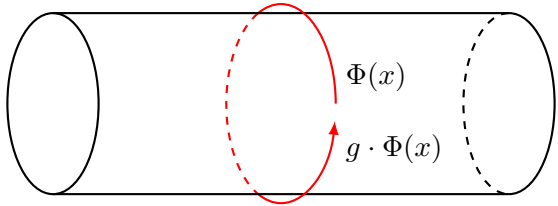

Naively, one can avoid the over-counting by picking a representative of $[\Phi]$ in the path integral and only integrating over those representatives. This would effectively result in dividing the path integral of the CFT $M$ by an overall factor of $1/|G|$. However, picking a unique representative of $[\Phi]$ is only possible locally, and globally one needs to be more careful. If $\Sigma$ has a non-contractible loop based at a reference point $x\in\Sigma$, then it is possible that $\Phi(x)$ obtains a monodromy $g$ upon being transported around this loop (see Figure \ref{fig:twisted-bc-orbifold}). This is a perfectly allowed field configuration, since $\Phi(x)$ and $g\cdot\Phi(x)$ represent the same point in the target space orbifold $M/G$, yet such a configuration does not allow one to smoothly choose a unique representative of the equivalence class $[\Phi]$ at every point.

Twisted boundary conditions like those in Figure \ref{fig:twisted-bc-orbifold} are characterized by assigning a monodromy $g$ to each loop $\gamma$ based at a point $x$, such that transporting the field $\Phi$ around $\gamma$ returns $g(\gamma)\cdot\Phi(x)$. The assignment of an element $g\in G$ to each loop satisfies the following properties:
\begin{itemize}
    \item If $\gamma$ is the trivial loop, then $g(\gamma)=\text{id}$.

    \item The monodromy $g(\gamma)$ depends only on the homotopy class of the loop $\gamma$, i.e. $g(\gamma)$ is invariant under smooth deformations of the loop $\gamma$.

    \item Given two loops $\gamma_1,\gamma_2$ based at the same point, $g(\gamma_1\circ\gamma_2)=g(\gamma_1)g(\gamma_2)$, where $\gamma_1\circ\gamma_2$ is the composition of the loops $\gamma_1$ and $\gamma_2$.
\end{itemize}

The above three properties are equivalent to specifying a group homomorphism $g:\pi_1(\Sigma)\to G$.\footnote{We assume that $\Sigma$ is connected, so that $\pi_1(\Sigma)$ is independent of the chosen basepoint.} Each such twisted boundary condition should in principle appear in the path integral of $M/G$, and so the path integral should sum over them. The result is that the path integral of $M/G$ on $\Sigma$ can be expressed as
\begin{equation}
Z=\frac{1}{|G|}\sum_{g:\pi_1(\Sigma)\to G}\int_{g}\mathcal{D}\Phi\,e^{-S[\Phi]}\,,
\end{equation}
where the subscript $g$ in the path integral instructs us to integrate over field configurations which obey the twisted boundary conditions. The factor of $1/|G|$ is again included so that physically equivalent fields are not overcounted.

\subsubsection*{Permutation orbifolds}

Consider a 2D CFT $X$ whose fundamental fields are labeled collectively by $\Phi$. If $X$ has central charge $c$, we can construct a CFT with arbitrarily large central charge $Nc$ by considering the $N^{\text{th}}$ tensor power of $X$
\begin{equation}
X^{\otimes N}:=\underbrace{X\otimes\cdots\otimes X}_{N\text{ times}}\, .
\end{equation}
The theory $X^{\otimes N}$ contains, as its fundamental fields, $N$-tuples $\boldmath{\Phi}=(\Phi_{1},\ldots,\Phi_{N})$ of fundamental fields of $X$. Since $X^{\otimes N}$ is constructed from $N$ copies of an identical seed theory the Hilbert space is an $N$ times tensor product of the original Hilbert space, and there is an obvious symmetry of permuting the individual copies. Let $\Omega\subset S_N$ be a permutation group acting on the letters $\{1,\ldots,N\}$. Then for any permutation $\pi\in\Omega$, there is a natural action on the fundamental fields of $X^{\otimes N}$ given by
\begin{equation}\label{eq:permutations}
\pi\cdot\boldsymbol{\Phi}=(\Phi_{\pi(1)},\ldots,\Phi_{\pi(N)})\,.
\end{equation}
We can define a new CFT, known as the permutation orbifold or $X\wr \Omega$, by gauging the action of $\Omega$ on $X^{\otimes N}$. That is, we define the orbifold theory
\begin{equation}
X\wr\Omega:=X^{\otimes N}/\Omega\,.
\end{equation}
The use of the wreath product symbol $\wr$ to denote permutation orbifolds will be clarified later when we discuss Chern-Simons theories with permutation symmetry.

A special case of a permutation orbifold comes from taking the permutation group $\Omega$ to be the full symmetric group $S_N$. In this case, the permutation orbifold $X\wr S_N$ is called the \textit{symmetric orbifold} and is often denoted by $\text{Sym}^N(X)$. We will mostly focus on symmetric orbifold theories in this paper, but most statements we make generalize in a straightfoward manner to generic permutation orbifolds.

\subsubsection*{Partition functions}

Let $\Sigma$ be a Riemann surface of genus one. Its cycles are denoted by $A$ and $B$, and given a point $z\in\Sigma$, we let $A\cdot z$ denote the operation of transporting $z$ along a cycle homotopic to $A$. Within the permutation orbifold, as with any orbifold, we impose the gauging of the discrete group $\Omega$ by allowing the fundamental fields $\boldsymbol{\Phi}$ to have non-trivial monodromies when transported around non-contractible loops on $\Sigma$. That is, given permutations $\pi_A$ and $\pi_B$, we allow the twisted boundary conditions
\begin{equation}
\begin{split}
\boldsymbol{\Phi}(A\cdot z)&=\pi_A\cdot\boldsymbol{\Phi}(z)\,,\\
\boldsymbol{\Phi}(B\cdot z)&=\pi_B\cdot\boldsymbol{\Phi}(z)\,.\\
\end{split}
\end{equation}
Now, given that $A\cdot B\cdot A^{-1}\cdot B^{-1}$ is a contractible cycle on the torus (see Figure \ref{fig:ABAB-contractible}), we should not pick up a monodromy when traversing it. Thus, we have
\begin{equation}
\boldsymbol{\Phi}(z)=\boldsymbol{\Phi}(A\cdot B\cdot A^{-1}\cdot B^{-1}\cdot z)=(\pi_A\pi_B\pi_A^{-1}\pi_B^{-1})\cdot\boldsymbol{\Phi}(z)\,,
\end{equation}
which is only consistent if
\begin{equation}
\pi_A\pi_B=\pi_B\pi_A\,,
\end{equation}
i.e. if the permutations $\pi_A$ and $\pi_B$ commute. This is precisely the requirement that the permutations $\pi_A,\pi_B$ define a group homomorphism $\pi_1(\Sigma)\cong\mathbb{Z}\oplus\mathbb{Z}\to S_N$. Now, from the general theory of orbifolds, we know that in the path integral of $X\wr\Omega$ on $\Sigma$ we are required to sum over all twisted boundary conditions $\pi_A,\pi_B$ which commute. That is, the partition function is given by
\begin{equation}\label{eq:permtuation-partition-function}
Z_\Omega(\Sigma)=\frac{1}{|\Omega|}\sum_{[\pi_A,\pi_B]=0}Z_{\pi_A,\pi_B}(\Sigma)\,,
\end{equation}
where $Z_{\pi_A,\pi_B}(\Sigma)$ is the path integral of $X^{\otimes N}$ on $\Sigma$ with the twisted boundary conditions imposed by $\pi_A,\pi_B$. For a general choice of $\pi_A,\pi_B$ the fields are not single valued, they permute amongst themselves as we travel around different cycles of the torus.

\begin{figure}
\centering
\ifdraft
\else
\begin{tikzpicture}[scale = 1.5]
\begin{scope}
\draw[ultra thick] (0,0) -- (0,2) -- (2,2) -- (2,0) -- (-0.018,0);
\draw[thick, red] (0,0) -- (0,2) -- (2,2) -- (2,0) -- (-0.01,0);
\draw[thick, red, -latex] (0,0.8) -- (0,1.2);
\draw[thick, red, -latex] (0.8,2) -- (1.2,2);
\draw[thick, red, latex-] (2,0.8) -- (2,1.2);
\draw[thick, red, -latex] (1.2,0) -- (0.8,0);
\node[left] at (-0.1,1) {$A$};
\node[below] at (1,-0.1) {$B$};
\node[above] at (1,2.1) {$B$};
\node[right] at (2.1,1) {$A$};
\end{scope}
\draw[thick, -latex] (3,1) -- (4,1);
\begin{scope}[xshift = 5cm]
\draw[ultra thick] (0,0) -- (0,2) -- (2,2) -- (2,0) -- (-0.018,0);
\draw[thick, red] (0,0) to[out = 90, in = 135] (1,1);
\draw[thick, red] (1,1) to[out = -45, in = 0] (0,0);
\draw[thick, red, -latex] (0.99+0.0775,1.01-0.1) -- (1.0+0.0775,0.99-0.1);
\node[left] at (-0.1,1) {$A$};
\node[below] at (1,-0.1) {$B$};
\node[above] at (1,2.1) {$B$};
\node[right] at (2.1,1) {$A$};
\end{scope}
\end{tikzpicture}
\fi
\caption{The loop $A\cdot B\cdot A^{-1}\cdot B^{-1}$ on a torus is contractible. Thus, twisted boundary conditions $\pi_A,\pi_B$ must satisfy $\pi_A\pi_B=\pi_B\pi_A$.}
\label{fig:ABAB-contractible}
\end{figure}
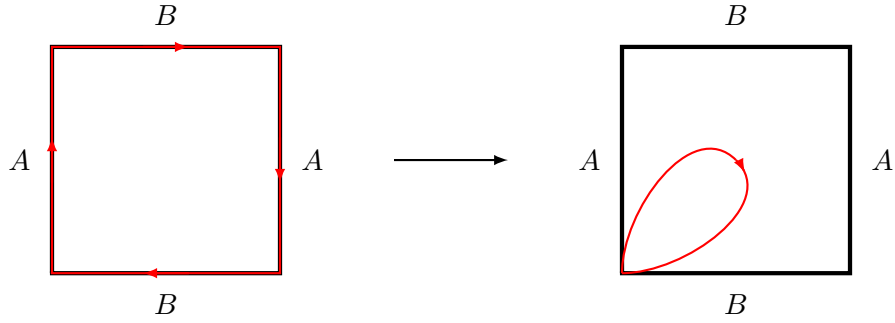

Let us now specialize to the case $\Omega=S_N$. In order to evaluate the individual summands in equation \eqref{eq:permtuation-partition-function}, we use a standard trick in the theory of orbifolds by considering a covering space $\widetilde{\Sigma}$ on which the fields become single-valued \cite{Dixon:1986qv}. The space $\widetilde{\Sigma}$  is constructed by taking $N$ copies of $\Sigma$, letting the field $\Phi_i$ live on the $i^{\text{th}}$ copy of $\Sigma$, and `stitching' together the copies of $\Sigma$ via the twisted boundary conditions $\pi_A,\pi_B$.

The resulting surface $\widetilde{\Sigma}$ is an $N$-fold covering space of $\Sigma$ in the topological sense (this process is easy to visualize in the case of a one-dimensional theory on a circle, see Figure \ref{fig:circle-covering}). The partition function with twisted boundary conditions reduces to a partition function of the seed theory $X$ on $\widetilde{\Sigma}$, i.e.
\begin{equation}
Z_{\pi_A,\pi_B}(\Sigma)=Z(\widetilde{\Sigma})\,,
\end{equation}
where $Z$ denotes the partition function of $X$. Therefore, we naively write
\begin{equation}
Z_{S_N}(\Sigma)\stackrel{?}{=}\frac{1}{N!}\sum_{\widetilde{\Sigma}\to\Sigma}Z(\widetilde{\Sigma})\,,
\end{equation}
where $Z(\widetilde{\Sigma})$ is the partition function of $X$ on $\widetilde{\Sigma}$.

\begin{figure}
\centering
\ifdraft
\else
\begin{tikzpicture}[scale = 0.7]
\begin{scope}
\draw[very thick] (0,-2) ellipse (3 and 0.9);
\draw[line width = 0.2cm, -latex, white] (0,1) -- (0,-2);
\draw[very thick, -latex] (0,1) -- (0,-2);
\draw[line width = 0.2cm, white] (0,1) [partial ellipse = -60:240:3 and 0.9];
\draw[line width = 0.2cm, white] (-1.51,0.2225) to[out = -10, in = 190] (1.51,1.2225);
\draw[very thick] (0,1) [partial ellipse = -60:240:3 and 0.9];
\draw[very thick] (-1.51,0.2225) to[out = -10, in = 190] (1.51,1.2225);
\draw[line width = 0.2cm, white] (0,2) [partial ellipse = -60:240:3 and 0.9];
\draw[line width = 0.2cm, white] (-1.51,1.2225) to[out = -10, in = 190] (1.51,0.2225);
\draw[very thick] (0,2) [partial ellipse = -60:240:3 and 0.9];
\draw[very thick] (-1.51,1.2225) to[out = -10, in = 190] (1.51,0.2225);
\draw[line width = 0.2cm, white] (0,3) ellipse (3 and 0.9);
\draw[very thick] (0,3) ellipse (3 and 0.9);
\node[right] at (3.2,-2) {base};
\node[right] at (3.2,1) {$1$};
\node[right] at (3.2,2) {$2$};
\node[right] at (3.2,3) {$3$};
\end{scope}
\begin{scope}[xshift = 9cm]
\draw[very thick] (0,-2) ellipse (3 and 0.9);
\draw[line width = 0.2cm, -latex, white] (0,1) -- (0,-2);
\draw[very thick, -latex] (0,1) -- (0,-2);
\draw[line width = 0.2cm, white] (0,1) [partial ellipse = -60:240:3 and 0.9];
\draw[line width = 0.2cm, white] (-1.51,0.2225) to[out = -10, in = 190] (1.51,1.2225);
\draw[very thick] (0,1) [partial ellipse = -60:240:3 and 0.9];
\draw[very thick] (-1.51,0.2225) to[out = -10, in = 190] (1.51,1.2225);
\draw[line width = 0.2cm, white] (0,2) [partial ellipse = -60:240:3 and 0.9];
\draw[line width = 0.2cm, white] (-1.51,1.2225) to[out = -10, in = 190] (1.51,2.2225);
\draw[very thick] (0,2) [partial ellipse = -60:240:3 and 0.9];
\draw[very thick] (-1.51,1.2225) to[out = -10, in = 190] (1.51,2.2225);
\draw[line width = 0.2cm, white] (0,3) [partial ellipse = -60:240:3 and 0.9];
\draw[line width = 0.2cm, white] (-1.51,2.2225) to[out = -10, in = 190] (1.51,0.2225);
\draw[very thick] (0,3) [partial ellipse = -60:240:3 and 0.9];
\draw[very thick] (-1.51,2.2225) to[out = -10, in = 190] (1.51,0.2225);
\node[right] at (3.2,-2) {base};
\node[right] at (3.2,1) {$1$};
\node[right] at (3.2,2) {$2$};
\node[right] at (3.2,3) {$3$};
\end{scope}
\end{tikzpicture}
\fi
\caption{Twisted boundary conditions on the circle as 3-fold covering spaces. Left: the fields $\{\Phi_1,\Phi_2,\Phi_3\}$ satisfy twisted boundary conditions $\Phi_1(2\pi)=\Phi_2(0)$, while $\Phi_3$ is single-valued. Right: The fields satisfy boundary conditions $\Phi_1(2\pi)=\Phi_2(0)$, $\Phi_2(2\pi)=\Phi_3(0)$, and $\Phi_3(2\pi)=\Phi_1(0)$.}
\label{fig:circle-covering}
\end{figure}
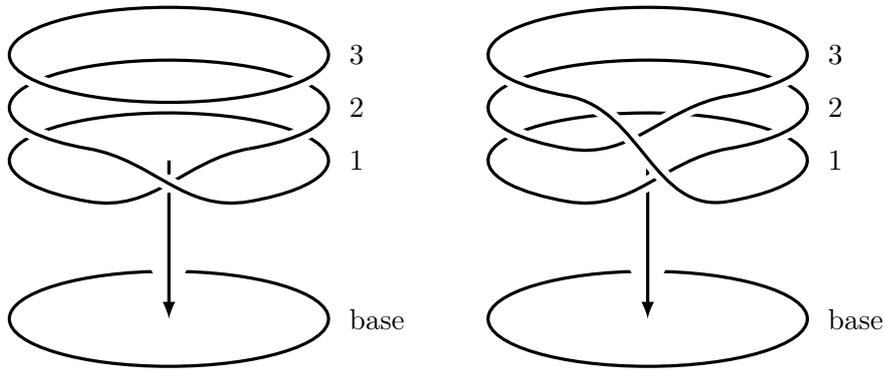

The above equation is actually not quite right. This is because not all pairs of boundary conditions $(\pi_A,\pi_B)$ give topologically inequivalent covering spaces. Indeed, if we define $(\pi'_A,\pi'_B)=(\pi\pi_A\pi^{-1},\pi\pi_B\pi^{-1})$, the resulting covering space is the same, since the effect of conjugating by $\pi$ it simply permutes the sheets of $\widetilde{\Sigma}\to\Sigma$ (the copies of $\Sigma$), which is a homeomorphism. If we let
\begin{equation}
\mathcal{O}_{\pi_A,\pi_B}=\left\{(\pi\pi_A\pi^{-1},\pi\pi_B\pi^{-1})|\pi\in S_N\right\}\,,
\end{equation}
where we do not double-count equal pairs of permutations, then each covering space $\widetilde{\Sigma}\to\Sigma$ occurs precisely $|\mathcal{O}_{\pi_A,\pi_B}|$ times in the sum \eqref{eq:permtuation-partition-function}.\footnote{The set $\mathcal{O}_{\pi_A,\pi_B}$ is the orbit set of the element $(\pi_A,\pi_B)\in\text{Hom}(\pi_1(\Sigma),S_N)$ under the $S_N$ action which acts as conjugation. Covering spaces are in one-to-one correspondence with the coset $\text{Hom}(\pi_1(\Sigma),S_N)/S_N$ of this action. Readers familiar with mathematical aspects of gauge theory will recognize this coset as the space of principal $S_N$ bundles over $\Sigma$, which is just a fancy word for a covering space.} We define the `symmetry factor' of a covering space $\widetilde{\Sigma}\to \Sigma$ to be the quotient $|\text{Aut}(\widetilde{\Sigma}\to\Sigma)|=N!/|\mathcal{O}_{\pi_A,\pi_B}|$. Thus,
\begin{equation}\label{eq:symmetric-orbifold-partition-general}
Z_{S_N}(\Sigma)=\sum_{\widetilde{\Sigma}\to\Sigma}\frac{Z(\widetilde{\Sigma})}{|\text{Aut}(\widetilde{\Sigma}\to\Sigma)|}\,.
\end{equation}
The factor $|\text{Aut}(\widetilde{\Sigma}\to\Sigma)|$ is precisely the degree of the group of \textit{deck transformations}: homeomorphisms of the covering space $\widetilde{\Sigma}$ which preserve the projection $\widetilde{\Sigma}\to\Sigma$.\footnote{Since a covering space can be considered a homomorphism $\phi:\pi_1(\Sigma)\to S_N$, we can equivalently define a deck transformation to be an automorphism $\psi:S_N\to S_N$ which leaves $\phi$ invariant, i.e. for which $\psi\circ\phi=\phi$. Group theoretically, conjugation by elements of $S_N$ defines an action on $\text{Hom}(\pi_1(\Sigma),S_N)$. The group of deck transformations of a covering space $\phi\in\text{Hom}(\pi_1(\Sigma),S_N)$ is the stabilizer $\text{Stab}(\phi)$ under the $S_N$ action. By the orbit-stabilizer theorem, we have $|\text{Stab}(\phi)||\mathcal{O}(\phi)|=|S_N|=N!$, or $N!/|\mathcal{O}(\phi)|=|\text{Stab}(\phi)|=|\text{Aut}(\widetilde{\Sigma}\to\Sigma)|$. Note that we do not require $\Sigma$ to be a torus, and this statement works for any topological space $\Sigma$.}

For a base space which is a torus, it is a topological fact that the covering spaces $\widetilde{\Sigma}\to\Sigma$ considered above are always given by disjoint unions of tori. That is, each covering space we want to consider is given by
\begin{equation}
\widetilde{\Sigma}=\Sigma_1\sqcup\cdots\sqcup\Sigma_n\to\Sigma\,,
\end{equation}
where $\Sigma_i$ is a torus with modular parameter $\tau_i$ that is not necessarily equal to the initial modular parameter. Since the partition function of a CFT on a disjoint union of spaces is the product of the partition functions, we have, for each covering space $\widetilde{\Sigma}\to\Sigma$,
\begin{equation}
Z(\widetilde{\Sigma})=\prod_{i=1}^{n}Z(\Sigma_i)\,.
\end{equation}
Thus, in order to calculate the partition function $Z_{S_N}(\Sigma)$ of the symmetric orbifold theory $X\wr S_N$ on a torus $\Sigma$, one only needs to know the generic torus partition function $Z(\Sigma)$ for the seed theory $X$ -- all of the other data is contained in the combinatorics of the covering spaces. This simplification does not occur for partition functions of $X\wr S_N$ on higher-genus surfaces: as we will see later, if $\Sigma$ has genus $g$, calculating the partition function $Z_{S_N}(\Sigma)$ requires knowing the partition functions of the seed theory $X$ on surfaces of many different genera.

\subsubsection*{\boldmath Example: \texorpdfstring{$N=2$}{N=2}}

For $N=2$, the above discussion can be made very concrete. The only two permutations in $S_2$ are the identity $e$ and the two-cycle $\pi$. Since $S_2$ is abelian, all permutations commute among each other, and we can immediately write down the sum \eqref{eq:permtuation-partition-function} as
\begin{equation}\label{eq:s2-partition-function-twist}
Z_2(\Sigma)=\frac{1}{2}\left(Z_{e,e}(\Sigma)+Z_{\pi,e}(\Sigma)+Z_{e,\pi}(\Sigma)+Z_{\pi,\pi}(\Sigma)\right)\,.
\end{equation}
If we realize the torus $\Sigma$ as a parallelogram in the complex plane $\mathbb{C}/\{m+n\tau\}$, we can choose the $A$-cycle to act as $A\cdot z=z+\tau$ and the $B$-cycle to act as $B\cdot z=z+1$. Then $Z_{\pi,e}$ is the partition function of fields $(\Phi_1,\Phi_2)$ with $\Phi_{1}(z+\tau)=\Phi_2(z)$. This is single valued on the torus obtained by making the $A$-cycle  twice as long, i.e.
\begin{equation}
\mathbb{C}/\{m+2n\tau\}\,.
\end{equation}
This is a torus with modular parameter $2\tau$, and so
\begin{equation}
Z_{\pi,e}(\tau)=Z(2\tau)\,.
\end{equation}
Similarly,
\begin{equation}
Z_{e,\pi}(\tau)=Z(\tfrac{\tau}{2})\,,\quad Z_{\pi,\pi}(\tau)=Z(\tfrac{\tau+1}{2})\,.
\end{equation}
All of the above covering tori are depicted in Figure \ref{fig:n2-coverings}. The partition function $Z_{e,e}(\tau)$ is just the partition function of $X^{\otimes 2}$ with no twisted boundary conditions, and so
\begin{equation}
Z_{e,e}(\tau)=Z(\tau)^2\,.
\end{equation}

\begin{figure}
\centering
\ifdraft
\else
\begin{tikzpicture}[scale = 0.7]
\begin{scope}[xshift = 5.5cm]
\begin{scope}
\draw[thick] (0,0) -- (1,1) -- (3,1) -- (2,0) -- (0,0);
\draw[thick] (1,1) -- (2,2) -- (4,2) -- (3,1) -- (1,1);
\draw[thick] (2,0) -- (3,1) -- (5,1) -- (4,0) -- (2,0);
\draw[thick] (3,1) -- (4,2) -- (6,2) -- (5,1) -- (3,1);
\end{scope}
\begin{scope}[xshift = 2cm, yshift = 2cm]
\draw[thick] (0,0) -- (1,1) -- (3,1) -- (2,0) -- (0,0);
\draw[thick] (2,0) -- (3,1) -- (5,1) -- (4,0) -- (2,0);
\end{scope}
\begin{scope}[xshift = 4cm, yshift = 0cm]
\draw[thick] (0,0) -- (1,1) -- (3,1) -- (2,0) -- (0,0);
\draw[thick] (1,1) -- (2,2) -- (4,2) -- (3,1) -- (1,1);
\end{scope}
\begin{scope}[xshift = 6cm, yshift = 2cm]
\draw[thick] (0,0) -- (1,1) -- (3,1) -- (2,0) -- (0,0);
\end{scope}
\node at (1.5,0.5) {\Large $1$};
\node at (3.5,0.5) {\Large $2$};
\node at (5.5,0.5) {\Large $1$};
\node at (2.5,1.5) {\Large $1$};
\node at (4.5,1.5) {\Large $2$};
\node at (6.5,1.5) {\Large $1$};
\node at (3.5,2.5) {\Large $1$};
\node at (5.5,2.5) {\Large $2$};
\node at (7.5,2.5) {\Large $1$};
\draw[ultra thick, red] (0,0) -- (1,1) -- (5,1) -- (4,0) -- (0,0);
\end{scope}
\begin{scope}[xshift = 0cm, yshift = -5cm]
\begin{scope}
\draw[thick] (0,0) -- (1,1) -- (3,1) -- (2,0) -- (0,0);
\draw[thick] (1,1) -- (2,2) -- (4,2) -- (3,1) -- (1,1);
\draw[thick] (2,0) -- (3,1) -- (5,1) -- (4,0) -- (2,0);
\draw[thick] (3,1) -- (4,2) -- (6,2) -- (5,1) -- (3,1);
\end{scope}
\begin{scope}[xshift = 2cm, yshift = 2cm]
\draw[thick] (0,0) -- (1,1) -- (3,1) -- (2,0) -- (0,0);
\draw[thick] (2,0) -- (3,1) -- (5,1) -- (4,0) -- (2,0);
\end{scope}
\begin{scope}[xshift = 4cm, yshift = 0cm]
\draw[thick] (0,0) -- (1,1) -- (3,1) -- (2,0) -- (0,0);
\draw[thick] (1,1) -- (2,2) -- (4,2) -- (3,1) -- (1,1);
\end{scope}
\begin{scope}[xshift = 6cm, yshift = 2cm]
\draw[thick] (0,0) -- (1,1) -- (3,1) -- (2,0) -- (0,0);
\end{scope}
\node at (1.5,0.5) {\Large $1$};
\node at (3.5,0.5) {\Large $1$};
\node at (5.5,0.5) {\Large $1$};
\node at (2.5,1.5) {\Large $2$};
\node at (4.5,1.5) {\Large $2$};
\node at (6.5,1.5) {\Large $2$};
\node at (3.5,2.5) {\Large $1$};
\node at (5.5,2.5) {\Large $1$};
\node at (7.5,2.5) {\Large $1$};
\draw[ultra thick, red] (0,0) -- (2,2) -- (4,2) -- (2,0) -- (0,0);
\end{scope}
\begin{scope}[xshift = 9cm, yshift = -5cm]
\begin{scope}
\draw[thick] (0,0) -- (1,1) -- (3,1) -- (2,0) -- (0,0);
\draw[thick] (1,1) -- (2,2) -- (4,2) -- (3,1) -- (1,1);
\draw[thick] (2,0) -- (3,1) -- (5,1) -- (4,0) -- (2,0);
\draw[thick] (3,1) -- (4,2) -- (6,2) -- (5,1) -- (3,1);
\end{scope}
\begin{scope}[xshift = 2cm, yshift = 2cm]
\draw[thick] (0,0) -- (1,1) -- (3,1) -- (2,0) -- (0,0);
\draw[thick] (2,0) -- (3,1) -- (5,1) -- (4,0) -- (2,0);
\end{scope}
\begin{scope}[xshift = 4cm, yshift = 0cm]
\draw[thick] (0,0) -- (1,1) -- (3,1) -- (2,0) -- (0,0);
\draw[thick] (1,1) -- (2,2) -- (4,2) -- (3,1) -- (1,1);
\end{scope}
\begin{scope}[xshift = 6cm, yshift = 2cm]
\draw[thick] (0,0) -- (1,1) -- (3,1) -- (2,0) -- (0,0);
\end{scope}
\draw[ultra thick, red] (1,1) -- (2,0) -- (5,1) -- (4,2) -- (1,1);
\node at (1.5,0.5) {\Large $1$};
\node at (3.5,0.5) {\Large $2$};
\node at (5.5,0.5) {\Large $1$};
\node at (2.5,1.5) {\Large $2$};
\node at (4.5,1.5) {\Large $1$};
\node at (6.5,1.5) {\Large $2$};
\node at (3.5,2.5) {\Large $1$};
\node at (5.5,2.5) {\Large $2$};
\node at (7.5,2.5) {\Large $1$};
\end{scope}
\end{tikzpicture}
\fi
\caption{The connected covering spaces for the torus $N=2$ symmetric orbifold. Individual cells represent the base torus (with modular parameter $\tau$), and the numbers label which copy of the seed theory lives on which sheet of the base torus. The permutations $\pi_A,\pi_B$ prescribe how to stitch together the copies of the seed theory onto the covering space. The covering space itself is the fundamental domain for which the arrangement of labels $1,2$ is periodic (shown in red). The fundamental domains in the above examples have modular parameter $\tau/2$, $2\tau$, and $(\tau+1)/(1-\tau)\sim (\tau+1)/2$, respectively.}
\label{fig:n2-coverings}
\end{figure}
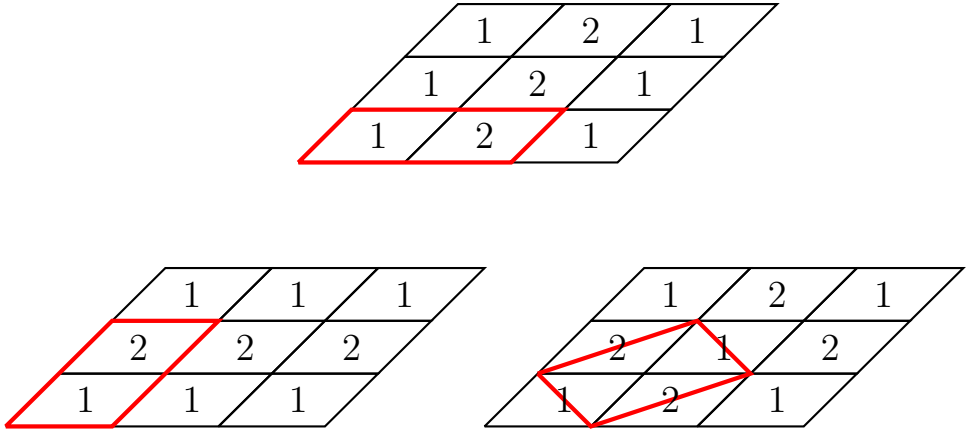

Putting the above discussion together, we can write the full $X\wr S_2$ symmetric orbifold partition function \eqref{eq:s2-partition-function-twist} as
\begin{equation}
Z_2(\tau)=\frac{1}{2}Z(\tau)^2+\frac{1}{2}Z(\tfrac{\tau}{2})+\frac{1}{2}Z(2\tau)+\frac{1}{2}Z(\tfrac{\tau+1}{2})\,.
\end{equation}
\begin{figure}
\centering
\ifdraft
\else
\begin{tikzpicture}[scale = 1.45]
\begin{scope}[scale = 0.3]
\draw[thick] (0,0) [partial ellipse = 0:360:2 and 3];
\draw[thick] (0.2,0) [partial ellipse = 100:260:0.5 and 1];
\draw[thick] (-0.2,0) [partial ellipse = -70:70:0.4 and 0.9];
\node[above] at (0,3) {$\tau$};
\end{scope}
\begin{scope}[xshift = 1.5cm, scale = 0.3]
\draw[thick] (0,0) [partial ellipse = 0:360:2 and 3];
\draw[thick] (0.2,0) [partial ellipse = 100:260:0.5 and 1];
\draw[thick] (-0.2,0) [partial ellipse = -70:70:0.4 and 0.9];
\node[above] at (0,3) {$\tau$};
\end{scope}
\node at (2.5,0) {$+$};
\begin{scope}[xshift = 3.5cm, scale = 0.3]
\draw[thick] (0,0) [partial ellipse = 0:360:2 and 3];
\draw[thick] (0.2,0) [partial ellipse = 100:260:0.5 and 1];
\draw[thick] (-0.2,0) [partial ellipse = -70:70:0.4 and 0.9];
\node[above] at (0,3) {$2\tau$};
\end{scope}
\node at (4.5,0) {$+$};
\begin{scope}[xshift = 5.5cm, scale = 0.3]
\draw[thick] (0,0) [partial ellipse = 0:360:2 and 3];
\draw[thick] (0.2,0) [partial ellipse = 100:260:0.5 and 1];
\draw[thick] (-0.2,0) [partial ellipse = -70:70:0.4 and 0.9];
\node[above] at (0,3) {$\tau/2$};
\end{scope}
\node at (6.5,0) {$+$};
\begin{scope}[xshift = 7.5cm, scale = 0.3]
\draw[thick] (0,0) [partial ellipse = 0:360:2 and 3];
\draw[thick] (0.2,0) [partial ellipse = 100:260:0.5 and 1];
\draw[thick] (-0.2,0) [partial ellipse = -70:70:0.4 and 0.9];
\node[above] at (0,3) {$(\tau+1)/2$};
\end{scope}
\end{tikzpicture}
\fi
\caption{All covering space geometries contributing to the $S_2$ partition function. The Riemann-Hurwitz formula guarantees that these covering spaces are also tori.}
\label{fig:n=2-boundary-covering}
\end{figure}
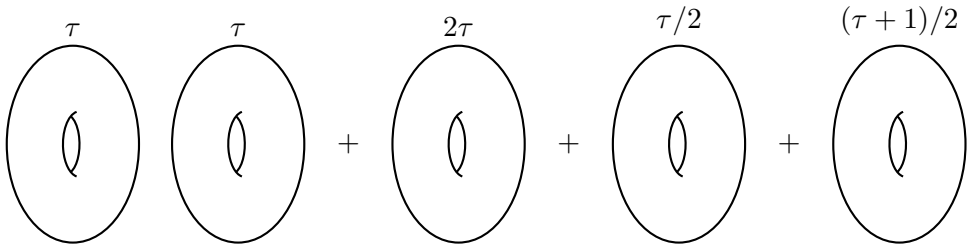 
Each term in this sum can be seen as the partition function of the seed theory $X$ evaluated on a covering space $\widetilde\Sigma \to\Sigma$, see Figure \ref{fig:n2-coverings}. A similar, more complicated discussion can be done for $N=3$, which requires 18 pairs of commuting permutations in $S_3$. The end result is
\begin{equation} \label{eqn:N=3Z}
\begin{split}
Z_3(\tau)=&\frac{1}{6}Z(\tau)^3+\frac{1}{2}Z(\tau)Z(2\tau)+\frac{1}{2}Z(\tau)Z(\tfrac{\tau}{2})+\frac{1}{2}Z(\tau)Z(\tfrac{\tau+1}{2})\\
&\hspace{3cm}+\frac{1}{3}Z(3\tau)+\frac{1}{3}Z(\tfrac{\tau}{3})+\frac{1}{3}Z(\tfrac{\tau+1}{3})+\frac{1}{3}Z(\tfrac{\tau+2}{3})\,.
\end{split}
\end{equation}

\subsubsection*{General \texorpdfstring{$N$}{N}}

A general expression for the $S_N$ symmetric orbifold partition function can be found by working in the grand canonical ensemble 
\begin{equation}\label{eq:grand-canonical-oplus}
\text{Sym}(X):=\bigoplus_{N=0}^{\infty}(X\wr S_N)\,.
\end{equation}
Working with $\text{Sym}(X)$ allows us to work with all symmetric orbifold theories at once. We can keep track of the specific orbifold $X\wr S_N$ by introducing a chemical potential $p$ which keeps track of $N$. In analogy to second-quantized statistical mechanics, we define the `grand canonical' partition function $\mathfrak{Z}(p,\tau)$ by
\begin{equation}
\mathfrak{Z}(p,\tau)=\sum_{N=0}^{\infty}p^NZ_N(\tau)\,.
\end{equation}
In the case of the torus, it can be shown \cite{Dijkgraaf:1996xw,Bantay:2001ay} that the grand canonical partition function has a simple expression in terms of Hecke operators, namely
\begin{equation}\label{eq:symm-general-partition}
\mathfrak{Z}(p,\tau)=\exp\left(\sum_{k=1}^{\infty}p^kT_kZ(\tau)\right)\,,
\end{equation}
where the $k^{\text{th}}$ Hecke operator is given by
\begin{equation}\label{eq:hecke-operator}
T_kZ(\tau)=\frac{1}{k}\sum_{ad=k}\sum_{b=0}^{d-1}Z\left(\frac{a\tau+b}{d}\right)\,.
\end{equation}
The relationship between Hecke operators and permutation orbifolds is that \eqref{eq:hecke-operator} sums over all \textit{connected} covering spaces of the original torus with degree $k$. The integers $a,d$ in the sum indicate how many times the covering space wraps around the $A$ and $B$ cycles, respectively. The integer $b$ then indicates a Dehn twist on the base torus, which is a Dehn twist around the $B$ cycle of angle $2\pi b/d$ on the covering torus. The exponential \eqref{eq:symm-general-partition} then can be expanded to include all disconnected covering spaces. Taking the degree $N$ coefficient then produces the partition function of the symmetric orbifold $X\wr S_N$, which can be written as 
\be \label{eqn:SymNZFull}
Z_N (\tau) = \sum_{\text{parititons of $N$}} \prod_{k=1}^N \frac{1}{N_k!} \left(T_k Z(\tau) \right)^{N_k},
\ee
where partitions of $N$ sums over $N_k$ such that $\sum_{k=1}^N k N_k = N$.

Finally, we list an equivalent definition of the Hecke operators in terms of modular transformations. Let $M_k$ denote the space of $2\times 2$ integer matrices with determinant $k$, and let $\Gamma=\text{SL}(2,\mathbb{Z})$ denote the modular group. The coset $\Gamma\backslash M_k$ is given by
\begin{equation}
\Gamma\backslash M_k=
\left\{
\begin{pmatrix}
a & b\\
0 & d
\end{pmatrix}\Bigg|\,ad=k\,,\quad b=0,\ldots,d-1\right\}\,.
\end{equation}
Thus, we can write
\begin{equation}\label{eq:Hecke-operator-modular}
T_kZ(\tau)=\frac{1}{k}\sum_{\gamma\in\text{SL}(2,\mathbb{Z})\backslash M_k}Z\left(\gamma\cdot\tau\right)\,,
\end{equation}
where $2\times 2$ matrices are taken to act on $\tau$ in the usual way, i.e.
\begin{equation}
\begin{pmatrix}a & b\\ c & d\end{pmatrix}\cdot\tau=\frac{a\tau+b}{c\tau+d}\,.
\end{equation}

\subsubsection*{Higher genus}\label{sec:higher-genus-symm}

Finally, we mention the situation for permutation CFTs formulated on surfaces $\Sigma$ with genus $g>1$. We will focus on the case $\Omega=S_N$.

Recall that the uniformization theorem states that every surface $\Sigma_g$ with genus $g>1$ can be expressed as a quotient $\mathbb{H}^2/G$ of the upper-half-plane by a Fuchsian group $G$.\footnote{A Fuchsian group is a discrete subgroup of $\text{SL}(2,\mathbb{R})$, which acts on the upper-half-plane in the usual way, i.e. 
\begin{equation*}
\begin{pmatrix}
a & b\\ c & d
\end{pmatrix}\cdot z=\frac{a z+b}{c z+d}\,.
\end{equation*}} Consider a (connected) covering space $\Sigma_{g'}\to\Sigma_g$ of order $N$. By the Riemann-Hurwitz formula, the genus $g'$ of $\Sigma_{g'}$ is related to the genus $g$ of $\Sigma$ by
\begin{equation}
g'=N(g-1)+1\,.
\end{equation}
One can classify such covering spaces in the following way: let $H$ be a subgroup of $G$ with index $[H:G]=N$.\footnote{The index of a subgroup $H$ of $G$ is the number of left cosets or equivalently the number $|G/H|$ of right cosets of $H$ in the group $G$.} Then the quotient $\mathbb{H}^2/H$ is a covering space of $\Sigma=\mathbb{H}^2/G$. The covering map is given by mapping the equivalence class $[z]_H=\{h\cdot z|h\in H\}$ to $[z]_G=\{g\cdot z|g\in G\}$, and this map is $N$-to-1. It can be shown that all connected covering spaces of $\Sigma$ can be found in the following way, and the covering spaces are determined uniquely by the subgroups $H$ up to conjugation by elements of $G$.

Using the above construction, it has been shown that the grand canonical partition function $\mathfrak{Z}(p,\Sigma)$ of the grand canonical ensemble \eqref{eq:grand-canonical-oplus} on a higher-genus Riemann surface $\Sigma_g=\mathbb{H}^2/G$ can be written as \cite{Bantay:2000eq}
\begin{equation}\label{eq:higher-genus-partition-function}
\mathfrak{Z}(p,\mathbb{H}^2/G)=\exp\left(\sum_{\substack{H\subset G\\\text{up to conjugation}}}\frac{p^{[H:G]}}{[H:G]}Z(\mathbb{H}^2/H)\right)\,,
\end{equation}
where $Z(\mathbb{H}^2/H)$ is the partition function of the seed theory $X$ on the covering surface $\Sigma_{g'}=\mathbb{H}^2/H$. The sum in the exponential with fixed value of $[H:G]$ can be thought of as a higher-genus generalization of the Hecke operators \eqref{eq:hecke-operator}.

Expanding out \eqref{eq:higher-genus-partition-function} will generally give terms of the form
\begin{equation}
\prod_{i=1}^{n}Z(\mathbb{H}^2/H_1)\cdots Z(\mathbb{H}^2/H_n)\,,
\end{equation}
where $H_1,\ldots,H_n\subset G$ have degrees $N_i=[H_i:G]$, with some combinatorial factors we don't care about. Isolating the $p^N$ coefficient (i.e. the terms contributing to the $S_N$ partition function) requires
\begin{equation}
\sum_{i=1}^{n}[H_i:G]=\sum_{i=1}^{n}N_i=N\,.
\end{equation}
Since $\mathbb{H}^2/H_i$ defines a Riemann surface $\Sigma_{g'_i}$ of genus $g'_i-1=N_i(g-1)$, we can equivalently write the above product as
\begin{equation}
Z(\Sigma_{g_1'})\cdots Z(\Sigma_{g_n'})\,,
\end{equation}
where the genera $g'_i$ are constrained by the requirements
\begin{equation}
\quad\sum_{i=1}^{n}(g_i'-1)=N(g-1)\quad\text{and}\quad(g_i'-1)|(g-1)\,.
\end{equation}
Thus, calculating the $S_N$ orbifold partition function on a surface of genus $g$ requires knowledge of the seed theory partition functions of surfaces of various genera.

\subsubsection*{Spin structures}

We now generalize the above discussion to CFTs with fermionic degrees of freedom. In this, case, in addition to a surface $\Sigma$, one must make a choice of spin structure. A surface of genus $g$ has $2g$ non-contractible cycles, and a choice of spin structure on $\Sigma$ is a choice of periodic or antiperiodic boundary conditions for fermions on $\Sigma$ around each cycle.\footnote{A spin structure can also be defined as a homomorphism $\phi:\pi_1(\Sigma)\to\mathbb{Z}_2$. Since $\mathbb{Z}_2$ is abelian, if $\Sigma$ is connected this is equivalently a homomorphism $\phi:H_1(\Sigma,\mathbb{Z})\to\mathbb{Z}_2$.} There are $2^{2g}$ such choices of spin structure, and the path integral of a fermionic CFT on $\Sigma$ is highly dependent on the choice of spin structure.

Specializing to genus $g=1$, the spin structure is labeled by two half integers $\alpha,\beta$, such that a fermion $\psi$ satisfies
\begin{equation}
\psi(A\cdot z)=e^{2\pi i\alpha}\psi(z)\,,\quad\psi(B\cdot z)=e^{2\pi i\beta}\psi(z)\,.
\end{equation}
Explicitly, if we choose $A$ to be the cycle along the $\tau$-direction and $B$ to be the cycle along the $1$-direction, we write
\begin{equation}
\psi(z+\tau)=e^{2\pi i\alpha}\psi(z)\,,\quad\psi(z+1)=e^{2\pi i\beta}\psi(z)\,.
\end{equation}
We will denote the partition function of a (fermionic) CFT $X$ with spin structure $(\alpha\,,\beta)$ and conformal structure $\tau$ as
\begin{equation}
Z\begin{bmatrix}\alpha\\\beta\end{bmatrix}(\tau) \quad\text{or}\quad Z_{\vec{e}}\,(\tau)\,,
\end{equation}
where $\vec{e}=(\alpha\,\beta)^T$ is a column vector.

If we consider the symmetric product of $X$, a formula similar to \eqref{eq:symm-general-partition} can be given which takes into account spin structures. We now must sum over permutations of the $N$ fermions such that $\psi (A \cdot z) = e^{2\pi i\alpha} \pi_A \cdot \psi(z)$ and $\psi (B \cdot z) =e^{2\pi i\beta} \pi_B \cdot \psi(z)$. The effect of this is that the spin structure on the covering space can differ from the base space. The end result is given by
\begin{equation} \label{eqn:supersymmetricgrandZ}
\mathfrak{Z} \begin{bmatrix}\alpha\\\beta\end{bmatrix}(p,\tau)=\exp\left(\sum_{k=1}^{\infty}p^k\mathcal{T}_kZ \begin{bmatrix}\alpha\\\beta\end{bmatrix} (\tau)\right)\,,
\end{equation}
where the fermionic Hecke operator $\mathcal{T}_k$ acts as
\begin{equation}
\mathcal{T}_k Z\begin{bmatrix}\alpha\\\beta\end{bmatrix}(\tau)=\frac{1}{k}\sum_{ad=k}\sum_{b=0}^{d-1}Z \begin{bmatrix}a \alpha + b \beta \\ d \beta \end{bmatrix}  \left(\frac{a\tau+b}{d}\right)\,.
\end{equation}
Here, we understand the parameters of the spin structure to be added modulo 2. The spin structure $(a\alpha+b\beta,d\beta)$ is the spin structure $(\alpha\,,\beta)$ pulled back to the covering torus. Note that if we take $\gamma\in\Gamma\backslash M_k$, then the spin structure on the covering space is written as
\begin{equation}
\begin{bmatrix}
\alpha'\\\beta'
\end{bmatrix}=\gamma\cdot
\begin{bmatrix}
\alpha\\\beta
\end{bmatrix}\,,
\end{equation}
and so, writing the spin structure as a column vector $\vec{e}=(\alpha\,\beta)^T$, we can write the Hecke operator compactly as
\begin{equation}
\mathcal{T}_kZ_{\vec{e}}=\frac{1}{k}\sum_{\gamma\in\Gamma\backslash M_k}Z_{\gamma\cdot\vec{e}}\,(\gamma\cdot\tau)\,.
\end{equation}

\section{Averaging the symmetric orbifold}\label{sec:averaging-orbifold}

As mentioned in the introduction, the fundamental objects of study in this paper are the permutation orbifolds $\mathbb{T}^D\wr\Omega$ of Narain CFTs. In this section, we compute the averaged torus partition functions of these orbifolds over the moduli space $\mathcal{M}_D$ of $\mathbb{T}^D$ targets. We focus primarily on the symmetric orbifold case $\Omega=S_N$. We begin by computing the averaged partition function of the $\mathbb{T}^D\wr S_2$ orbifold, which is simple enough to write down explicitly, but still complicated enough to exhibit general features which persist at larger $N$. We then turn to the case of $N>2$. Finally, we briefly comment on the generalization to higher-genus surfaces.

\subsection{Generalities}

Before jumping straight into averaging symmetric orbifold partition functions, let us briefly comment on the general structure one might expect to see. Given a surface $\Sigma$, we know that the symmetric orbifold partition function of some theory $X$ on $\Sigma$ is expressed by summing the partition function of the seed theory $X$ on all possible $N$-fold covering spaces $\widetilde{\Sigma}\to\Sigma$, weighted by an appropriate automorphism factor. We again quote the result:
\begin{equation}
Z_N(\Sigma)=\sum_{\widetilde{\Sigma}\to\Sigma}\frac{Z(\widetilde{\Sigma})}{|\text{Aut}(\widetilde{\Sigma}\to\Sigma)|}\,,
\end{equation}
where the sum is restricted to $N$-sheeted covering surfaces.

Now, let us assume that $X$ is not a single theory, but an element of some moduli space $m\in\mathcal{M}$. Furthermore, we assume that, as in the case of $\text{U}(1)$-gravity, there is a sense in which averaging the partition function $Z(\Sigma,m)$ over the moduli space $\mathcal{M}$ admits a holographic interpretation in terms of `filling in' the manifold $\Sigma$. That is, we assume
\begin{equation}
\int_{\mathcal{M}}\mathrm{d}\mu(m)\,Z(\Sigma,m)=\sum_{\partial M=\Sigma}Z_{\text{grav}}(M)\,,
\end{equation}
where we sum over three-dimensional `bulk manifolds' $M$ with boundary $\Sigma$, weighted by some sort of gravitational path integral $Z_{\text{grav}}$ evaluated on $M$.\footnote{We allow the notion of a `bulk manifold' to be rather vague. In the case of $\text{U}(1)$ gravity, this role is played by Lagrangian sublattices $\Gamma_0\subset H_1(\Sigma,\mathbb{Z})$.} With this assumption in mind, we can automatically compute the average of the symmetric orbifold partition function $Z_N(\Sigma,m)$ over the moduli space $\mathcal{M}$, and the result will take the schematic form
\begin{equation}\label{eq:symmetric-general-average-sum}
\int_{\mathcal{M}}\mathrm{d}\mu(m)\,Z_N(\Sigma,m)=\int_{\mathcal{M}}\mathrm{d}\mu(m)\,\sum_{\widetilde{\Sigma}\to\Sigma}\frac{Z(\widetilde{\Sigma},m)}{|\text{Aut}(\widetilde{\Sigma}\to\Sigma)|}=\sum_{\widetilde{\Sigma}\to\Sigma}\sum_{\partial\widetilde{M}=\widetilde{\Sigma}}\frac{Z_{\text{grav}}(\widetilde{M})}{|\text{Aut}(\widetilde{\Sigma}\to\Sigma)|}\,.
\end{equation}
That is, to average the symmetric orbifold partition function, we first sum over covering spaces $\widetilde{\Sigma}\to\Sigma$, and then sum over bulk manifolds $\widetilde{M}$ with boundary $\widetilde{\Sigma}$.

In principle, the above procedure is straightforward (although most likely analytically intractable). However, it leaves much to be desired in terms of the standard holographic dictionary. Specifically, each term in equation \eqref{eq:symmetric-general-average-sum} is calculated on a bulk manifold which has boundary $\widetilde{\Sigma}$, while holographically one expects gravitational quantities dual to CFT data to be computed on manifolds with boundary $\Sigma$. Moreover, the automorphism factors $|\text{Aut}(\widetilde{\Sigma}\to\Sigma)|$ don't have an immediate interpretation in terms of a gravitational path integral on $\widetilde{M}$.

A natural solution to both of the above problems arises if we think of $\widetilde{M}$ not as a true gravitational bulk manifold, but as a covering space of a gravitational bulk manifold $M$ with boundary $\Sigma$. As we will discuss in more detail below, if the bulk is described by a gauge theory which has a discrete $S_N$ factor, then the path integral of the gravitational theory on the bulk $M$ is naturally calculated by passing to a covering space $\widetilde{M}$. In direct analogy to the symmetric orbifold, we would expect the three-dimensional partition function on $M$ of a theory with $S_N$ gauge symmetry to take the schematic form
\begin{equation}
\sum_{\widetilde{M}\to M}\frac{Z_{\text{grav}}(\widetilde{M})}{|\text{Aut}(\widetilde{M}\to M)|}\,,
\end{equation}
where the sum is over all covering spaces $\widetilde{M}\to M$ of degree $N$. The full gravitational path integral would then be given by summing over all appropriate bulk geometries $M$ with boundary $\Sigma$. In order for this procedure to reproduce the averaged symmetric orbifold partition function \eqref{eq:symmetric-general-average-sum}, we would demand
\begin{equation}\label{eq:proposed-wrong-duality}
\sum_{\widetilde{M}\to M}\sum_{\partial M=\Sigma}\frac{Z_{\text{grav}}(\widetilde{M})}{|\text{Aut}(\widetilde{M}\to M)|}\stackrel{!}{=}\sum_{\widetilde{\Sigma}\to\Sigma}\sum_{\partial\widetilde{M}=\widetilde{\Sigma}}\frac{Z_{\text{grav}}(\widetilde{M})}{|\text{Aut}(\widetilde{\Sigma}\to\Sigma)|}\,.
\end{equation}
Unfortunately, such a strict prescription has little chance of working. This is because there are many ways to fill a covering surface $\widetilde{\Sigma}$ with a three-manifold $\widetilde{M}$ which itself does not cover any three manifold $M$ whose boundary is $\Sigma$. In other words, given a surface $\Sigma$, a covering $\widetilde{\Sigma}$, and a bulk manifold $\widetilde{M}$ bounded by $\widetilde{\Sigma}$, the bottom-right corner of the diagram in Figure \ref{fig:geoemtry-commutative diagram} does not always exist. However, the opposite statement is always true: given a surface $\Sigma$, a three-manifold $M$ with $\partial M=\Sigma$, and a covering space $\widetilde{M}$ of $M$, the  boundary of $\widetilde{M}$ is always a covering space of $\widetilde{\Sigma}$. This means that the left-hand side of \eqref{eq:proposed-wrong-duality} is strictly contained in the right-hand side, and so at the very least some information of the averaged symmetric orbifold CFT can be recovered from a bulk theory with gauge group $S_N$.

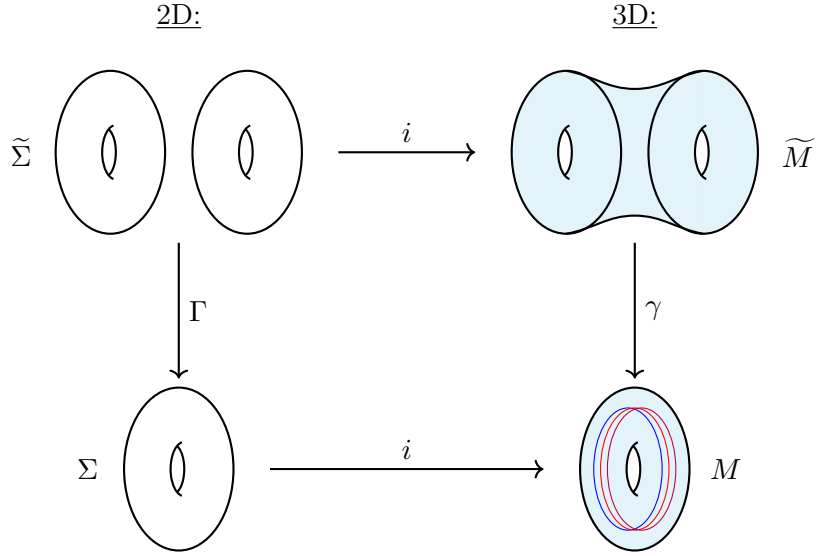
\begin{figure}
\centering
\ifdraft
\else
\begin{tikzpicture}[scale = 1.2]
\begin{scope}[scale = 0.3]
\draw[thick] (0,0) [partial ellipse = 0:360:2 and 3];
\draw[thick] (0.2,0) [partial ellipse = 100:260:0.5 and 1];
\draw[thick] (-0.2,0) [partial ellipse = -70:70:0.4 and 0.9];
\end{scope}
\draw[thick, ->] (1,0) -- (4,0);
\node[above] at (2.5,0) {$i$};
\begin{scope}[xshift = 5cm, scale = 0.3]
\fill[CornflowerBlue, opacity = 0.15] (0,0) [partial ellipse = 0:360:2 and 3];
\fill[white] (0.2,0) [partial ellipse = 115:245:0.5 and 1];
\fill[white] (-0.2,0) [partial ellipse = -70:70:0.4 and 0.9];
\draw[thick] (0,0) [partial ellipse = 0:360:2 and 3];
\draw[thick] (0.2,0) [partial ellipse = 100:260:0.5 and 1];
\draw[thick] (-0.2,0) [partial ellipse = -70:70:0.4 and 0.9];
\draw[blue] (-0.25,0) [partial ellipse = 0:360:1.25 and 2.25];
\draw[red] (0,0) [partial ellipse = 0:360:1.25 and 2.25];
\draw[purple] (0.25,0) [partial ellipse = 0:360:1.25 and 2.25];
\end{scope}
\draw[thick, <-] (0,1) -- (0,2.5);
\node[right] at (0,1.75) {$\Gamma$};
\draw[thick, <-] (5,1) -- (5,2.5);
\node[right] at (5,1.75) {$\gamma$};
\begin{scope}[yshift = 3.5cm]
\begin{scope}[xshift = -0.75cm, scale = 0.3]
\draw[thick] (0,0) [partial ellipse = 0:360:2 and 3];
\draw[thick] (0.2,0) [partial ellipse = 100:260:0.5 and 1];
\draw[thick] (-0.2,0) [partial ellipse = -70:70:0.4 and 0.9];
\end{scope}

\begin{scope}[xshift = 0.75cm, scale = 0.3]
\draw[thick] (0,0) [partial ellipse = 0:360:2 and 3];
\draw[thick] (0.2,0) [partial ellipse = 100:260:0.5 and 1];
\draw[thick] (-0.2,0) [partial ellipse = -70:70:0.4 and 0.9];
\end{scope}
\end{scope}
\begin{scope}[yshift = 3.5cm, xshift = 5cm]
\begin{scope}[xshift = -0.75cm]
\fill[CornflowerBlue, opacity = 0.15] (0,0.9) to[out = -10, in = 180] (0.75,0.7) to[out = 0, in = -170] (1.5,0.9) -- (1.5,-0.9) to[out = 170, in = 0] (0.75,-0.7) to[out = 180, in = 10] (0,-0.9);
\end{scope}
\begin{scope}[xshift = -0.75cm, scale = 0.3]
\fill[CornflowerBlue, opacity = 0.15] (0,0) [partial ellipse = 90:270:2 and 3];
\fill[white] (0.2,0) [partial ellipse = 115:245:0.5 and 1];
\fill[white] (-0.2,0) [partial ellipse = -70:70:0.4 and 0.9];
\draw[thick] (0,0) [partial ellipse = 0:360:2 and 3];
\draw[thick] (0.2,0) [partial ellipse = 100:260:0.5 and 1];
\draw[thick] (-0.2,0) [partial ellipse = -70:70:0.4 and 0.9];
\end{scope}
\draw[thick] (-0.75,0.9) to[out = -10, in = 180] (0,0.7) to[out = 0, in = -170] (0.75,0.9);
\draw[thick] (-0.75,-0.9) to[out = 10, in = 180] (0,-0.7) to[out = 0, in = 170] (0.75,-0.9);
\begin{scope}[xshift = 0.75cm, scale = 0.3]
\fill[CornflowerBlue, opacity = 0.15] (0,0) [partial ellipse = -90:90:2 and 3];
\fill[white] (0.2,0) [partial ellipse = 115:245:0.5 and 1];
\fill[white] (-0.2,0) [partial ellipse = -70:70:0.4 and 0.9];
\draw[thick] (0,0) [partial ellipse = 0:360:2 and 3];
\draw[thick] (0.2,0) [partial ellipse = 100:260:0.5 and 1];
\draw[thick] (-0.2,0) [partial ellipse = -70:70:0.4 and 0.9];
\end{scope}
\end{scope}
\draw[thick, ->] (1.75,3.5) -- (3.25,3.5);
\node[above] at (2.5,3.5) {$i$};
\node[left] at (-1.5,3.5) {$\widetilde{\Sigma}$};
\node[right] at (6.5,3.5) {$\widetilde{M}$};
\node at (-1,0) {$\Sigma$};
\node at (6,0) {$M$};
\node at (0,5) {\underline{2D:}};
\node at (5,5) {\underline{3D:}};
\end{tikzpicture}
\fi
\caption{We construct a bulk manifold $\widetilde{M}$ by `filling in' the covering space $\widetilde{\Sigma}$. If $\widetilde{M}$ is a (branched) covering space of another bulk manifold $M$ with boundary $\Sigma$, then $M$ is interpreted as a bulk geometry in our theory of quantum gravity, and the branch loci of the covering $\gamma:\widetilde{M}\to M$ are interpreted as `vortices' for the bulk gauge theory. For bulks $\widetilde{M}$ which do not cover a bulk $M$ with boundary $\Sigma$, a semiclassical interpretation of that contribution to the gravitational path integral is less clear.}
\label{fig:geoemtry-commutative diagram}
\end{figure}

We will return to the bulk/boundary duality later in Section \ref{sec:bulk-theory}, where we explicitly calculate the two sides of \eqref{eq:proposed-wrong-duality}. We will find that, although the two sides do not match, one is contained within the other and we find a match between the \textit{connected parts} of both sides in the case of a single torus boundary. We will refer to the terms that match as ``semiclassical'' contributions, since they can be obtained through the usual rules of the bulk gravity path integral. We also postulate that the ``extra'' terms appearing in the right-hand-side of \eqref{eq:proposed-wrong-duality} represent quantum gravity corrections that go beyond the usual semiclassical picture, since it seems difficult to interpret these extra terms as standard bulk geometries.

\subsection[Example: \texorpdfstring{$N=2$}{N=2}]{\boldmath Example: \texorpdfstring{$N=2$}{N=2}}
Let us make some of the general statements made above more explicit by considering the case of the $\mathbb{T}^D\wr S_2$ orbifold. As discussed in Section \ref{subsec:permutation-orbifolds}, the partition function of the orbifold $X\wr S_2$ on a torus of modular parameter $\tau$ is given by
\begin{equation}
Z_{\mathbb{T}^D \wr S_2}(\tau)=\frac{1}{2}Z_{\mathbb{T}^D}(\tau)^2+\frac{1}{2}Z_{\mathbb{T}^D}(2\tau)+\frac{1}{2}Z_{\mathbb{T}^D}(\tfrac{\tau}{2})+\frac{1}{2}Z_{\mathbb{T}^D}(\tfrac{\tau+1}{2})\,,
\end{equation}
where $Z(\tau)$ is the partition function of $X$. We will split up the partition function into a ``connected'' and a ``disconnected'' part
\be
Z_{\mathbb{T}^D \wr S_2}(\tau) = Z_{\mathbb{T}^D\wr S_2 \text{, conn.}}(\tau) + Z_{\mathbb{T}^D\wr S_2\text{, dis.}}(\tau),
\ee
where connectedness refers to whether the covering space is connected or not. That is, the connected part only contains terms with a single partition function $Z$. If $X$ is a Narain theory $X=\mathbb{T}^D$, then we can write the $\mathbb{T}^D\wr S_2$ partition function as
\begin{equation}
Z_{\mathbb{T}^D \wr S_2}(m,\tau)=\frac{1}{2}\frac{\Theta(m,2\tau)}{|\eta(2\tau)|^{2D}}+\frac{1}{2}\frac{\Theta(m,\tfrac{\tau}{2})}{|\eta(\tfrac{\tau}{2})|^{2D}}+\frac{1}{2}\frac{\Theta(m,\tfrac{\tau+1}{2})}{|\eta(\tfrac{\tau+1}{2})|^{2D}}+\frac{1}{2}\left(\frac{\Theta(m,\tau)}{|\eta(\tau)|^{2D}}\right)^2\,,
\end{equation}
where $\Theta(m,\tau)$ is the Narain theta function evaluated at the point $m\in\mathcal{M}_D$ of the Narain moduli space. The ``connected'' and ``disconnected'' parts are given by
\begin{equation}
\begin{split}
Z_{\mathbb{T}^D\wr S_2 \text{, conn.}}(m,\tau)&=\frac{1}{2}\frac{\Theta(m,2\tau)}{|\eta(2\tau)|^{2D}}+\frac{1}{2}\frac{\Theta(m,\tfrac{\tau}{2})}{|\eta(\tfrac{\tau}{2})|^{2D}}+\frac{1}{2}\frac{\Theta(m,\tfrac{\tau+1}{2})}{|\eta(\tfrac{\tau+1}{2})|^{2D}}\,,\\
Z_{\mathbb{T}^D\wr S_2 \text{, dis.}}(m,\tau)&=\frac{1}{2}\left(\frac{\Theta(m,\tau)}{|\eta(\tau)|^{2D}}\right)^2\,.
\end{split}
\end{equation}
These correspond to contributions to the symmetric orbifold partition function from double-covers of the torus which are connected and disconnected, respectively. Using the Siegel-Weil formula \eqref{eqn:SiegelWeilTorus} we have
\begin{equation}
\int_{\mathcal{M}_D}\mathrm{d}\mu\,\frac{\Theta(m,\tau)}{|\eta(\tau)|^{2D}}=\sum_{\gamma\in\Gamma_{\infty}\backslash\text{SL}(2,\mathbb{Z})}\frac{1}{|\eta(\gamma\cdot\tau)|^{2D}}\,,
\end{equation}
and we can immediately write down the average of the connected piece:
\begin{equation}\label{eq:n=2-conn-average}
\braket{Z_{\mathbb{T}^D\wr S_2 \text{, conn.}}(m,\tau)}=\frac{1}{2}\sum_{\gamma\in\Gamma_{\infty}\backslash\text{SL}(2,\mathbb{Z})}\left(\frac{1}{|\eta(\gamma\cdot 2\tau)|^{2D}}+\frac{1}{|\eta(\gamma\cdot(\tfrac{\tau}{2}))|^{2D}}+\frac{1}{|\eta(\gamma\cdot\left(\tfrac{\tau+1}{2})\right)|^{2D}}\right)\,.
\end{equation}
The average for the disconnected part was worked out in Section \ref{sec:preliminaries}, and requires the Siegel-Weil formula for disconnected surfaces \eqref{eq:generic-siegel-weil-disconnected}. The period matrix $\Omega$ of the disjoint union $\Sigma\sqcup\Sigma$ is given by $\Omega = \operatorname{diag}(\tau, \tau)$, and following the rules of \cite{Maloney:2020nni}, we can write the average of the disconnected component as a sum over Lagrangian sublattices $\Gamma_0$ of the total homology lattice $H_1(\Sigma\sqcup\Sigma,\mathbb{Z})\cong H_1(\Sigma,\mathbb{Z})\oplus H_1(\Sigma,\mathbb{Z})$, namely
\begin{equation} \label{eq:n=2-dis-average}
\braket{Z_{\mathbb{T}^D\wr S_2 \text{, dis.}}(m,\tau)}=\frac{1}{2 \operatorname{Im}(\tau)^{ D}|\eta(\tau)|^{4D}}\sum_{\Gamma_0\subset H_1(\Sigma\sqcup\Sigma,\mathbb{Z})}(\det\text{Im}(\Omega_{\Gamma_0}))^{D/2}\,,
\end{equation}
where $\Omega_{\Gamma_0}$ is the period matrix $\Omega$ evaluated on the sublattice $\Gamma_0$ as explained in Section \ref{sec:preliminaries}. Each Lagrangian sublattice $\Gamma_0$ is to be associated with a bulk geometry with boundary $\Sigma\sqcup\Sigma$ such that the boundary cycles $\Gamma_0$ are contractible in the bulk. Choices of $\Gamma_0$ correspond to either ``Wormhole'' geometries, or completely disconnected bulk geometries. 

\begin{figure}
\centering
\ifdraft
\else
\begin{tikzpicture}[scale = 1.5]
\fill[CornflowerBlue, opacity = 0.15] (0,0.9) to[out = -10, in = 180] (0.75,0.7) to[out = 0, in = -170] (1.5,0.9) -- (1.5,-0.9) to[out = 170, in = 0] (0.75,-0.7) to[out = 180, in = 10] (0,-0.9);
\draw[thick] (0,0.9) to[out = -10, in = 180] (0.75,0.7) to[out = 0, in = -170] (1.5,0.9);
\draw[thick] (0,-0.9) to[out = 10, in = 180] (0.75,-0.7) to[out = 0, in = 170] (1.5,-0.9);
\begin{scope}[scale = 0.3]
\fill[CornflowerBlue, opacity = 0.15] (0,0) [partial ellipse = 90:270:2 and 3];
\fill[white] (0.2,0) [partial ellipse = 115:245:0.5 and 1];
\fill[white] (-0.2,0) [partial ellipse = -70:70:0.4 and 0.9];
\draw[thick] (0,0) [partial ellipse = 0:360:2 and 3];
\draw[thick] (0.2,0) [partial ellipse = 100:260:0.5 and 1];
\draw[thick] (-0.2,0) [partial ellipse = -70:70:0.4 and 0.9];
\node[above] at (0,3) {$\tau$};
\end{scope}
\begin{scope}[xshift = 1.5cm, scale = 0.3]
\fill[CornflowerBlue, opacity = 0.15] (0,0) [partial ellipse = -90:90:2 and 3];
\fill[white] (0.2,0) [partial ellipse = 115:245:0.5 and 1];
\fill[white] (-0.2,0) [partial ellipse = -70:70:0.4 and 0.9];
\draw[thick] (0,0) [partial ellipse = 0:360:2 and 3];
\draw[thick] (0.2,0) [partial ellipse = 100:260:0.5 and 1];
\draw[thick] (-0.2,0) [partial ellipse = -70:70:0.4 and 0.9];
\node[above] at (0,3) {$\tau$};
\end{scope}
\node at (2.5,0) {$+$};
\begin{scope}[xshift = 3.5cm, scale = 0.3]
\fill[CornflowerBlue, opacity = 0.15] (0,0) [partial ellipse = 0:360:2 and 3];
\fill[white] (0.2,0) [partial ellipse = 115:245:0.5 and 1];
\fill[white] (-0.2,0) [partial ellipse = -70:70:0.4 and 0.9];
\draw[thick] (0,0) [partial ellipse = 0:360:2 and 3];
\draw[thick] (0.2,0) [partial ellipse = 100:260:0.5 and 1];
\draw[thick] (-0.2,0) [partial ellipse = -70:70:0.4 and 0.9];
\node[above] at (0,3) {$2\tau$};
\end{scope}
\node at (4.5,0) {$+$};
\begin{scope}[xshift = 5.5cm, scale = 0.3]
\fill[CornflowerBlue, opacity = 0.15] (0,0) [partial ellipse = 0:360:2 and 3];
\fill[white] (0.2,0) [partial ellipse = 115:245:0.5 and 1];
\fill[white] (-0.2,0) [partial ellipse = -70:70:0.4 and 0.9];
\draw[thick] (0,0) [partial ellipse = 0:360:2 and 3];
\draw[thick] (0.2,0) [partial ellipse = 100:260:0.5 and 1];
\draw[thick] (-0.2,0) [partial ellipse = -70:70:0.4 and 0.9];
\node[above] at (0,3) {$\tau/2$};
\end{scope}
\node at (6.5,0) {$+$};
\begin{scope}[xshift = 7.5cm, scale = 0.3]
\fill[CornflowerBlue, opacity = 0.15] (0,0) [partial ellipse = 0:360:2 and 3];
\fill[white] (0.2,0) [partial ellipse = 115:245:0.5 and 1];
\fill[white] (-0.2,0) [partial ellipse = -70:70:0.4 and 0.9];
\draw[thick] (0,0) [partial ellipse = 0:360:2 and 3];
\draw[thick] (0.2,0) [partial ellipse = 100:260:0.5 and 1];
\draw[thick] (-0.2,0) [partial ellipse = -70:70:0.4 and 0.9];
\node[above] at (0,3) {$(\tau+1)/2$};
\end{scope}
\end{tikzpicture}
\fi
\caption{The ``bulk geometries'' contributing to the averaged $\mathbb{T}^D\wr S_2$ partition function. The geometry on the left is understood to be a connected three-manifold whose boundary is the disjoint union of two tori.}
\label{fig:n=2-bulk-covering}
\end{figure}
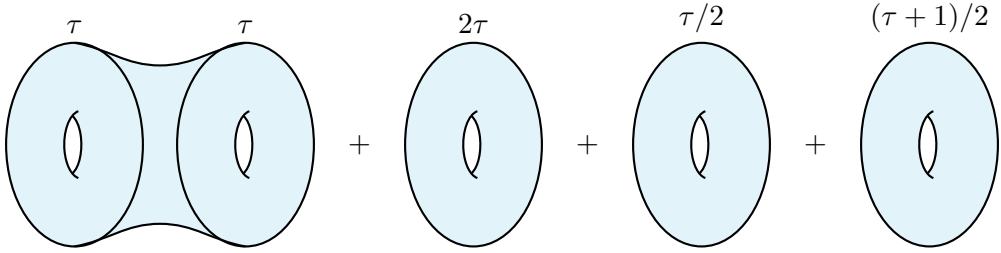

In both the connected and disconnected case, the averaged partition function has an interpretation as a sum over geometries shown in Figure \ref{fig:n=2-bulk-covering}. For the connected piece, $\braket{Z_{\text{conn}}}$ is a sum over $\text{U}(1)^{2D}\times\text{U}(1)^{2D}$ Chern-Simons partition functions on geometries with boundary tori of modular parameter $2\tau$, $\tfrac{\tau}{2}$, or $\tfrac{\tau+1}{2}$. The disconnected piece $\braket{Z_{\text{dis}}}$ is written as a sum over both disconnected handlebody geometries and ``wormhole'' geometries whose boundary is the disjoint union $\Sigma\sqcup\Sigma$. From a holographic perspective this sum over geometries is unsatisfactory for two reasons:
\begin{itemize}

    \item Each geometry comes with a symmetry factor of $\frac{1}{2}$, which, from the point of view of a bulk Chern-Simons theory has no reason to be there.
    
    \item Holographically, we expect the bulk dual of a CFT on a surface $\Sigma$ to consist of a gravitational theory on bulk manifold $\mathcal{M}$ whose boundary is $\Sigma$. However, none of the geometries mentioned above have boundary $\Sigma$, but rather their boundaries are double covers of $\Sigma$.

\end{itemize}
Thus, we need a way to interpret the contributions of \eqref{eq:n=2-conn-average} and \eqref{eq:n=2-dis-average} in terms of geometries whose boundary is given by $\Sigma$. We now explain under what conditions this can be done. 

We will first make some formal statements before giving a more intuitive picture towards the end of the paragraph. Let us denote one of the bulk geometries contributing to \eqref{eq:n=2-conn-average} and \eqref{eq:n=2-dis-average}  by $\widetilde{\mathcal{M}}$ and its boundary by $\widetilde{\Sigma}$. Suppose that $\widetilde{\mathcal{M}}$ has symmetry group $G$ such that if we consider the quotient geometry $\widetilde{\mathcal{M}} / G$ we obtain a manifold with desired asymptotic boundary $\partial \widetilde{\mathcal{M}} / G = \Sigma$. Then we can holographically interpret the geometry $\widetilde{\mathcal{M}} / G$ as contributing to the averaged partition function through the usual rules of AdS/CFT. It turns out that we can give such an interpretation to certain terms in the sum, as we now explain for the case of $N=2$. This is achieved by recalling that the boundaries of all of the above geometries are covering spaces of the original torus $\Sigma$. There exists a 2-to-1 map
\begin{equation}
\widetilde{\Sigma}\to\Sigma
\end{equation}
which is the covering map of $\Sigma$ by $\widetilde{\Sigma}$. This map can be thought of as the quotient map of a $\mathbb{Z}_2$ automorphism (deck transformation) $\iota:\widetilde{\Sigma}\to\widetilde{\Sigma}$ which projects $\widetilde{\Sigma}$ to the quotient space $\widetilde{\Sigma}/\iota\cong\Sigma$.\footnote{The $S_2$ orbifold on the torus is a special case since all covering spaces $\widetilde{\Sigma}\to\Sigma$ satisfy $\Sigma\cong\widetilde{\Sigma}/\mathbb{Z}_2$. In general the relationship between covering and base spaces is not a simple quotient.}
One might hope that the bulk geometries $\widetilde{\mathcal{M}}$ with boundary $\widetilde{\Sigma}$ also inherit a $\mathbb{Z}_2$ automorphism that restricts to $\iota$ on the boundary. For many of the bulk geometries this is indeed the case, such as with the solid tori which fill the connected covering spaces, shown in Figure \ref{fig:n=2-bulk-covering}.  For these geometries we can define a (possibly singular) bulk manifold $\mathcal{M}=\widetilde{\mathcal{M}}/\mathbb{Z}_2$ which has boundary $\Sigma$, and thus provides a promising candidate for the bulk manifold which should appear in the sum over geometries contributing to the averaged $\mathbb{T}^D\wr S_2$ partition function.
However, as we will explore more in detail later, many of the bulk geometries appearing in the averaged $\mathbb{T}^D\wr S_2$ do not inherit a $\mathbb{Z}_2$ automorphism compatible with the automorphism of the boundary $\widetilde{\Sigma}$. Specifically, the geometries in Figure \ref{fig:n=2-bulk-covering} with connected boundary all inherit the $\mathbb{Z}_2$ automorphism of the boundary, and so they can indeed be thought of as covering spaces of well-defined three-manifolds. However, for the geometries whose boundaries are disconnected, the $\mathbb{Z}_2$ symmetry of the boundary can be `broken' by specific details of the bulk. 

 Let us also explain this in a more pictorial way. Consider the concrete case of the potential wormhole contribution depicted in the leftmost panel of Figure \ref{fig:n=2-bulk-covering}. The corresponding boundary contribution, the leftmost panel of figure \ref{fig:n=2-boundary-covering}, is invariant under swapping of the two tori. This symmetry of the two coverings leads to the factor $\frac{1}{2}$ as explained in Section \ref{subsec:permutation-orbifolds}. Analogously, we can also consider the second to left panel in Figure \ref{fig:n=2-boundary-covering}. The upper and lower half of the torus correspond to the two covering sheets. Again, we observe a symmetry by exhanging these two sheets. Similar statements can be made about the remaining two covering surfaces pictured in \ref{fig:n=2-boundary-covering}. In order to define a theory on a potential quotient three manifold, we must be able to extend these symmetries of the covering surfaces into the bulk. It is then clear that for the 3 connected contributions pictured in Figure \ref{fig:n=2-bulk-covering} this can be done. However, for the wormhole contribution it is also clear that this may only be possible if the two boundary tori are imbued with the same complex structure.

Geometries for which this is not possible can be avoided by either only considering the average of the connected contribution to the $\mathbb{T}^D\wr S_2$ partition function, for which such a $\mathbb{Z}_2$ automorphism always exists, or by trying to find a suitable generalization of what is meant by a `bulk geometry with boundary $\Sigma$'. We will return to this point in Section \ref{sec:bulk-theory} in the context of the bulk gauge theory description.

\subsection[General \texorpdfstring{$N$}{N}]{\boldmath General \texorpdfstring{$N$}{N}}
For general $N$ the theory is given by the $\mathbb{T}^D\wr S_N$ orbifold. The partition function for fixed $N$ can be extracted from the grand canonical partition function in equation \eqref{eq:symm-general-partition} by extracting the term proportional to $p^N$ in the series expansion. For general $N$ the form of the partition function is quite complicated and is given by \eqref{eqn:SymNZFull},
\be
Z_{\mathbb{T}^D\wr S_N } (m,\tau) = \sum_{\text{parititons of $N$}} \prod_{k=1}^N \frac{1}{N_k!} \left(T_k Z_{\mathbb{T}^D}(m,\tau) \right)^{N_k},
\ee
where again partitions of $N$ sum over $N_k$ such that $\sum_{k=1}^N k N_k = N$. We can immediately obtain the ensemble average by applying the Siegel-Weil formula \eqref{eq:generic-siegel-weil-disconnected} to the above expression
\be \label{eqn:sec3AverageZFinal}
\langle Z_{\mathbb{T}^D\wr S_N } (m,\tau) \rangle = \sum_{\text{parititons of $N$}} \left\langle \prod_{k=1}^N \frac{1}{N_k!} \left(T_k Z_{\mathbb{T}^D}(m,\tau) \right)^{N_k} \right\rangle.
\ee
However, this expression is quite formal since it is a complicated sum over various wormhole and non-wormhole geometries with up to $N$ asymptotic boundaries. It is useful to split the partition function into a sum over ``connected'' and ``disconnected'' contributions
\be
Z_{\mathbb{T}^D\wr S_N}(m, \tau) = Z_{\text{conn}}(m, \tau) + Z_{\text{dis}}(m, \tau).
\ee
The ``connected'' part of the partition function only includes contributions from connected covering spaces, and it has a simple expression since it can be extracted from \eqref{eq:symm-general-partition} by keeping contributions proportional to $p^N$ and a single copy of $Z$, giving
\be
Z_{\text{conn}}(m, \tau) = T_N Z_{\mathbb{T}^D}(m, \tau) = \frac{1}{N} \sum_{ad=k}\sum_{b=0}^{d-1} Z_{\mathbb{T}^D}\left(m,  \frac{a \tau + b}{d}\right).
\ee
The ensemble average over the connected part is simple since there are no wormhole contributions. We find
\begin{align}
\braket{Z_{\text{conn}}(m, \tau)} &= \frac{1}{N} \sum_{ad=k}\sum_{b=0}^{d-1} \Braket{Z_{\mathbb{T}^D}\left(m,  \frac{a \tau + b}{d}\right)},\\
&= \frac{1}{N} \sum_{ad=k}\sum_{b=0}^{d-1} \sum_{\gamma\in\Gamma_{\infty}\backslash\text{SL}(2,\mathbb{Z})}\frac{1}{|\eta(\gamma\cdot \left(\frac{a \tau + b}{d}\right) )|^{2D}}\, .
\end{align}
where the quantity on the right is simply the torus average given in \eqref{eqn:TorusAverage}. We will later see that we can give this term a holographic interpretation as a Chern-Simons path integral with vortices, similar to the case of $N=2$.

The ``disconnected'' contribution $Z_{\text{dis}}(m, \tau)$ is more difficult to write out explicitly, but it would contain sums over multiple copies of partition functions with different modular parameters. As an explicit example, from \eqref{eqn:N=3Z} we see that for $N=3$ the ``disconnected'' piece would be given by
\be \label{eqn:Zdis_N=3}
Z_{\text{dis}}(\tau)=\frac{1}{6}Z_{\mathbb{T}^D}(\tau)^3+\frac{1}{2}Z_{\mathbb{T}^D}(\tau)Z_{\mathbb{T}^D}(2\tau)+\frac{1}{2}Z_{\mathbb{T}^D}(\tau)Z_{\mathbb{T}^D}(\tfrac{\tau}{2})+\frac{1}{2}Z_{\mathbb{T}^D}(\tau)Z_{\mathbb{T}^D}(\tfrac{\tau+1}{2}).
\ee
We can again apply the Siegel-Weil formula for disconnected surfaces \eqref{eq:generic-siegel-weil-disconnected} to perform the average over the ``disconnected'' partition function. This provides a bulk interpretation for the disconnected piece as a sum over wormhole geometries with up to $N$ asymptotic boundary tori. Similar to the discussion for $N=2$, we are unable to give a holographic interpretation to the disconnected piece as a sum over bulk geometries with a single asymptotic boundary torus of modular parameter $\tau$.

\subsection{Averaging at higher genus}\label{sec:averaging-at-higher-genus}

Finally, we make some brief comments on how to generalize the above discussion to symmetric orbifold partition functions on higher genus surfaces, only focusing on the connected component of the partition function. In Section \ref{sec:higher-genus-symm} we mentioned that the symmetric orbifold partition function on higher genus surface $\Sigma_g=\mathbb{H}^2/G$ can be neatly packaged in the grand canonical ensemble as a sum over subgroups $H\subset G$ where $G$ is the Fuchsian group acting on the upper-half-plane. Specifically,
\begin{equation}
\mathfrak{Z}(p,\mathbb{H}^2/G)=\exp\left(\sum_{\substack{H\subset G\\\text{up to conjugation}}}\frac{p^{[H:G]}}{[H:G]}Z(\mathbb{H}^2/H)\right)\,.
\end{equation}
The connected component of the $X\wr S_N$ orbifold can then be easily extracted by isolating the $p^N$ coefficient in the exponential, namely
\begin{equation}
Z_{N,\text{conn}}(\mathbb{H}^2/G)=\frac{1}{N}\sum_{\substack{H\subset G,\,\,[H:G]=N\\\text{up to conjugation}}}Z(\mathbb{H}^2/H)\,.
\end{equation}
This sum is precisely what one would expect: a sum over all connected covering spaces $\mathbb{H}^2/H$ of the surface $\Sigma\cong\mathbb{H}^2/G$. An equivalent yet slightly less algebraic notation would be to write the connected contribution as
\begin{equation}
Z_{N,\text{conn}}(\Sigma_g)=\frac{1}{N}\sum_{\widetilde{\Sigma}\to\Sigma_g}Z(\widetilde{\Sigma})\,,
\end{equation}
where the sum is over all genus $g'=N(g-1)+1$ surfaces which cover $\Sigma_g$.

We can now consider the ensemble average of this quantity by specifying $X=\mathbb{T}^D$. By the dictionary of \cite{Maloney:2020nni}, we know that the average of the partition function $Z(\widetilde{\Sigma})$ is given as a sum over all handlebodies bounded by $\widetilde{\Sigma}$ (or equivalently as a sum over all Lagrangian sublattices $\Gamma_0\subset H_1(\widetilde{\Sigma},\mathbb{Z})$) weighted by the Chern-Simons action on that handlebody. Explicitly,
\begin{equation}\label{eq:higher-genus-averaged-partition}
\Braket{Z_{N,\text{conn}}(m,\Sigma_g)}=\frac{1}{N}\sum_{\widetilde{\Sigma}\to\Sigma_g}\sum_{\partial \widetilde{M}=\widetilde{\Sigma}}Z_{\text{CS}}(\widetilde{M})\,.
\end{equation}
While this expression is short and conceptually simple, its practical computation is very difficult. Specifically, computing the Chern-Simons path integral on the handlebody $\widetilde{M}$ requires knowledge of the period matrix $\widetilde{\Omega}$ of $\widetilde{\Sigma}$, which is difficult to generically compute \cite{Hidalgo2011:xyz}. That said, the averaged partition function \eqref{eq:higher-genus-averaged-partition} is in principle well-defined, up to divergence issues in the resulting Eisenstein series.\footnote{The $\mathbb{T}^D$ averaged partition function on a surface of genus $g'$ diverges when $D\leq g'+1$. Since each term in \eqref{eq:higher-genus-averaged-partition} is evaluated on a surface of genus $g'=N(g-1)+1$, where $g$ is the genus of $\Sigma_g$, we require
\begin{equation}
D > N(g-1)+2
\end{equation}
for convergence.}

\section{The bulk theory}\label{sec:bulk-theory}

The torus theory $\mathbb{T}^D$ possesses affine $\text{U}(1)^D_L\times\text{U}(1)^D_R$ currents, whose zero-modes generate a global symmetry algebra. These currents are given by the left- and right-moving derivatives of the worldsheet scalars, namely
\begin{equation}
J^m=\partial X^m\,,\quad\overline{J}^m=\overline{\partial} X^m\,,\quad m=1,\ldots,D\,.
\end{equation}
By the standard holographic dictionary, these currents are dual to gauge fields in the bulk,
\begin{equation}
J^m\to A^m\,,\quad \overline{J}^m\to B^m\,,
\end{equation}
where $A$ transforms under the left $\text{U}(1)^D$ and $B$ transforms under the right $\text{U}(1)^D$\,. In the Narain ensemble, this identification of global symmetries on the boundary with gauge symmetries in the bulk is all that needs to be done, and we can write the bulk action as a $\text{U}(1)^D\times\text{U}(1)^D$ Chern-Simons theory, namely
\begin{equation}
S=\int_{M}(A\wedge\mathrm{d}A-B\wedge\mathrm{d}B)\,.
\end{equation}
Here, we implicitly perform a sum over the $\text{U}(1)^D$ indices $m$, see \eqref{eqn:CSAction}. The rules of Narain averaging then also tell us that the path integral must be summed over bulk geometries $\mathcal{M}$.

For the permutation orbifold $\mathbb{T}^D\wr\,\Omega$, things are a bit different. We now have $N$ copies of each current, and so in the bulk we expect $N$ copies of the gauge fields $A,B$. Thus, we postulate the following action:
\begin{equation} \label{eqn:CSNAction}
S=\sum_{i=1}^{N}\int_{M}\left(A_{(i)}\wedge\mathrm{d}A_{(i)}-B_{(i)}\wedge\mathrm{d}B_{(i)}\right)\,.
\end{equation}
Where again the $U\lr{1}^D$ indices are implicit. Since the boundary theory contains $N$ copies of the original $\text{U}(1)^D\times\text{U}(1)^D$ symmetry the bulk theory has $N$ copies of the $\text{U}(1)^D\times\text{U}(1)^D$ gauge symmetry
\begin{equation}\label{eq:n-u(1)-transforamtions}
A_{(i)}\to A_{(i)}+\mathrm{d}\lambda^A_{(i)}\,,\quad B_{(i)}\to B_{(i)}+\mathrm{d}\lambda^B_{(i)}\,.
\end{equation}
We must now understand how quotienting by $\Omega$ modifies the bulk theory. Before we quotient the boundary theory by $\Omega$ we have a permutation symmetry given by $J^m 
\to J^{\pi (m)}$, \, $\ol{J}^m 
\to \ol{J}^{\pi (m)}$. This immediately maps into a symmetry of the bulk Chern-Simons theory $A_{(i)}\to A_{(\pi(i))}, \, B_{(i)}\to B_{(\pi(i))}$. After gauging the permutation symmetry $\Omega$, it is natural to expect the bulk theory to also carry this gauge symmetry. This promotes the permutation symmetry of Chern-Simons fields to a gauge symmetry

\begin{equation}\label{eq:n-perm-transformations}
A_{(i)}\to A_{(\pi(i))}\,,\quad B_{(i)}\to B_{(\pi(i))}\,.
\end{equation}
The full gauge group of the bulk dual of the $\mathbb{T}^D\wr\,\Omega$ permutation orbifold should then be the group generated by combinations of the $\text{U}(1)^D\times\text{U}(1)^D$ transformations \eqref{eq:n-u(1)-transforamtions} and the permutations \eqref{eq:n-perm-transformations}. The group generated by these two transformations is a semidirect product of $(\text{U}(1)^D\times\text{U}(1)^D)^N$ with the permutation group $\Omega$, and in the mathematical literature is denoted by the \textit{wreath product}\footnote{The wreath product is just mathematical notation meaning the gauge group which is generated by composing \eqref{eq:n-u(1)-transforamtions} and \eqref{eq:n-perm-transformations}. The relationship between the wreath product symbol $\wr$ and permutation orbifolds $X\wr \Omega$ is that if a CFT $X$ has symmetry group $G$, then the permutation orbifold $X\wr\Omega$ has symmetry group $G\wr\Omega$.}
\begin{equation}
\text{U}(1)^D\times\text{U}(1)^D\wr\,\Omega\,.
\end{equation}

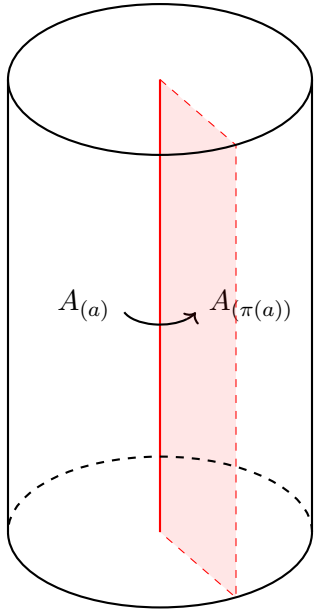
\begin{figure}
\centering
\ifdraft
\else
\begin{tikzpicture}
\node at (1.2,0) {$A_{(\pi(a))}$};
\draw[thick, ->] (0,0) [partial ellipse = 315:340:0.5 and 0.25];
\fill[red, opacity = 0.1] (0,-3) -- (0,3) -- (1,2.13) -- (1,-3.87) -- (0,-3);
\draw[thick, red] (0,-3) -- (0,3);
\draw[red, dashed] (0,3) -- (1,2.13) -- (1,-3.87) -- (0,-3);
\draw[thick, ] (0,0) [partial ellipse = 200:315:0.5 and 0.25];
\node at (-1,0) {$A_{(a)}$};
\draw[thick, dashed] (0,-3) [partial ellipse = 0:180:2 and 1];
\draw[thick] (0,-3) [partial ellipse = 180:360:2 and 1];
\draw[thick] (0,3) [partial ellipse = 0:360:2 and 1];
\draw[thick] (-2,-3) -- (-2,3);
\draw[thick] (2,-3) -- (2,3);

\end{tikzpicture}
\fi
\caption{The effect of including a vortex associated to a permutation $\pi$ in the bulk dual of $\mathbb{T}^D\wr\,\Omega$.}
\label{fig:vortex}
\end{figure}
Armed with the above discussion, a natural duality to propose would be:
\begin{equation}\label{eq:proposed-duality-sec-4}
\begin{gathered}
\text{Narain-ensemble of }\mathbb{T}^D\wr\,\Omega\text{ orbifolds}\\
\Longleftrightarrow\\
\text{U}(1)^D\times\text{U}(1)^D\wr\,\Omega\text{ Chern-Simons coupled to topological gravity}
\end{gathered}
\end{equation}
Although this gauge group is not discrete, it contains a discrete factor\footnote{ Indeed, $\text{U}(1)^D\times\text{U}(1)^D\wr\,\Omega$ is equivalent to $(\text{U}(1)^D\times\text{U}(1)^D)^N\times\Omega$ as a topological space (but not as a group).} of $\Omega$. Thus, the above theory will exhibit behaviors universal to all discrete gauge theories. One of these is the existence of \textit{twist operators} in the bulk, which in three-dimensional discrete gauge theories take the form of one-dimensional vortices \cite{PhysRevLett.62.1221,Preskill:1990bm}. We now explain the bulk partition function of this theory and how gauging by $\Omega$ introduces new features such as vortices.

The Chern-Simons path integral with action \eqref{eqn:CSNAction} and gauge group $G \wr S_N$, where $G$ is some Lie group, on a bulk manifold $M$ is given by
\be \label{eqn:CSpathintegralG}
Z_{G \wr S_N} = \frac{1}{N!}\sum_{\mathrm{bundles}} \int \mathcal{D}A \mathcal{D}B~ e^{- S_{\mathrm{CS}}[A,B]} \,.
\ee
The factor of $N!$ comes from taking permutations of the fields to be gauge equivalent. We take this into account by integrating over all possible fields $A,B$ but divide by $N!$ not to overcount. In the path integral we are to integrate over gauge connections on all $G \wr S_N$ bundles over $M$. Typically, the total gauge group is a Lie group and so there exists only a single trivial bundle. However, if the gauge group contains a discrete factor, such as $S_N$, then there are additional bundles that must be included in the path integral \cite{dijkgraaf1989geometrical,Dijkgraaf:1989pz}. This is one of the new features when dealing with gauge theories with discrete groups. As we will see slightly later, the effect of including the sum over bundles is that we must include gauge field configurations where the fields $A, B$ are twisted as we travel around the non-contractible cycles in $M$. 

We will find that to match to the boundary ensemble average we will need to include gauge field configurations that also have non-trivial monodromies around the contractible cycles in $M$, but these are not reproduced by the sum over bundles above. However, there is a way to include such configurations by including codimension two ``vortices'' in the path integral. We thus claim that the correct bulk path integral of interest is given by
\be \label{eqn:bulkCSPI}
Z_{\text{Bulk}} \equiv \frac{1}{N!} \sum_{M}\sum_{\substack{\mathrm{bundles,} \\ \mathrm{vortices}}} \int \mathcal{D}A \mathcal{D}B~ e^{- S_{\mathrm{CS}}[A,B]} \mathcal{V} \,,
\ee
where we have modified the path integral by inserting an additional vortex operator $\mathcal{V}$. We specify what the summation over vortices means slightly later. A vortex is a line operator embedded into the manifold $M$ which enforces that the gauge fields $A, B$ pick up certain monodromies as they travel around the vortex. The sum over vortices needs to be put in by hand, and we take the philosophy that the summation over vortex configurations should be chosen to match the boundary answer. In the bulk path integral we also sum over a set of bulk manifolds $M$ with given asymptotic boundary structure, and the specific choice of bulk manifolds $M$ will be clarified later.

In the remainder of this section we go into additional details regarding the summation over bundles and vortices. We then explain the bulk path integral calculation in the case of $N=2$ with an asymptotic torus boundary, and we clarify which terms in the boundary ensemble average are reproduced by the bulk calculation. We discuss the case of larger $N$ and higher genus boundaries, and we comment on bulk non-handlebody contributions. 

\subsection{Topological theories with finite gauge group} \label{sec:4.1}
Before fully exploring our proposed bulk theory, let us first make some general remarks about topological field theories with finite gauge group \cite{Dijkgraaf:1989pz}, see \cite{Teleman2016} for an expository introduction.

A field theory with a discrete gauge group $\Omega$ is formally very similar to 2D orbifold CFTs. Indeed, an orbifold CFT can be formulated as a 2D theory with discrete gauge group. In three-dimensions, we can similarly consider field theories whose gauge groups are discrete. Let $M$ be some three-manifold. If the field content of the theory is $\Phi$, then gauging the group $\Omega$ amounts to identifying
\begin{equation}
\Phi(p)\sim g\cdot\Phi(p)
\end{equation}
at all points $p$. Again, just as in the orbifold CFT case, this leads to interesting behavior when $M$ is not simply-connected. In this case, we can imagine transporting $\Phi$ around some non-contractible loop $\gamma$ based at $p$. It is perfectly fine if $\Phi$ itself is not single-valued, but rather picks up a monodromy $\phi(\gamma)\in \Omega$ upon being transported around $\gamma$. That is,
\begin{equation}
\Phi(\gamma\cdot p)=\phi(\gamma)\cdot\Phi(p)\,,
\end{equation}
where $\Phi(\gamma\cdot p)$ is shorthand for the value of $\Phi$ after being transported along $\gamma$. Since $\Phi$ and $\phi(\gamma)\cdot\Phi$ are physically equivalent, the above should be a perfectly allowed configuration in the path integral.

Similarly to the discussion in Section \ref{subsec:permutation-orbifolds}, the assignment of a twisted boundary condition to each loop $\gamma$ must be consistent with the process of concatenation of loops. That is, the twisted boundary conditions must form a homomorphism $\phi:\pi_1(M)\to \Omega$. Moreover, for any $g\in \Omega$, the homomorphisms $\phi(\gamma)$ and $g^{-1}\phi(\gamma)g$ represent the same physical field configuration, since it is related to $\phi$ simply by a global field redefinition $\Phi\to g\cdot\Phi$. That is, contributions to the path integral are defined only up to conjugation by elements of $\Omega$ (this is the $\Omega$ group action on the map $\phi$). Mathematically, the set of such twisted boundary conditions are in one-to-one correspondence with the representation variety
\begin{equation}
\text{Hom}(\pi_1(M),\Omega)/\Omega\,,
\end{equation}
which is equivalently the moduli space of flat $\Omega$-bundles on $M$.\footnote{Since $\Omega$ is discrete, all $\Omega$-bundles are flat.}

We can also consider a fully topological theory whose only content is the discrete gauge group $\Omega$. The field content of the theory is trivial, and the only nontrivial observables are the partition functions on $M$. It is given as a sum over all homomorphisms $\phi:\pi_1(M)\to \Omega$ (i.e. twisted boundary conditions) weighted by the automorphism group $\text{Aut}(\phi)$ of that homomorphism. That is, the partition function of the TQFT with gauge group $\Omega$ is simply:
\begin{equation}
Z_\Omega(M)= \sum_{\substack{\phi:\pi_1(M)\to \Omega \\ \text{up to conjugation}}}\frac{1}{|\text{Aut}(\phi)|}\,.
\end{equation}
As a trivial example, if $M\cong\text{S}^3$ is the three-sphere, then $\pi_1(\text{S}^3)$ is trivial, hence the only choice for $\phi$ is $\phi(\text{id}_{\pi_1(\text{S}^3)})=\text{id}_{\Omega}$ and we simply have
\begin{equation}
Z_\Omega(\text{S}^3)=\frac{1}{|\Omega|}\,.
\end{equation}

Now, we are specifically interested in the case of $\Omega=S_N$. As we discussed in Section \ref{sec:preliminaries}, a set of twisted boundary conditions $\phi:\pi_1(M)\to S_N$ can be identified with a covering space $\widetilde{M}\to M$, and the group of automorphisms becomes the group of deck transformations $\text{Aut}(\widetilde{M}\to M)$. Thus, the $S_N$ topological gauge theory on $M$ computes a weighted sum over all covering spaces of $M$ of degree $N$, weighted by the order of the group of deck transformations. Explicitly,
\begin{equation}
Z_{S_N}(M)=\sum_{\substack{\widetilde{M}\to M\\\text{degree }N}}\frac{1}{|\text{Aut}(\widetilde{M}\to M)|}\,.
\end{equation}

The above construction can be rather nicely extended to the case of not a finite gauge group, but a gauge group $G\wr S_N$ where $G$ is some Lie group and $\wr$ is the wreath product defined above. In this case, the path integral is over all gauge connections on $G\wr S_N$ bundles over $M$. Now, topologically, a $G\wr S_N$ bundle $E\to M$ is equivalent to a $G$ bundle $E\to\widetilde{M}$ over the $S_N$ bundle $\widetilde{M}\to M$ (i.e. a degree $N$ covering space of $M$).\footnote{This statement is already implicit in \cite{Bantay:1997ek}.} Thus we can trade the path integral of a $G\wr S_N$ gauge theory for a sum over path integrals of a $G$ gauge theory on the covering spaces of $M$. The final result is
\begin{equation}\label{eq:G-wreath-sn-partition function}
Z_{G\,\wr S_N}(M)=\sum_{\substack{\widetilde{M}\to M\\\text{degree }N}}\frac{Z_G(\widetilde{M})}{|\text{Aut}(\widetilde{M}\to M)|}\,.
\end{equation}
Another way to view the above construction is to take $N$ copies of a gauge theory on $G$, and then to gauge the $S_N$ permutation symmetry of the resulting product theory. This is equivalent to defining a gauge theory with gauge group $G\wr S_N$, and the arguments for computing path integrals in symmetric orbifold theories from Section \ref{subsec:permutation-orbifolds} carries over, and we arrive at \eqref{eq:G-wreath-sn-partition function}, in direct analogy to the logic that allowed us to derive equation \eqref{eq:symmetric-orbifold-partition-general}.\footnote{In fact, given any TQFT $Z$, not just gauge theories, one can construct a `symmetric product' theory by formally averaging $Z$ over covering spaces, see \cite{Gunningham_2016}. Such TQFTs are useful, for example, in computing generalizations of Hurwitz numbers, see also \cite{alexeevski2004noncommutative}.}

To summarize the above discussion, the summation over non-trivial bundles appearing in equation \eqref{eqn:bulkCSPI} induces a summation over non-trivial boundary conditions on the gauge fields around the non-contractible cycles in $M$. This can be rewritten as a sum over covering spaces of $M$ given by equation \eqref{eq:G-wreath-sn-partition function}. Note that this does not induce twisted boundary conditions around the contractible cycle. To include such configurations we need to include bulk vortices, which we now discuss. 

\subsection*{Vortices}

Another property of gauge theories with discrete gauge groups is the presence of so-called vortices. As explained above, vortices are codimension-2 objects which impose twisted boundary conditions on gauge fields transported around them. These are the 3-dimensional analogues of twist fields in orbifold CFTs, and as we will see in Section \ref{sec:correlators} are holographically dual to twist fields.

Informally, a vortex (often also called a \textit{monodromy defect} \cite{Witten:2011zz} or a \textit{Gukov-Witten operator} \cite{Gukov:2008sn}) is a codimension-2 extended object in a gauge theory which has the property that being transported around it induces a monodromy of the gauge group $\Omega$. This is analogous to the Aharonov-Bohm effect, in which the wavefunction of a charged particle picks up a phase upon being transported around a solenoid. In the case of a gauge theory with a discrete gauge group, a field $\Phi$ picks up a monodromy $\Phi\to g\cdot\Phi$ upon being transported around the vortex. 

 Formally, a vortex is a codimension-2 sublocus $L$ of a three-manifold $M$ which carries charge $[g]$, where $[g]$ is some conjugacy class of elements of the discrete gauge group $\Omega$.\footnote{We assign a charge to $\mathcal{V}$ in terms of conjugacy classes of $\Omega$ because otherwise the vortex $\mathcal{V}$ would not be a gauge-invariant object. This is because gauge transformations would act on a charge $g$ as $hgh^{-1}$. This generically changes the group element $g$, but leaves it in the conjugacy class. This is spelled out more concretely in Section \ref{sec:correlators} in the context of orbifold twist fields. For abelian groups like those considered in \cite{Benjamin:2021wzr} this problem does not arise.} We can formally write a vortex operator $\mathcal{V}_{[g], L}$ of charge $[g]$ associated to a sublocus $L$ by the operator
\be
\mathcal{V}_{[g], L}=\sum_{\sigma \in [g]}\mathcal{V}_{\sigma, L}\,,
\ee
where we sum over permutations $\sigma$ in the conjugacy class $[g]$, which makes the operator gauge invariant. In discrete gauge theories the vortex operator cannot typically be represented in terms of fundamental fields of the theory. It's action is implemented by imposing twisted boundary conditions on the fields as they are transported around the vortex: transporting the fields $A_I,B_I$ around $\mathcal{V}_{\sigma,L}$ enforces the twisted boundary condition $A_I \to A_{\sigma(I)}$ and $B_I \to B_{\sigma(I)}$.

We can combine vortex operators with the non-trivial sum over bundles to implement twisted boundary conditions around multiple cycles. Suppose our bulk manifold is a solid torus with boundary cycles $a,b$, and that a vortex operator imposes twisted boundary conditions around the $a$ cycle given by $\pi_a$, while a non-trivial bundle imposes twisted boundary conditions $\pi_b$ around the $b$ cycle. There will only exist non-trivial gauge field configurations if $[\pi_a,\pi_b]=0$. That is, going around the same set of cycles in different orders gives the same boundary conditions for the fields. The end result is that when summing over bundles and vortices the only contributions come from combinations with consistent boundary conditions. For a general bulk manifold $M$ this condition is formalized as follows.

The partition function of a discrete gauge theory on $M$ in the presence of a vortex $L$ of charge $[g]$ is defined similarly to the partition function on $M$ alone. Let $\ell\in\pi_1(M\setminus L)$ be a generator of the fundamental group of $M\setminus L$ which winds once around $L$. Then the path integral in the presence of the vortex is defined by summing over all homomorphisms $\phi:\pi_1(M\setminus L)\to \Omega$ (up to conjugation) such that $\phi(\ell)$ lies in the conjugacy class $[g]$. Specifically,
\begin{equation}
Z_\Omega(M;L,[g])=\sum_{\substack{\phi:\pi_1(M\setminus L)\to \Omega\\ \text{up to conjugation} \\ \phi(\ell)\in[g]}}\frac{1}{|\text{Aut}(\phi)|}\,.
\end{equation}

Again, we can specialize to the case where $\Omega$ is the permutation group $S_N$. Conjugacy classes of $S_N$ are labeled uniquely by cycle-types of permutations. Let us denote the charge associated to $L$ with $[\pi]$ and let us assume that $[\pi]$ is the conjugacy class of permutations with cycle-type $w_1,\ldots,w_k$, where
\begin{equation}
w_1+\cdots+w_k=N\,.
\end{equation}
Then a homomorphism $\phi:\pi_1(M\setminus L)\to S_N$ with $\phi(\ell)\in[\pi]$ defines a covering space of $M$ which is \textit{branched} over $L$ with branching structure given by the cycle type $w_1,\ldots,w_k$. In the language of Figure \ref{fig:vortex}, this branched covering space $\widetilde{M}\to M$ is the $N$-fold cover such that the gauge fields are single-valued. Thus,
\begin{equation}
Z_{S_N}(M;L,[\pi])=\sum_{\substack{\widetilde{M}\to M\\\text{branched over }L}}\frac{1}{|\text{Aut}(\widetilde{M}\to M)|}\,,
\end{equation}
where the branching over $L$ has branching structure $w_1,\ldots,w_k$.

Just as in the case without vortices, we can also enrich the above partition function by considering a topological gauge theory with group $G\wr S_N$, where $G$ is any Lie group. The partition function then becomes a sum over $G$ partition functions on branched covers of $M$ over $L$, i.e.
\begin{equation}
Z_{G\wr S_N}(M;L,[\pi])=\sum_{\substack{\widetilde{M}\to M\\\text{branched over }L}}\frac{Z_G(\widetilde{M})}{|\text{Aut}(\widetilde{M}\to M)|}\,.
\end{equation}
Note that this is essentially the three-dimensional generalization of the covering space construction for calculating correlation functions of twist fields in two-dimensional CFTs \cite{Dixon:1986qv,Lunin:2000yv}.

The upshot of the above construction is that the introduction of a vortex into a topological $G\wr S_N$ gauge theory amounts to computing partition functions on the branched cover $\widetilde{M}$ of $M$ over the branching locus $L$. This is completely analogous to the case in 2D orbifold CFTs, where correlation functions of twist fields are computed by passing to the branched cover, branched at the points where the twist fields are inserted \cite{Dixon:1986qv}. In fact, as we will see, this analogy is made precise in the holographic setting, and we will find that vortices in $G\wr S_N$ gauge theory which intersect the boundary $\partial M$ are dual to twist fields in the symmetric orbifold CFT. We delay this discussion to Section \ref{sec:correlators}.

Below we will perform some calculations of the bulk $\text{U}(1)^D\times\text{U}(1)^D\wr S_N$ Chern-Simons theory. We will start by taking the simple case of a torus boundary and $N=2$, where we can be quite explicit. We then move on to the case of a torus boundary but for generic $N$. Finally, we will make comments about the calculation of bulk partition functions on 3-manifolds with higher-genus boundaries.

\subsection[Example: \texorpdfstring{$N=2$}{N=2}]{\boldmath Example: \texorpdfstring{$N=2$}{N=2}}\label{sec:cs-calculations-n=2}

Let us now turn to calculating the $G \wr S_N$ Chern-Simons partition functions on bulk manifolds where $G=\text{U}(1)^D\times\text{U}(1)^D$. For now, let us work with the simple case $N=2$. We fix an asymptotic boundary $\Sigma$ which is a torus of modular parameter $\tau$. As discussed above, we can reduce the problem of calculating $\text{U}(1)^D\times\text{U}(1)^D\wr S_2$ partition functions to the task of computing $\text{U}(1)^D\times\text{U}(1)^D$ partition functions on degree 2 covering spaces $\widetilde{M}$ of $M$.

Let us start by letting $M$ be a handlebody, i.e. a solid torus. Then $\pi_1(M)\cong\mathbb{Z}$, and there are precisely two covering spaces for a given $M$: the covering space which is simply two disconnected copies of $M$, and the handlebody whose asymptotic boundary is a torus of modular parameter $2\tau$. We know that the Chern-Simons partition function on a handlebody with contractible spatial cycle and modular parameter $\tau$ is simply
\begin{equation}
Z_{G}(M)=\frac{1}{|\eta(\tau)|^{2D}}\,,
\end{equation}
and so we have
\begin{equation}
Z_{G\wr S_2}(M)=\frac{1}{2}\frac{1}{|\eta(\tau)|^{4D}}+\frac{1}{2}\frac{1}{|\eta(2\tau)|^{2D}}\,,
\end{equation}
where the factors of two come from the automorphism factors of the covering spaces.

Now, we can also introduce a nontrivial vortex. Let $L$ run along the non-contractible cycle of $M$. The only nontrivial conjugacy class of $S_2$ is $[(1\,2)]$, and there are precisely two covering spaces of $M$ branched over $L$ with that structure: a handlebody with modular parameter $\tau/2$ and a handlebody with modular parameter $(\tau+1)/2$. Thus, the the partition function of the $G\wr S_2$ Chern-Simons theory on $M$ with vortex $L$ is given by
\begin{equation}
Z_{G\wr S_2}(M;L)=\frac{1}{2}\frac{1}{\left|\eta(\tfrac{\tau}{2})\right|^{2D}}+\frac{1}{2}\frac{1}{\left|\eta(\tfrac{\tau+1}{2})\right|^{2D}}\,.
\end{equation}

In a theory of quantum gravity it is natural to sum over bulk manifolds with fixed asymptotic boundary, which in the case of the handlebody $M$ is given by a sum over modular images of the boundary torus. The natural partition function after coupling $G\wr S_2$ Chern-Simons theory to topological gravity is then given by
\begin{equation}\label{eq:ZBulk_S2}
Z_{\mathrm{Bulk}}=\frac{1}{2}\sum_{\gamma\in\Gamma_{\infty}\backslash\text{SL}(2,\mathbb{Z})} \left(\underbrace{\frac{1}{|\eta(\gamma\cdot\tau)|^{4D}}}_{\text{disconnected}}+\frac{1}{|\eta(2\,\gamma\cdot\tau)|^{2D}}+\underbrace{\frac{1}{\left|\eta(\tfrac{\gamma\cdot\tau}{2})\right|^{2D}}+\frac{1}{\left|\eta(\tfrac{\gamma\cdot\tau+1}{2})\right|^{2D}}}_{\text{vortex}}\right)\,.
\end{equation}
In the above we have identified the term associated with a disconnected covering space of the torus, as well as the contributions arising from including the vortex. The last three terms represent covering spaces $\widetilde{M}$ which are connected, the last two of which are branched over the vortex $L$. All of the covering spaces, both with and without vortex, are shown in Figure \ref{fig:3d-covering-spaces-s2}.

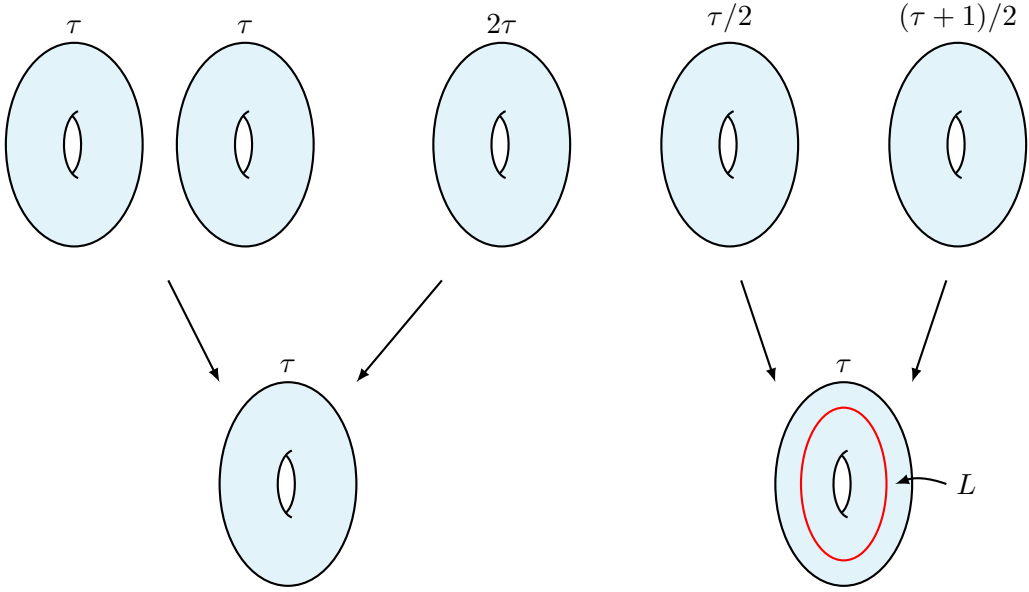
\begin{figure}
\centering
\ifdraft
\else
\begin{tikzpicture}[scale = 1.5]
\begin{scope}[xshift = 1.625cm, yshift = -3cm, scale = 0.3]
\fill[CornflowerBlue, opacity = 0.15] (0,0) [partial ellipse = 0:360:2 and 3];
\fill[white] (0.2,0) [partial ellipse = 115:245:0.5 and 1];
\fill[white] (-0.2,0) [partial ellipse = -70:70:0.4 and 0.9];
\draw[thick] (0,0) [partial ellipse = 0:360:2 and 3];
\draw[thick] (0.2,0) [partial ellipse = 100:260:0.5 and 1];
\draw[thick] (-0.2,0) [partial ellipse = -70:70:0.4 and 0.9];
\node[above] at (0,3) {$\tau$};
\draw[thick, latex-] (-2,3) -- (-3.5,6);
\draw[thick, latex-] (2,3) -- (4.5,6);
\end{scope}
\begin{scope}[xshift = 6.5cm, yshift = -3cm, scale = 0.3]
\fill[CornflowerBlue, opacity = 0.15] (0,0) [partial ellipse = 0:360:2 and 3];
\fill[white] (0.2,0) [partial ellipse = 115:245:0.5 and 1];
\fill[white] (-0.2,0) [partial ellipse = -70:70:0.4 and 0.9];
\draw[thick] (0,0) [partial ellipse = 0:360:2 and 3];
\draw[thick] (0.2,0) [partial ellipse = 100:260:0.5 and 1];
\draw[thick] (-0.2,0) [partial ellipse = -70:70:0.4 and 0.9];
\draw[thick, red] (0,0) [partial ellipse = 0:360:1.25 and 2.25];
\node[right] at (3,0) {$L$};
\draw[thick, -latex] (3,0) to[out = 160, in = 20] (1.5,0);
\node[above] at (0,3) {$\tau$};
\draw[thick, latex-] (-2,3) -- (-3,6);
\draw[thick, latex-] (2,3) -- (3,6);
\end{scope}
\begin{scope}[xshift = -0.25cm, scale = 0.3]
\fill[CornflowerBlue, opacity = 0.15] (0,0) [partial ellipse = 0:360:2 and 3];
\fill[white] (0.2,0) [partial ellipse = 115:245:0.5 and 1];
\fill[white] (-0.2,0) [partial ellipse = -70:70:0.4 and 0.9];
\draw[thick] (0,0) [partial ellipse = 0:360:2 and 3];
\draw[thick] (0.2,0) [partial ellipse = 100:260:0.5 and 1];
\draw[thick] (-0.2,0) [partial ellipse = -70:70:0.4 and 0.9];
\node[above] at (0,3) {$\tau$};
\end{scope}
\begin{scope}[xshift = 1.25cm, scale = 0.3]
\fill[CornflowerBlue, opacity = 0.15] (0,0) [partial ellipse = 0:360:2 and 3];
\fill[white] (0.2,0) [partial ellipse = 115:245:0.5 and 1];
\fill[white] (-0.2,0) [partial ellipse = -70:70:0.4 and 0.9];
\draw[thick] (0,0) [partial ellipse = 0:360:2 and 3];
\draw[thick] (0.2,0) [partial ellipse = 100:260:0.5 and 1];
\draw[thick] (-0.2,0) [partial ellipse = -70:70:0.4 and 0.9];
\node[above] at (0,3) {$\tau$};
\end{scope}
\begin{scope}[xshift = 3.5cm, scale = 0.3]
\fill[CornflowerBlue, opacity = 0.15] (0,0) [partial ellipse = 0:360:2 and 3];
\fill[white] (0.2,0) [partial ellipse = 115:245:0.5 and 1];
\fill[white] (-0.2,0) [partial ellipse = -70:70:0.4 and 0.9];
\draw[thick] (0,0) [partial ellipse = 0:360:2 and 3];
\draw[thick] (0.2,0) [partial ellipse = 100:260:0.5 and 1];
\draw[thick] (-0.2,0) [partial ellipse = -70:70:0.4 and 0.9];
\node[above] at (0,3) {$2\tau$};
\end{scope}
\begin{scope}[xshift = 5.5cm, scale = 0.3]
\fill[CornflowerBlue, opacity = 0.15] (0,0) [partial ellipse = 0:360:2 and 3];
\fill[white] (0.2,0) [partial ellipse = 115:245:0.5 and 1];
\fill[white] (-0.2,0) [partial ellipse = -70:70:0.4 and 0.9];
\draw[thick] (0,0) [partial ellipse = 0:360:2 and 3];
\draw[thick] (0.2,0) [partial ellipse = 100:260:0.5 and 1];
\draw[thick] (-0.2,0) [partial ellipse = -70:70:0.4 and 0.9];
\node[above] at (0,3) {$\tau/2$};
\end{scope}
\begin{scope}[xshift = 7.5cm, scale = 0.3]
\fill[CornflowerBlue, opacity = 0.15] (0,0) [partial ellipse = 0:360:2 and 3];
\fill[white] (0.2,0) [partial ellipse = 115:245:0.5 and 1];
\fill[white] (-0.2,0) [partial ellipse = -70:70:0.4 and 0.9];
\draw[thick] (0,0) [partial ellipse = 0:360:2 and 3];
\draw[thick] (0.2,0) [partial ellipse = 100:260:0.5 and 1];
\draw[thick] (-0.2,0) [partial ellipse = -70:70:0.4 and 0.9];
\node[above] at (0,3) {$(\tau+1)/2$};
\end{scope}
\end{tikzpicture}
\fi
\caption{The four double covering spaces of a solid torus $M$ without a vortex (on the left) and with a vortex (on the right). The connected covering spaces are also solid tori with modified modular parameters, while the disconnected covering space is simply $M\sqcup M$.}
\label{fig:3d-covering-spaces-s2}
\end{figure}

\subsection*{Comparing to the symmetric orbifold}
We would like to compare the above bulk calculation to the Narain-averaged symmetric orbifold result, which we recall takes the form

\begin{align} \label{eq:section-4-symmetric-s2-partition-function}
\braket{Z_{\mathbb{T}^D\wr S_2}(m,\tau)}&=\frac{1}{2}\sum_{\gamma\in\Gamma_{\infty}\backslash\text{SL}(2,\mathbb{Z})}\left(\frac{1}{|\eta(\gamma\cdot(2\tau))|^{2D}}+\frac{1}{\left|\eta(\gamma\cdot(\tfrac{\tau}{2}))\right|^{2D}}+\frac{1}{\left|\eta(\gamma\cdot(\tfrac{\tau+1}{2}))\right|^{2D}}\right) \nonumber\\
&+\frac{1}{2}\braket{Z_{\mathbb{T}^D}(\tau,m)^2}\,,
\end{align}
where the second line is the disconnected part of the partition function. Comparing the connected parts of \eqref{eq:ZBulk_S2} and \eqref{eq:section-4-symmetric-s2-partition-function}, we see that the modular parameters in the sum do not quite match. However, it is an algebraic fact that these two sums actually coincide,\footnote{See, for example, Theorem 6.9 of \cite{Iwaniec1997TopicsIC}.} i.e.
\begin{equation}
\begin{split}
\sum_{\gamma\in\Gamma_{\infty}\backslash\text{SL}(2,\mathbb{Z})}&\left(\frac{1}{|\eta(2\,\gamma\cdot\tau)|^{2D}}+\frac{1}{\left|\eta(\tfrac{\gamma\cdot\tau}{2})\right|^{2D}}+\frac{1}{\left|\eta(\tfrac{\gamma\cdot\tau+1}{2})\right|^{2D}}\right)\\
&=\sum_{\gamma\in\Gamma_{\infty}\backslash\text{SL}(2,\mathbb{Z})}\left(\frac{1}{|\eta(\gamma\cdot(2\tau))|^{2D}}+\frac{1}{\left|\eta(\gamma\cdot(\tfrac{\tau}{2}))\right|^{2D}}+\frac{1}{\left|\eta(\gamma\cdot(\tfrac{\tau+1}{2}))\right|^{2D}}\right)\,.
\end{split}
\end{equation}
Thus, the connected part of the $\text{U}(1)^D\times\text{U}(1)^D\wr S_2$ Chern-Simons theory coupled to 3D gravity precisely reproduces the Narain average of the connected part of the symmetric orbifold theory $\mathbb{T}^D\wr S_2$.

\subsection*{The disconnected part}

Now that we have shown that the connected parts of the Narain-averaged symmetric orbifold and topological gravity partition functions agree, let us move on to the disconnected part. The disconnected part of the symmetric orbifold partition function is given by
\begin{equation}\label{eq:s2-symmetric-orbifold-disconnected}
\Braket{Z_{\mathbb{T}^D\wr S_2,\,\text{dis.}}(\tau)}=\frac{1}{2}\Braket{Z_{\mathbb{T}^D}(m,\tau)^2}=\frac{1}{2|\eta(\tau)|^{4D}\operatorname{Im}(\tau)^{D}}\sum_{\Gamma_0}\left(\det\text{Im}\,\Omega_{\Gamma_0}\right)^{D/2}\,,
\end{equation}
where
\begin{equation}
\Omega=
\begin{pmatrix}
\tau & 0\\
0 & \tau
\end{pmatrix}
\end{equation}
is the period matrix of the double cover $\Sigma\sqcup\Sigma$ of the boundary torus $\Sigma$, and the sum is over Lagrangian sublattices of $H_1(\Sigma\sqcup\Sigma,\mathbb{Z})$ (see Section \ref{sec:narain-review}). Intuitively, the sum is over bulk manifolds $\widetilde{M}$ with boundary $\Sigma\sqcup\Sigma$ such that the sublattice $\Gamma_0$ is contractible in $\widetilde{M}$. On the other hand, the disconnected piece of the bulk $G \wr S_2$ partition function \eqref{eq:ZBulk_S2} is
\begin{equation}\label{eq:s2-chern-simons-disconnected}
Z_{\text{Bulk},\,\text{dis.}}=\frac{1}{2}\sum_{\gamma\in\Gamma_{\infty}\backslash\text{SL}(2,\mathbb{Z})}\frac{1}{|\eta(\gamma\cdot\tau)|^{4D}}=\frac{1}{2|\eta(\tau)|^{4D}\operatorname{Im}(\tau)^{D}}\sum_{\gamma\in\Gamma_{\infty}\backslash\text{SL}(2,\mathbb{Z})} \operatorname{Im}(\gamma\cdot\tau)^{D}\,,
\end{equation}
where we have used the fact that $|\eta(\tau)|^4 \operatorname{Im}(\tau)$ is modular invariant.

\begin{figure}
\centering
\ifdraft
\else
\begin{tikzpicture}[scale = 0.9]
\begin{scope}
\draw[dashed] (0,-1.15) [partial ellipse = 90:270:0.25 and 0.86];
\draw[thick, red] (0,-0.2) [partial ellipse = 0:360:2.25 and 1.25];
\draw[line width = 0.1cm, white] (0,0.2) [partial ellipse = 0:360:2.25 and 1.25];
\draw[thick, red] (0,0.2) [partial ellipse = 0:360:2.25 and 1.25];
\begin{scope}[scale = 1, rotate = 90]
\fill[CornflowerBlue, opacity = 0.2] (0,0) [partial ellipse = 0:360:2 and 3];
\fill[white] (-0.2,0) [partial ellipse = -70:70:0.4 and 0.9];
\fill[white] (0.2,0) [partial ellipse = 120:240:0.5 and 1];
\draw[thick] (0,0) [partial ellipse = 0:360:2 and 3];
\draw[thick] (0.2,0) [partial ellipse = 100:260:0.5 and 1];
\draw[thick] (-0.2,0) [partial ellipse = -70:70:0.4 and 0.9];
\end{scope}

\draw (0,-1.15) [partial ellipse = 90:-90:0.25 and 0.86];
\end{scope}

\node at (2.75,1.75) {\Large $\tau$};

\end{tikzpicture}
\fi
\caption{A vortex configuration for $\text{U}(1)^D\times\text{U}(1)^D\wr S_2$ Chern-Simons theory on a solid torus. The double-cover of this geometry is topologically $\Sigma\times I$.}
\label{fig:permutation-wormhole}
\end{figure}
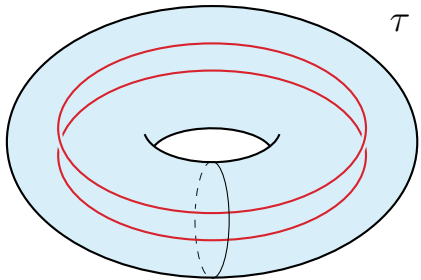

Clearly, \eqref{eq:s2-symmetric-orbifold-disconnected} and \eqref{eq:s2-chern-simons-disconnected} are not equal. However, the sum in \eqref{eq:s2-symmetric-orbifold-disconnected} actually contains the full sum \eqref{eq:s2-chern-simons-disconnected}. Let $\mathcal{A}^{(1)}$ and $\mathcal{A}^{(2)}$ be the A-cycles of the two boundaries, while $\mathcal{B}^{(1)}$ and $\mathcal{B}^{(2)}$ are their B-cycles. Then the Lagrangian sublattices of the form
\begin{equation}\label{eq:lagrangian-sublattice-disconnected-tori}
\Gamma_0=\text{Span}_{\mathbb{Z}}\left(\gamma(\mathcal{A}^{(1)}),\gamma(\mathcal{A}^{(2)})\right)\,,
\end{equation}
where $\gamma\in\Gamma_{\infty}\backslash\text{SL}(2,\mathbb{Z})$ is a modular transformation which acts on the homology cycles of $\Sigma$ in the usual way, contribute
\begin{equation}
(\det\text{Im}\,\Omega_{\Gamma_0})^{D/2}=\operatorname{Im}(\gamma\cdot\tau)^D\,,
\end{equation}
and thus reproduce the elements in the sum \eqref{eq:s2-chern-simons-disconnected}. This makes sense, given that sublattices \eqref{eq:lagrangian-sublattice-disconnected-tori} correspond to manifolds $\widetilde{M}$ which are the disjoint union $M\sqcup M$ of two handlebodies of modular parameter $\gamma\cdot\tau$, which are precisely the covering spaces that appeared in the disconnected part of the Chern-Simons partition function. Thus, although \eqref{eq:s2-symmetric-orbifold-disconnected} is not precisely reproduced by \eqref{eq:s2-chern-simons-disconnected}, we have the inclusion
\begin{equation}
\Braket{Z_{\mathbb{T}^D\wr S_2,\,\text{dis.}}(\tau)} \supset Z_{\text{Bulk},\,\text{dis.}}\,,
\end{equation}
where by $\supset$ we mean that the sum on the left-hand-side contains all elements of the sum on the right-hand-side.

So what about the other geometries contributing to \eqref{eq:s2-symmetric-orbifold-disconnected} which aren't reproduced by the Chern-Simons calculation? It turns out that at least the $T^2 \times I$ wormhole can be recovered from the Chern-Simons theory if we include more complicated vortex configurations.\footnote{Strictly speaking, the Chern-Simons calculation on the $T^2 \times I$ wormhole has not yet been carried out and matched to the expected boundary answer.} For example, \eqref{eq:s2-symmetric-orbifold-disconnected} contains the Lagrangian sublattice
\begin{equation} \label{eqn:wormholesublattice}
\Gamma_0=\text{Span}_{\mathbb{Z}}\left(\mathcal{A}^{(1)}+\mathcal{A}^{(2)},\mathcal{B}^{(1)}-\mathcal{B}^{(2)}\right)\,,
\end{equation}
which geometrically corresponds to a bulk manifold $\widetilde{M}\cong\Sigma\times I$. This manifold can actually be included in the Chern-Simons calculation if we include a two vortex configuration as shown in Figure \ref{fig:permutation-wormhole}. The two vortices in the figure individually act on the fields by the swap $[(1\, 2)]$. The branched covering space of this vortex configuration has two boundaries, since one does not pick up a monodromy upon being transported along a cycle at the boundary, and it is not difficult to see that the topology of the covering space is indeed $\Sigma\times I$ (see Figure \ref{fig:permutation-wormhole-cover}). One can also come up with stranger covering spaces which are not topologically $\Sigma\times I$ by, for example, applying Dehn twists to the vortex in Figure \ref{fig:permutation-wormhole}.

\begin{figure}
\centering

    \includegraphics[width=0.5\textwidth]{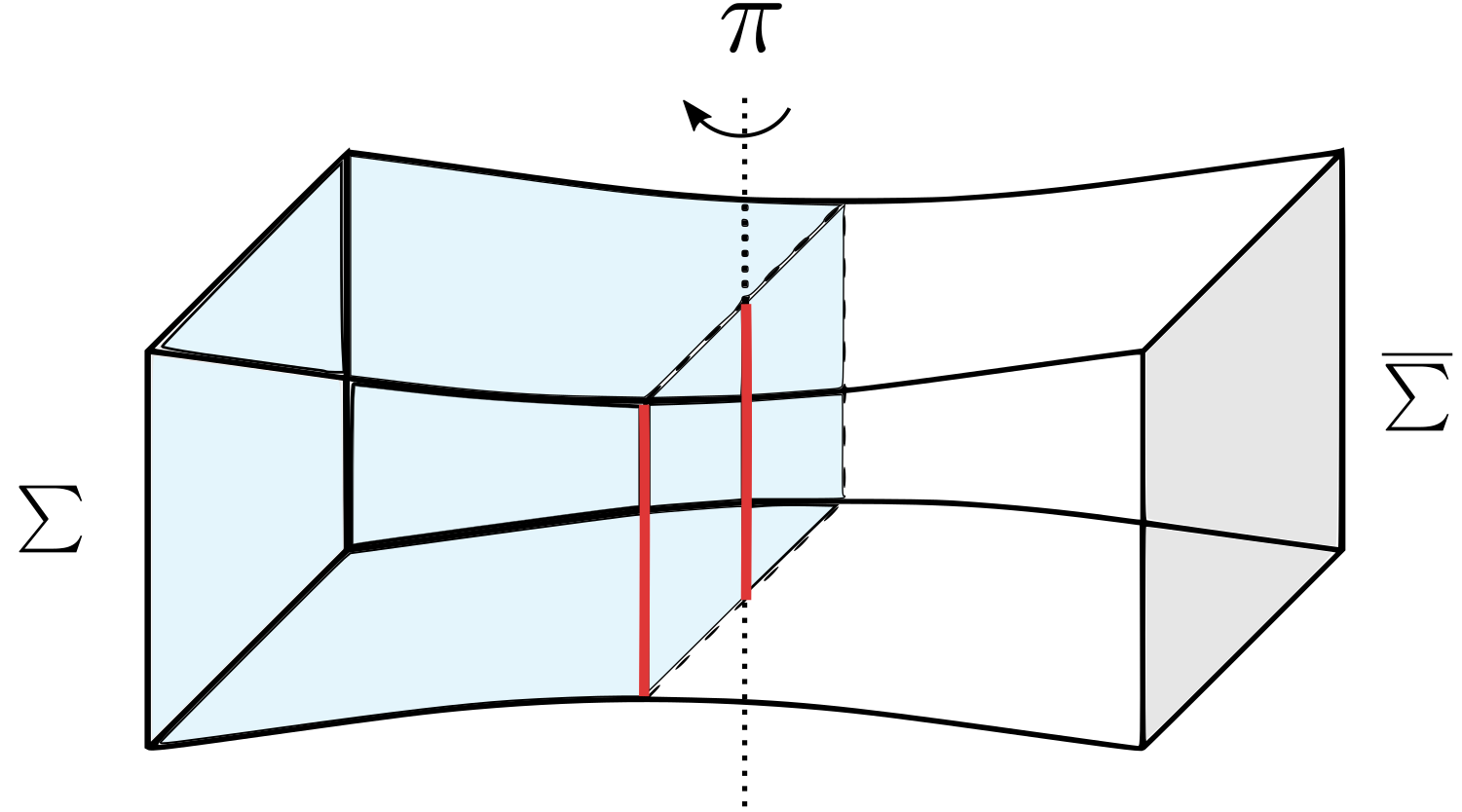}
\caption{The three-manifold $\widetilde{M}=\Sigma\times I$ has an orientation-preserving involution given by rotating the geometry around it's center axis by angle $\pi$. The fixed points of this involution are the two circles shown in red, which run through the bulk. The quotient $\widetilde{M}/\mathbb{Z}_2$ (light blue) is a solid torus with two vortices running through the non-contractible cycle.}
\label{fig:permutation-wormhole-cover}
\end{figure}

So now the natural question is: are there contributions to \eqref{eq:s2-symmetric-orbifold-disconnected} which cannot be recovered by a sufficiently complicated vortex configuration on the bulk Chern-Simons theory? As we will show, the answer is actually yes. Both the averaged symmetric orbifold and the Chern-Simons theory are summing over geometries in two different ways. The symmetric orbifold theory is summing over covering spaces $\widetilde{\Sigma}$ of the boundary manifold $\Sigma$, and then summing over fillings (Lagrangian sublattices) of $\widetilde{M}$. The Chern-Simons theory is summing fillings (Lagrangian sublattices) of $M$ and (branched) covering spaces $\widetilde{M}$ of $M$. Every contribution to the Chern-Simons theory, so long as $\widetilde{M}$ is specified by a Lagrangian sublattice of its boundary, computes a contribution to the symmetric orbifold partition function, since $\widetilde{\Sigma}:=\partial\widetilde{M}$ is always a covering space of $\Sigma$. However, given a covering space $\widetilde{\Sigma}$ of $\Sigma$ and a bulk three-manifold $\widetilde{M}$ specified by a Lagrangian sublattice of $\widetilde{\Sigma}$, it is not always true that $\widetilde{M}$ is a covering space (branched or otherwise) of a three-manifold $M$ with boundary $\Sigma$. Put pictorially, the diagram
\begin{equation}
\begin{tikzcd}[row sep=large,column sep=huge]
\widetilde{\Sigma} \arrow[r, hook, "\widetilde{i}"] \arrow[d, "\Gamma"]
& \widetilde{M} \\
\Sigma  & 
\end{tikzcd}
\end{equation}
does not always have a completion of the form
\begin{equation}\label{eq:extend-commute}
\begin{tikzcd}[row sep=large,column sep=huge]
\widetilde{\Sigma} \arrow[r, hook, "\widetilde{i}"] \arrow[d, "\Gamma"]
& \widetilde{ M} \arrow[d, "\Gamma"] \\
\Sigma \arrow[r, hook, "i"] & M
\end{tikzcd}
\end{equation}
where $\Gamma:\widetilde{M}\to M$ is a (branched) covering map.

In order for $\widetilde{M}$ to be a covering space of a three-manifold $M$ whose boundary is $\Sigma$, we need the Lagrangian sublattice $\Gamma_0$ to be `compatible' with the covering space structure of $\widetilde{\Sigma}$. The covering space $\Gamma:\widetilde{\Sigma}\to\Sigma$ comes equipped with a group of automorphisms (deck transformations), which are self-homeomorphisms $\phi$ of $\widetilde{\Sigma}$ such that $\Gamma(\phi(p))=\phi(p)$  for all $p\in\widetilde{\Sigma}$. This group $\text{Aut}(\widetilde{\Sigma}\to\Sigma)$ can be thought of as the set of symmetries of the covering space $\widetilde{\Sigma}$. Now, we specify a bulk manifold $\widetilde{M}$ by picking a Lagrangian sublattice $\Gamma_0\subset H_1(\widetilde{\Sigma},\mathbb{Z})$, and it is clear that $\widetilde{M}$ can only inherit the symmetries of $\widetilde{\Sigma}$ if the group $\text{Aut}(\widetilde{\Sigma}\to\Sigma)$ leaves the Lagrangian sublattice $\Gamma_0$ invariant.\footnote{An element $\phi\in\text{Aut}(\widetilde{\Sigma}\to\Sigma)$ has a natural action $\phi_*:H_1(\widetilde{\Sigma},\mathbb{Z})\to H_1(\widetilde{\Sigma},\mathbb{Z})$ given by simply pushing-forward one-cycles with $\phi$.}

A natural condition for $\widetilde{M}\to M$ to be a covering space respecting the structure of $\widetilde{\Sigma}\to\Sigma$ is for the Lagrangian sublattice $\Gamma_0$ to be invariant under the action of the deck transformations $\text{Aut}(\widetilde{\Sigma}\to\Sigma)$.\footnote{In the case that $\widetilde{\Sigma}$ is a regular covering of $\Sigma$ and $\widetilde{M}\to M$ is not branched, this is straightfoward. A regular covering $\widetilde{\Sigma}\to\Sigma$ has precisely the structure of a principal $G=\text{Aut}(\widetilde{\Sigma}\to\Sigma)$ bundle. The diagram \ref{eq:extend-commute} requires that $\widetilde{M}\to M$ is also a principal $G$ bundle, i.e. the deck transformations of $\widetilde{M}$ should be those of $\widetilde{\Sigma}$. Put another way, if $\widetilde{M}$ does not admit an action of the deck transformations $G$, then there is an obstruction for a base space $M$ to exist. We suspect that a similar logic exists if $\widetilde{\Sigma}$ is not regular and $\widetilde{M}\to M$ is branched. We thank Ivano Basile for pointing this out to us.} We emphasize, however, that this is not a sufficient condition, and that there could be Lagrangian sublattices $\Gamma_0$ which are invariant under the group $\text{Aut}(\widetilde{\Sigma}\to\Sigma)$, but for which the desired covering space $\widetilde{M}\to M$ does not exist.

Let us now return to the $S_2$ symmetric orbifold example. The sum in equation \eqref{eq:s2-symmetric-orbifold-disconnected} is over all Lagrangian sublattices $\Gamma_0$ of $H_1(\Sigma\sqcup\Sigma,\mathbb{Z})$. The group of deck transformations of the covering map $\Sigma\sqcup\Sigma\to\Sigma$ is simply $\mathbb{Z}_2$, generated by swapping the two copies of the tori. In terms of the homology, this acts as:
\begin{equation}
\phi_*(\mathcal{A}^{(1)})=\mathcal{A}^{(2)}\,,\quad\phi_*(\mathcal{B}^{(1)})=\mathcal{B}^{(2)}\,.
\end{equation}
According to the discussion above, only Lagrangian sublattices which are left invariant under the action of $\phi_*$ have a chance of being computed by a Chern-Simons calculation on a bulk manifold $M$. As an example of a Lagrangian sublattice which does not work, take
\begin{equation}\label{eq: Ads3 eucl btz lattice}
\Gamma_0=\text{Span}_{\mathbb{Z}}\left(\mathcal{A}^{(1)},\mathcal{B}^{(2)}\right)\,.
\end{equation}
This sublattice corresponds to a three-manifold which is a disjoint union of thermal $\text{AdS}_3$ and the Euclidean BTZ black hole. Of course, there is no three-manifold with single torus boundary whose 2-fold cover is a disjoint union of thermal $\text{AdS}_3$ and Euclidean BTZ, and so this Lagrangian sublattice has no chance of being reproduced by a Chern-Simons calculation. Indeed, $\Gamma_0$ is not invariant under the deck transformation $\phi_*$, as
\begin{equation}
\phi_*(\Gamma_0)=\text{Span}_{\mathbb{Z}}\left(\mathcal{A}^{(2)},\mathcal{B}^{(1)}\right)\neq\Gamma_0\,.
\end{equation}
As another example, we can take the sublattice
\begin{equation}\label{eq: Sublattice for figure}
\Gamma_0=\text{Span}_{\mathbb{Z}}\left(\mathcal{A}^{(1)}+\gamma(\mathcal{A}^{(2)}),\mathcal{B}^{(1)}-\gamma(\mathcal{B}^{(2)})\right)\,,
\end{equation}
for some modular transformation $\gamma$. This sublattice corresponds to a wormhole with topology $\Sigma\times I$ for which one boundary has modular parameter $\tau$ and the other has modular parameter $\gamma(\tau)$. The deck transformation $\phi_*$ only leaves $\Gamma_0$ invariant if the modular parameter is trivial, i.e. $\gamma=\text{id}$, for which we simply recover the double-vortex configuration in Figure \ref{fig:permutation-wormhole}.
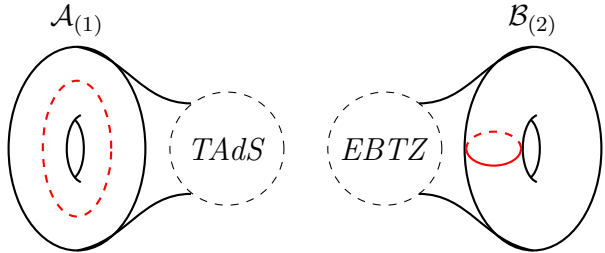
\begin{figure}
\centering
\ifdraft
\else
		\begin{center}
		\begin{tikzpicture}[scale = 1.5]
			
			\begin{scope}[scale = 0.3]
				\draw[thick] (0,0) [partial ellipse = 0:360:2 and 3];
				\draw[thick] (0.2,0) [partial ellipse = 100:260:0.5 and 1];
				\draw[thick] (-0.2,0) [partial ellipse = -70:70:0.4 and 0.9];
				\node[above] at (0,3) {$\mathcal{A}_{(1)}$};
				\draw[thick, red, dashed] (0,0) [partial ellipse = 0:360:1 and 2];
			\end{scope}
			
			\draw[thick] (0,0.9) to[out = -10, in = 180] (1,0.4);
			\draw[thick] (0,-0.9) to[out = 10, in = 180] (1.0,-0.4);
			
				\begin{scope}[scale = 0.5]
				\draw [dashed] (6.4,0) arc[x radius = 1 ,y radius = 1, start angle = 0, end angle =360];
				\node at (5.4,0) {\large\textit{EBTZ}};
		     	\end{scope}
		
			\begin{scope}[scale = 0.5]
				\draw [dashed] (3.63,0) arc[x radius = 1 ,y radius = 1, start angle = 0, end angle =360];
				\node at (2.63,0) {\large\textit{TAdS}};
			\end{scope}
			
			\draw[thick] (4,0.9) to[out =190, in = 0] (3,0.4);
			\draw[thick] (4,-0.9) to[out =-190, in = 0] (3.0,-0.4);
			
			\begin{scope}[xshift = 4cm, scale = 0.3]
				\draw[thick] (0,0) [partial ellipse = 0:360:2 and 3];
				\draw[thick] (0.2,0) [partial ellipse = 100:260:0.5 and 1];
				\draw[thick] (-0.2,0) [partial ellipse = -70:70:0.4 and 0.9];
				\node[above] at (0,3) {$\mathcal{B}_{(2)}$};
				\draw[thick, red, dashed] (-1.15,0) [partial ellipse = 0:180:0.8 and 0.5];
				\draw[thick, red] (-1.15,0) [partial ellipse = 180:360:0.8 and 0.5];
			\end{scope}
			
		\end{tikzpicture}
	\end{center}
\fi
\caption{The geometry corresponding to the choice of Lagrangian sublattice \eqref{eq: Ads3 eucl btz lattice} is a disjoint union of thermal $\text{AdS}_3$ and the BTZ black hole. Such a geometry does not have an interpretation when we have a single bulk manifold $M$, but it can be included in the gravitational path integral if we allow different gauge fields $A_I$ to live on independent bulk manifolds $M_I$ with the ``same'' asymptotic boundary.}
\label{fig:disconn-bdy-BTZ-AdS3}
\end{figure}

While the average of the disconnected component of the symmetric orbifold partition function contains bulk geometries which do not inherit the $\mathbb{Z}_2$ automorphism present in the 2D theory, we note that such bulk geometries come in pairs. For example, the two Lagrangian sublattices
\begin{equation} \label{eqn:sublatticebreaksreplicasym}
\Gamma_0^{(1)}=\text{Span}_{\mathbb{Z}}\left(\gamma(\mathcal{A}^{(1)}),\rho(\mathcal{A}^{(2)})\right)\,,\quad\Gamma_0^{(2)}=\text{Span}_{\mathbb{Z}}\left(\rho(\mathcal{A}^{(1)}),\gamma(\mathcal{A}^{(2)})\right)
\end{equation}
transform into one another under the $\mathbb{Z}_2$ automorphism swapping the two boundary tori. More generally, geometries contributing to \eqref{eq:s2-symmetric-orbifold-disconnected} either transform as singlets under the $\mathbb{Z}_2$ automorphism, or as doublets. Sublattices $\Gamma_0$ which transform as singlets correspond to bulk geometries $M$ which inherit the boundary automorphism, and thus can be realized as the double cover of a manifold $M/\mathbb{Z}_2$, which we think of as a bulk geometry with bulk vortices.

It is tempting to also think of pairs $(\Gamma_0^{(1)},\Gamma_0^{(2)})$ of sublattices which transform as $\mathbb{Z}_2$ doublets as also corresponding, in some abstract sense, to the `double cover' of a generalized bulk geometry (or `microgeometry', borrowing the terminology of \cite{Eberhardt:2021jvj}). Such a generalized notion of a bulk geometry is not entirely far-fetched, and has been considered before in the context of pure 3D gravity (see the discussion in Section 4.2 of \cite{Maloney:2007ud}, and also Section 3.1 of \cite{Yin:2007gv}). In that case, holomorphic factorization of the CFT dual to pure gravity required the introduction of geometries for which the left- and right-moving boundary gravitons lived on separate bulk geometries with the same boundary. Analogously, it is possible that the theory of quantum gravity dual to the Narain-averaged symmetric orbifold includes, in its sum over `geometries', contributions for which separate copies of the Chern-Simons gauge fields probe different classical bulks with the same boundary.

\subsection[General \texorpdfstring{$N$}{N}]{\boldmath General \texorpdfstring{$N$}{N}} \label{sec:4.3}

Now that we have seen the details of how the Narain-averaged symmetric orbifold and $\text{U}(1)^D\times\text{U}(1)^D\wr S_N$ Chern-Simons theory are related for $N=2$, let us turn our attention to generic $N$. We find it convenient to work in the grand canonical ensemble. We define the grand canonical partition function of the symmetric orbifold to be
\begin{equation}
\mathfrak{Z}(\tau,p)=\sum_{N=0}^{\infty}p^N Z_{\mathbb{T}^D \wr S_N} (\tau)\,,
\end{equation}
which, as we have seen, admits a nice expression in terms of Hecke operators
\begin{equation}
\mathfrak{Z}(\tau,p)=\exp\left(\sum_{n=1}^{\infty}p^nT_n Z_{\mathbb{T}^D}(\tau)\right)\,.
\end{equation}
We can also define the grand canonical ensemble of the bulk Chern-Simons theory by specifying a solid torus $M$ with boundary modular parameter $\tau$, as well as a vortex locus $L$, which we keep implicit, which runs along the non-contractible cycle of $M$
\begin{equation}
\mathfrak{Z}_{\text{CS}}(M,p)=\sum_{N=0}^{\infty}p^NZ_{G \wr S_N}(M).
\end{equation}
As in the case of $N=2$, we propose that the correct thing to do is to include the vortex $L$ in the definition of the bulk partition function and to allow it to take any charge. That is, the $\text{U}(1)^D\times\text{U}(1)^D\wr S_N$ partition function is computed by summing over all degree $N$ covering spaces of $M$ branched along $L$ with any allowed branching structure. Since $\pi_1(M\backslash L)\cong\pi_1(\partial M)$ (i.e. $M\backslash L$ retracts onto $\partial M$) the covering spaces of $M$ branched over $L$ are in one-to-one correspondence with the (unbranched) covering spaces of the boundary torus. This means the combinatorial counting of bulk and boundary covering spaces matches. The grand canonical partition function includes both connected and disconnected covering spaces, and through standard combinatorial arguments it is given by the exponential of the connected covering spaces
\begin{equation}
\mathfrak{Z}_{\text{CS}}(M,p)=\exp\left(\sum_{n=1}^{\infty}p^nT_n Z_{G}\lr{\tau}\right)\,.
\end{equation}
Indeed, the connected covering spaces are simply handlebodies whose boundaries are $N$-fold covering spaces of the boundary torus of $M$. From the above we can extract a formula for the partition function of the $G \wr S_N$ Chern-Simons theory with a vortex along the non-contractible cycle by keeping all terms with a power of $p^N$ which is
\be \label{eqn:SymNZFull_bulk}
Z_{G \wr S_N} (M) = \sum_{\text{parititons of $N$}} \prod_{k=1}^N \frac{1}{N_k!} \left(T_k Z_G(\tau) \right)^{N_k},
\ee
where the partitions of $N$ are $\sum_{k=1}^N k N_k = N$. This can be compared to the boundary partition function given in equation \eqref{eqn:SymNZFull}. The connected part of the partition function is given by
\begin{equation}
Z_{G\wr S_N,\text{conn.}}(M)=T_NZ_{G}(\tau)\,.
\end{equation}
Which can also be obtained by recalling that the Hecke operator $T_N$ sums over all connected covering spaces of the original torus. To obtain the bulk partition function we additionally need to sum over all bulk handlebodies $M$, which is implemented by the sum over modular images. 

We claim that the averaged free energy of the grand canonical symmetric orbifold exactly equals the free energy of the grand canonical Chern-Simons theory, summed over all solid tori $M$, i.e.
\begin{equation}\label{eq:connected-duality-claim}
\int_{\mathcal{M}_D}\mathrm{d}\mu\,\log\mathfrak{Z}(\tau,p)=\sum_{M}\log\mathfrak{Z}_{\text{CS}}(M,p)\,.
\end{equation}
This is equivalent to stating that the connected covering space contribution to the symmetric orbifold is exactly reproduced by the bulk Chern-Simons theory. To check this claim, note that the averaged symmetric orbifold free energy takes the form
\begin{equation}\label{eq:general-n-sym-averaged-free-energy}
\begin{split}
\int_{\mathcal{M}_D}\mathrm{d}\mu\,\log\mathfrak{Z}(\tau,p)&=\sum_{n=1}^{\infty}p^n\Braket{T_nZ(\tau)}\\
&=\sum_{n=1}^{\infty}\frac{p^n}{n}\sum_{\gamma\in\Gamma_{\infty}\backslash{SL}(2,\mathbb{Z})}\sum_{\gamma'\in\text{SL}(2,\mathbb{Z})\backslash M_n}\frac{1}{|\eta(\gamma\cdot\gamma'\cdot\tau)|^{2D}}\,,
\end{split}
\end{equation}
where $M_n$ is the set of all $2\times 2$ integer matrices and we have used the definition \eqref{eq:Hecke-operator-modular} of the $n^{\text{th}}$ Hecke operator. Now, the sum over geometries in the Chern-Simons free energy is implemented by a sum over modular images of the boundary torus. We have
\begin{equation}\label{eq:general-n-CS-free-energy}
\begin{split}
\sum_{M}\log\mathfrak{Z}_{\text{CS}}(M,p)&=\sum_{\gamma\in\Gamma_{\infty}\backslash\text{SL}(2,\mathbb{Z})}\sum_{n=1}^{\infty}p^nT_nZ_{G}(\gamma\cdot\tau)\\
&=\sum_{n=1}^{\infty}\frac{p^n}{n}\sum_{\gamma\in\Gamma_{\infty}\backslash{SL}(2,\mathbb{Z})}\sum_{\gamma'\in\text{SL}(2,\mathbb{Z})\backslash M_n}\frac{1}{|\eta(\gamma'\cdot\gamma\cdot\tau)|^{2D}}\,,
\end{split}
\end{equation}
where we have again used the definition of the Hecke operator $T_n$. Equations \eqref{eq:general-n-sym-averaged-free-energy} and \eqref{eq:general-n-CS-free-energy} appear to yield different results, since the summand of one includes the modular parameter $\gamma\cdot\gamma'\cdot\tau$, while the other includes $\gamma'\cdot\gamma\cdot\tau$. However, it turns out\footnote{See again Theorem 6.9 of \cite{Iwaniec1997TopicsIC}.} that the sums \eqref{eq:general-n-sym-averaged-free-energy} and \eqref{eq:general-n-CS-free-energy} are composed of all the same terms, simply shuffled around. That is,
\begin{equation}
\sum_{\gamma\in\Gamma_{\infty}\backslash{SL}(2,\mathbb{Z})}\sum_{\gamma'\in\text{SL}(2,\mathbb{Z})\backslash M_n}\frac{1}{|\eta(\gamma\cdot\gamma'\cdot\tau)|^{2D}}=\sum_{\gamma\in\Gamma_{\infty}\backslash{SL}(2,\mathbb{Z})}\sum_{\gamma'\in\text{SL}(2,\mathbb{Z})\backslash M_n}\frac{1}{|\eta(\gamma'\cdot\gamma\cdot\tau)|^{2D}}\,.
\end{equation}
This proves the claim of \eqref{eq:connected-duality-claim}.

Since the free energy of the grand canonical partition function (either in the case of the symmetric orbifold or of Chern-Simons) computes the connected contribution, we have the following result:
\begin{quote}
\textit{The ensemble average of the connected part of the $\mathbb{T}^D\wr S_N$ orbifold torus partition function is equal to the connected part of the $\text{U}(1)^D\times\text{U}(1)^D\wr S_N$ Chern-Simons partition function, summed over all handlebodies bounded by the CFT torus for all $N$.}
\end{quote}
\be
\Braket{Z_{\mathbb{T}^D\wr S_N,\,\text{conn.}}(\tau)} =Z_{\text{Bulk, conn.}}.
\ee
As we have seen in the $N=2$ example, however, the disconnected parts of the two theories cannot so easily be matched.

\subsection*{Disconnected Part}
We now briefly comment on what portion of the disconnected partition function the Chern-Simons calculation reproduces. We restrict to a torus handlebody $M$ with a single vortex operator running along the non-contractible cycle $L$. As explained earlier, since $\pi_1(M \backslash L) \cong \pi_1(\partial M)$ the branched covering spaces of $M$ exactly match the covering spaces of the boundary torus. The disconnected part of the bulk Chern-Simons partition function is given by discarding the connected covering spaces from the full answer in equation \eqref{eqn:SymNZFull_bulk} and summing over modular images
\be \label{eqn:SymNZFull_disconnected}
Z_{\text{Bulk, dis.}} = \sum_{\gamma\in\Gamma_{\infty}\backslash{SL}(2,\mathbb{Z})}\sum_{\text{parititons of $N$}} \prod_{k=1}^{N-1} \frac{1}{N_k!} \left(T_k Z_G(\gamma \cdot \tau) \right)^{N_k},
\ee
where now we sum over disconnected partitions of $N$ by imposing $\sum_{k=1}^{N-1} k N_k = N$. Similarly, the contribution of disconnected covering spaces to the boundary ensemble average is given by \eqref{eqn:SymNZFull}
\be \label{eqn:Boundary_Disconnected contribution_average}
\Braket{Z_{\mathbb{T}^D\wr S_N,\,\text{dis.}}(\tau)} = \sum_{\text{parititons of $N$}} \left\langle \prod_{k=1}^{N-1} \frac{1}{N_k!} \left(T_k Z(\tau) \right)^{N_k} \right\rangle,
\ee
where we must use the disconnected Siegel-Weil formula \eqref{eq:generic-siegel-weil-disconnected} to evaluate the average. One of the contributions to the above average will be given by filling in the ``same cycle'', specified by a modular parameter $\gamma$, on each disconnected covering torus in \eqref{eqn:Boundary_Disconnected contribution_average}. More precisely, these are the configurations where the preimage of the cycles on the covering tori map to the same cycle on the base torus.

This precisely matches the bulk computation in \eqref{eqn:SymNZFull_disconnected} after using Theorem 6.9 of \cite{Apostol:1989xyz} to commute the sum over modular images with the Hecke operator sum. However, the other terms appearing in the boundary average will not be reproduced by the bulk computation. We therefore have that the bulk Chern-Simons computation is strictly contained within the boundary average
\be
\Braket{Z_{\mathbb{T}^D\wr S_N,\,\text{dis.}}(\tau)} \supset Z_{\text{Bulk, dis.}}.
\ee

\subsection{Higher-genus boundaries}

Let us now make a few comments about the case of geometries whose boundaries have genus $g\geq 2$. We only comment on the connected components.

In the symmetric orbifold theory on a surface $\Sigma_g$ of genus $g$, the connected component of the partition function can be expressed as
\begin{equation}
Z_{\mathbb{T}^D\wr S_N, \text{conn.}}(m, \Sigma_g)=\frac{1}{N}\sum_{\widetilde{\Sigma}\to\Sigma_g}Z_{\mathbb{T}^D}(m, \widetilde{\Sigma})\,,
\end{equation}
where the sum is over all connected unramified covering surfaces $\widetilde{\Sigma}\to\Sigma_g$ of degree $N$. As in the case of the torus, these covering spaces are constructed by summing over all twisted boundary conditions around the cycles of $\Sigma_g$, with the requirement that around contractible cycles the resulting boundary conditions are trivial. As noted above, such surfaces have constrained topology, and specifically have genus
\begin{equation}
g'=N(g-1)+1\,,
\end{equation}
We argued in Section \ref{sec:averaging-at-higher-genus} that the average of this connected component over the Narain moduli space is given as a sum of the $\text{U}(1)^D\times\text{U}(1)^D$ Chern-Simons partition function on handlebodies whose boundaries are the covering surfaces $\widetilde{\Sigma}_{g'}$:
\begin{equation}\label{eq:symmetric-orbifold-higher-genus-average}
\Braket{Z_{\mathbb{T}^D\wr S_N, \text{conn.}}(m, \Sigma_g)}=\frac{1}{N}\sum_{\widetilde{\Sigma}_{g'}\to\Sigma_g}\sum_{\partial\widetilde{M}=\widetilde{\Sigma}}Z_{G}(\widetilde{M})\,.
\end{equation}

\begin{figure}
\centering
\ifdraft
\else
\begin{tikzpicture}[scale = 1.3]
\draw[dashed] (1,-0.725) [partial ellipse = 90:270:0.25 and 0.57];
\draw[dashed] (4,-0.725) [partial ellipse = 90:270:0.25 and 0.57];
\draw[smooth, line width = 0.1cm, white] (2.5,0.1) to[out = 180, in = 0] (1,-0.6) to[out = 180, in = -90] (0,0) to[out = 90, in = 180] (1,0.75) to[out = 0, in = 180] (2.5,0.1) to[out = 0, in = 180] (4,0.75) to[out = 0, in = 90] (5,0) to[out = -90, in = 0] (4,-0.6) to[out = 180, in = 0] (2.5,0.1);
\draw[smooth, thick, blue] (2.5,0.1) to[out = 180, in = 0] (1,-0.6) to[out = 180, in = -90] (0,0) to[out = 90, in = 180] (1,0.75) to[out = 0, in = 180] (2.5,0.1);
\draw[smooth, thick, blue] (2.5,0.1) to[out = 0, in = 180] (4,0.75) to[out = 0, in = 90] (5,0) to[out = -90, in = 0] (4,-0.6) to[out = 180, in = 0] (2.5,0.1);
\fill[CornflowerBlue, opacity = 0.2] (0,1) to[out=30,in=150] (2,1) to[out=-30,in=210] (3,1) to[out=30,in=150] (5,1) to[out=-30,in=30] (5,-1) to[out=210,in=-30] (3,-1) to[out=150,in=30] (2,-1) to[out=210,in=-30] (0,-1) to[out=150,in=-150] (0,1);
\fill[white] (1,-0.37) [partial ellipse = 47:133:0.75 and 0.5];
\fill[white] (1,0.49) [partial ellipse = -47:-133:0.75 and 0.65];
\fill[white] (4,-0.37) [partial ellipse = 47:133:0.75 and 0.5];
\fill[white] (4,0.49) [partial ellipse = -47:-133:0.75 and 0.65];
\draw[thick] (1,-0.37) [partial ellipse = 49:131:0.75 and 0.5];
\draw[thick] (1,0.49) [partial ellipse = -40:-140:0.75 and 0.65];
\draw[thick] (4,-0.37) [partial ellipse = 49:131:0.75 and 0.5];
\draw[thick] (4,0.49) [partial ellipse = -40:-140:0.75 and 0.65];
\node[above] at (2,0.3) {$L$};

\draw[smooth, thick] (0,1) to[out=30,in=150] (2,1) to[out=-30,in=210] (3,1) to[out=30,in=150] (5,1) to[out=-30,in=30] (5,-1) to[out=210,in=-30] (3,-1) to[out=150,in=30] (2,-1) to[out=210,in=-30] (0,-1) to[out=150,in=-150] (0,1);
\draw (1,-0.725) [partial ellipse = -90:90:0.25 and 0.57];
\draw (4,-0.725) [partial ellipse = -90:90:0.25 and 0.57];
\node[above right] at (4.5,1.25) {$\partial M=\Sigma_g$};
\end{tikzpicture}
\fi
\caption{A handlebody $M$ is a singular foliation of its boundary $\Sigma_g$. The leaves of the foliation become singular at the `center' $L$ of $M$, which roughly resembles $\Sigma_g$ with the contractible cycles collapsed to points. In the Chern-Simons theory dual to the Narain-averaged symmetric orbifold, we treat $L$ as the locus of permutation gauge vortices.}
\label{fig:higher-genus-vortex}
\end{figure}
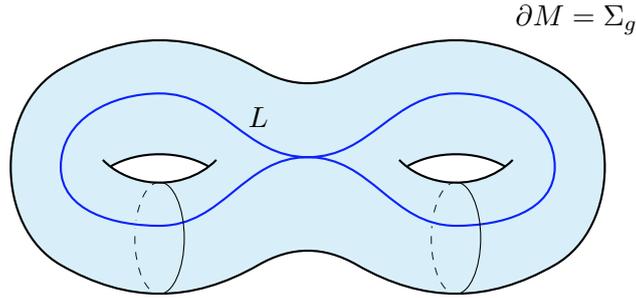

We can now try to interpret the above sum in terms of Chern-Simons theory. Let $M$ be a handlebody bounded by $\Sigma_g$. Next, let $L$ be the codimension 2 locus shown in Figure \ref{fig:higher-genus-vortex}. The fundamental group $\pi_1(M\backslash L)$ is isomorphic to the fundamental group of the boundary of $M$, i.e.
\begin{equation}
\pi_1(M\setminus L)\cong\pi_1(\Sigma_g)\,.
\end{equation}
Since covering spaces are classified by choices of consistent twisted boundary conditions around different cycles, that is homomorphisms $\phi:\pi_1(\Sigma_g) \to S_N$, we have that covering spaces of $M$ branched over $L$ have the same structure as covering spaces of $\Sigma_g$.\footnote{The isomorphism $\pi_1(M\setminus L)\cong\pi_1(\Sigma_g)$ is due to the fact that a handlebody can be foliated by copies of its boundary, up to a singular locus given by $L$. That is, $M\setminus L$ is homeomorphic to $\Sigma_g\times[0,1)$, see \cite{Porrati:2021sdc}.} Furthermore, the covering spaces $\widetilde{M}$ will also be handlebodies. There is one subtlety regarding the singular locus $L$. Earlier we demanded that vortex operators were defined along a dimension one submanifold of $M$, but $L$ is not a manifold so it is not obvious in what sense a vortex operator can be associated to $L$. However, it remains perfectly consistent to impose twisted boundary conditions on the bulk Chern-Simons fields as they travel around $L$. Since $\pi_1(\Sigma_g) \cong \pi_1(M\backslash L)$ a choice of twisted boundary conditions on $\Sigma_g$ descends to a consistent set of monodromies for the gauge fields as they travel around different cycles of $L$, and we define our bulk theory by demanding such monodromies.

Coupling our Chern-Simons theory to topological gravity results in summing over all handlebodies $M$ with boundary $\Sigma_g$. Summarizing, we have the bulk contribution
\begin{equation}\label{eq:chern-simons-higher-genus-sum}
\sum_{\partial M=\Sigma_g}Z_{\text{CS},\text{conn.}}(M)=\frac{1}{N}\sum_{\partial M=\Sigma_g}\sum_{\substack{\widetilde{M}\to M\\\text{branched over }L}}Z_{G}(\widetilde{M})\,.
\end{equation}
This formula, however, will not reproduce the full symmetric orbifold answer \eqref{eq:symmetric-orbifold-higher-genus-average}, even at the level of the connected parts. Algebraically, this is rather straightfoward to see: the sum over handlebodies $\widetilde{M}$ in \eqref{eq:symmetric-orbifold-higher-genus-average} amounts to summing over Lagrangian sublattices of $H_1(\widetilde{\Sigma},\mathbb{Z})$, which can be repackaged into a sum over the quotient space $P_{g'}\backslash\text{Sp}(2g',\mathbb{Z})$, where $P_{g'}$ is the parabolic subgroup of $\text{Sp}(2g',\mathbb{Z})$. On the other hand, the sum over handlebodies in \eqref{eq:chern-simons-higher-genus-sum} is over Lagrangian sublattices of $H_1(\Sigma_g,\mathbb{Z})$, which in turn is a sum over images of a fixed sublattice under the action of $P_g\backslash\text{Sp}(2g,\mathbb{Z})$. Since $g'>g$ when $g\geq 2$ (and $N\geq 2$), these sums can't possibly be equal. On the other hand, when $g=1$ (i.e. for the torus partition function), we always have $g'=g$, and the sums contain all of the same terms.

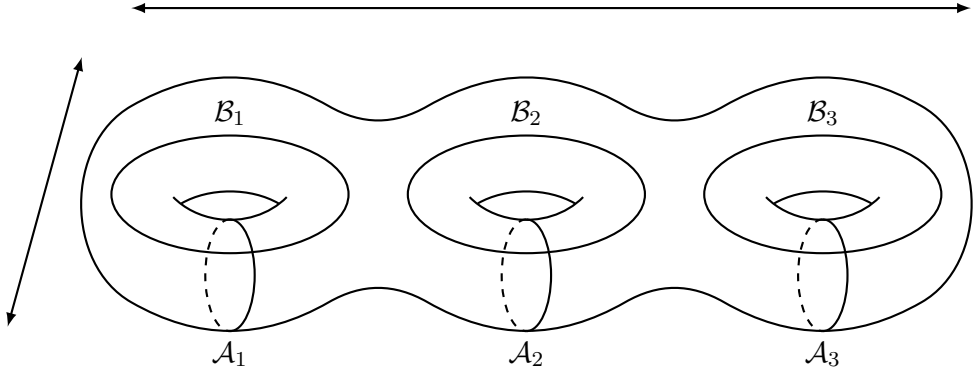
\begin{figure}
\centering
\ifdraft
\else
\begin{tikzpicture}[scale = 1.3]
\draw[thick] (1,-0.37) [partial ellipse = 49:131:0.75 and 0.5];
\draw[thick] (1,0.49) [partial ellipse = -40:-140:0.75 and 0.65];
\draw[thick] (4,-0.37) [partial ellipse = 49:131:0.75 and 0.5];
\draw[thick] (4,0.49) [partial ellipse = -40:-140:0.75 and 0.65];
\draw[thick] (7,-0.37) [partial ellipse = 49:131:0.75 and 0.5];
\draw[thick] (7,0.49) [partial ellipse = -40:-140:0.75 and 0.65];
\draw[smooth, thick] (0,1) to[out=30,in=150] (2,1) to[out=-30,in=210] (3,1) to[out=30,in=150] (5,1) to[out = -30, in = 210] (6,1) to[out = 30, in = 150] (8,1) to[out = -30, in = 30] (8,-1) to[out = 210, in = -30] (6,-1) to[out = 150, in = 30] (5,-1) to[out=210,in=-30] (3,-1) to[out=150,in=30] (2,-1) to[out=210,in=-30] (0,-1) to[out=150,in=-150] (0,1);
\draw[thick, dashed] (1,-0.725) [partial ellipse = 90:270:0.25 and 0.57];
\draw[thick] (1,-0.725) [partial ellipse = -90:90:0.25 and 0.57];
\draw[thick, dashed] (4,-0.725) [partial ellipse = 90:270:0.25 and 0.57];
\draw[thick] (4,-0.725) [partial ellipse = -90:90:0.25 and 0.57];
\draw[thick, dashed] (7,-0.725) [partial ellipse = 90:270:0.25 and 0.57];
\draw[thick] (7,-0.725) [partial ellipse = -90:90:0.25 and 0.57];
\draw[thick] (1,0.1) [partial ellipse = 0:360:1.2 and 0.6];
\draw[thick] (4,0.1) [partial ellipse = 0:360:1.2 and 0.6];
\draw[thick] (7,0.1) [partial ellipse = 0:360:1.2 and 0.6];
\node[below] at (1,-1.3) {$\mathcal{A}_1$};
\node[below] at (4,-1.3) {$\mathcal{A}_2$};
\node[below] at (7,-1.3) {$\mathcal{A}_3$};
\node[above] at (1,0.7) {$\mathcal{B}_1$};
\node[above] at (4,0.7) {$\mathcal{B}_2$};
\node[above] at (7,0.7) {$\mathcal{B}_3$};
\draw[thick, latex-latex] (0,2) -- (8.5,2);
\draw[thick, latex-latex] (-0.5,1.5) -- (-1.25,-1.25);
\end{tikzpicture}
\fi
\caption{A genus 3 surface which double covers a genus 2 surface. The deck transformation acts as a product of two reflections (shown), and maps the homology cycles as $\mathcal{A}_1,\mathcal{B}_1\to \mathcal{A}_3,\mathcal{B}_3$ and fixes $\mathcal{A}_2,\mathcal{B}_2$.}
\label{fig:genus-3-covers-genus-2}
\end{figure}

The more geometrical reason for the fact that not all geometries appearing in the average of the symmetric orbifold partition function can be recovered from the Chern-Simons theory was already discussed in Section \ref{sec:cs-calculations-n=2}. We can think of the symmetric orbifold calculation as finding a (connected) branched covering space $\widetilde{\Sigma}\to\Sigma_g$, and then choosing a Lagrangian sublattice $\Gamma_0$ of $H_1(\widetilde{\Sigma},\mathbb{Z})$, which in turn defines a handlebody $\widetilde{M}$. However, if $\Gamma_0$ does not fill in the cycles of $\widetilde{\Sigma}$ in a symmetric way (i.e. such that $\Gamma_0$ is invariant under the deck transformations $\text{Aut}(\widetilde{\Sigma}\to\Sigma_g)$), then $\widetilde{M}$ will not be the branched cover of some 3-manifold $M$ whose boundary is $\Sigma_g$. As a simple example, let us take $g=2$ and take the two-fold covering space shown in Figure \ref{fig:genus-3-covers-genus-2}. The covering surface has genus $g'=2g-1=3$, and its homology cycles are labeled by $\mathcal{A}_i,\mathcal{B}_i$ for $i=1,\ldots,3$ (also shown in Figure \ref{fig:genus-3-covers-genus-2}). The deck transformation group is $\text{Aut}(\widetilde{\Sigma}\to\Sigma_2)\cong\mathbb{Z}_2$, and simply consists of the operation of swapping the two sheets of the cover. On the homology elements, the nontrivial element $\phi\in\text{Aut}(\widetilde{\Sigma}\to\Sigma_2)$ acts as
\begin{equation}
\phi_*(\mathcal{A}_1)=\mathcal{A}_3\,,\quad\phi_*(\mathcal{A}_2)=\mathcal{A}_2\,,\quad\phi_*(\mathcal{B}_1)=\mathcal{B}_3\,,\quad\phi_*(\mathcal{B}_2)=\mathcal{B}_2\,.
\end{equation}
Any Lagrangian sublattice $\Gamma_0$ which is not invariant under this group of deck transformations has no hope of describing a bulk manifold $\widetilde{M}$ which is a covering space of a manifold $M$ with boundary $\partial M=\Sigma_2$. For example, the sublattice
\begin{equation}
\Gamma_0=\text{Span}_{\mathbb{Z}}\left(\mathcal{A}_1,\mathcal{A}_2,\mathcal{B}_3\right)
\end{equation}
is Lagrangian but is clearly not invariant under the deck transformations of $\widetilde{\Sigma}$.

In the case of the torus, this simply doesn't happen for connected covering spaces. This is because the set of deck transformations of a connected covering space acts trivially on homology, i.e.~all Lagrangian sublattices of the covering space are invariant under deck transformations. This explains, at least qualitatively, why we were able to get a match between the connected parts of the symmetric orbifold calculation and the bulk Chern-Simons theory in the case of a genus one boundary.

\subsection{Non-handlebody contributions}

Among the gravitational contributions considered so far, all have arisen either from handlebodies with genus-$g$ boundaries, potentially with vortices in the bulk. We now show that for surfaces of genus $g\geq 2$, the Narain-averaged partition function of $\mathbb{T}^D\wr S_N$ will also include \textit{smooth} bulk geometries which are not handlebodies. We work again with $N=2$ for simplicity.

Consider a surface $\Sigma_g$ of genus $g$, and consider the $\mathbb{T}^D\wr S_2$ partition function. It will take the form
\begin{equation}
Z_{\mathbb{T}^D\wr S_2}(m,\Sigma_g)=\frac{1}{2}Z_{\mathbb{T}^D}(m,\Sigma_g)Z_{\mathbb{T}^D}(m,\Sigma_g)+\cdots\,,
\end{equation}
where we are concentrating only on the contribution from the double cover $\Sigma_g\sqcup\Sigma_g\to\Sigma_g$ (of course, there will be other contributions, but we will not need them for our purposes). Upon averaging over the Narain moduli space, we have
\begin{equation}\label{eq:genus-g-disconnected-non-handlebody}
\Braket{Z_{\mathbb{T}^D\wr S_2}(m,\Sigma_g)}=\frac{1}{2}\Braket{Z_{\mathbb{T}^D}(m,\Sigma_g)Z_{\mathbb{T}^D}(m,\Sigma_g)}+\cdots\,.
\end{equation}
As we know, the average of $Z(\Sigma_g)Z(\Sigma_g)$ is computed by summing over all Lagrangian sublattices $\Gamma\subset H_1(\Sigma_g\sqcup\Sigma_g,\mathbb{Z})$, weighted by an appropriate one-loop determinant. A special class of sublattices $\Gamma$ are constructed by picking a basis $(\mathcal{A}_i,\mathcal{B}_i)$ for the homology group of $\Sigma_g$, and letting
\begin{equation}\label{eq:non-handlebody-lagrangian-sublattice}
\Gamma=\text{Span}_{\mathbb{Z}}\left(\mathcal{A}_i^{(1)}+\mathcal{A}_i^{(2)},\mathcal{B}_i^{(1)}-\mathcal{B}_{i}^{(2)}\right)\,,
\end{equation}
where the $(1)$ and $(2)$ superscripts differentiate between the two copies of $\Sigma_g$. This is the Lagrangian sublattice associated to a manifold which is topologically of the form $\widetilde{M}\cong\Sigma_g\times I$. For $g\geq 2$, such manifolds are hyperbolic with metric
\begin{equation}
\mathrm{d}s^2=\mathrm{d}\rho^2+\cosh^2\rho\,\mathrm{d}s_{\Sigma_g}^2\,,
\end{equation}
where the coordinate $\rho\in\mathbb{R}$ is the coordinate along the interval. Furthermore, for special choices of complex structure on $\Sigma_g$ the boundary admits a fixed point free $\mathbb{Z}_2$ involution which reverses the orientation of $\Sigma_g$. Combining this boundary involution with the bulk reversal $\rho\to -\rho$ we obtain a bulk involution $\iota$ \cite{Yin:2007gv,Yin:2007at}. This involution acts without fixed points, and we can consider the quotient manifold
\begin{equation}
M:=\widetilde{M}/\iota\,.
\end{equation}
As noted in \cite{Yin:2007at,Yin:2007gv}, this geometry is smooth and hyperbolic with boundary $\Sigma_g$, but is not a handlebody. Specifically, it has the following properties. Given the inclusion map $i:\Sigma_g\hookrightarrow\mathcal{M}$, the induced map $i_*:\pi_1(\Sigma_g)\hookrightarrow\pi_1(\mathcal{M})$:
\begin{itemize}

    \item is injective. This means that the non-contractible cycles of $\Sigma_g$ do not become contractible when viewed as cycles of $\mathcal{M}$.

    \item is not surjective. This means that there are non-contractible cycles of $\mathcal{M}$ which are not visible from the boundary $\Sigma_g$.

\end{itemize}
The first property means that $\mathcal{M}$ is not associated to \textit{any} Lagrangian sublattice $\Gamma_0\subset H_1(\Sigma_g,\mathbb{Z})$. The second property means that there are generators of the fundamental group which do not exist on the boundary. This is why $\mathcal{M}$ is able to be both smooth and have a connected double cover $\widetilde{M}$ which has a disconnected boundary.

By the standard $\text{U}(1)$-gravity dictionary, we should be able to associate the contribution of the sublattice \eqref{eq:non-handlebody-lagrangian-sublattice} in the averaged partition function \eqref{eq:genus-g-disconnected-non-handlebody} to the path integral of $\text{U}(1)^D\times\text{U}(1)^D$ Chern-Simons theory on the wormhole geometry $\widetilde{M}$. This, in turn, should be reproduced by the $\text{U}(1)^D\times\text{U}(1)^D\wr S_2$ partition function on $M$ with a nontrivial monodromy around the `internal' generator(s) of $\pi_1(M)$ (i.e. the generator(s) of $\pi_1(M)$ which are not inherited from $\pi_1(\Sigma_g)$).

We emphasize that the gravitational instanton associated to \eqref{eq:non-handlebody-lagrangian-sublattice} appears to be a \textit{smooth} bulk manifold $M$ with a connected boundary, but which is not a handlebody. This is worth emphasizing, since $\text{U}(1)$ gravity with a connected boundary (i.e. the bulk dual of the non-orbifolded Narain ensemble) includes only handlebodies in its sum over geometries.\footnote{Strictly speaking, $\text{U}(1)$ gravity only classifies bulk geometries by their associated Lagrangian sublattice, which for connected boundaries are in one-to-one correspondence with handlebodies.} It would be thus be interesting to explore further in what sense the ensemble average of $\mathbb{T}^D\wr S_N$ CFTs includes more generic gravitational instantons which are not visible in $\text{U}(1)$ gravity, but which are generally expected to be included in a more general theory of three-dimensional quantum gravity (for example, semiclassical gravity).

\section{Averaging correlation functions}\label{sec:correlators}

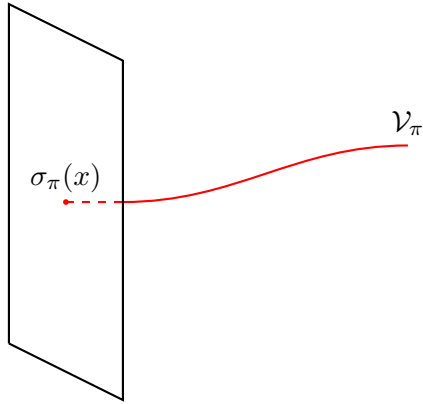
\begin{figure}
\centering
\begin{tikzpicture}[scale = 0.75]
\draw[thick, dashed, red] (1,2.5) -- (2,2.5);
\draw[thick, red] (2,2.5) to[out = 0, in = 180] (7,3.5);
\draw[thick] (0,0) -- (0,6) -- (2,5) -- (2,-1) -- (0,0);
\fill[red] (1,2.5) circle (0.05);
\node[above] at (1,2.5) {$\sigma_{\pi}(x)$};
\node[above] at (7,3.5) {$\mathcal{V}_{\pi}$};
\end{tikzpicture}
\caption{A twist field $\sigma_{\pi}(x)$ is holographically dual to a vortex $\mathcal{V}_{\pi}$ in the bulk which ends at the point $x$.}
\label{fig:bulk-boundary-twist-field}
\end{figure}

In the previous sections, we have considered the ensemble average of partition functions of permutation orbifolds $\mathbb{T}^D\wr\,\Omega$ (mostly for $\Omega=S_N$) and showed that a large family of contributions can be recovered from a Chern-Simons theory with gauge group $\text{U}(1)^D\times\text{U}(1)^D\wr\Omega$ coupled to topological gravity. In this section, we consider correlation functions of permutation orbifolds and explore their bulk interpretations.

Before jumping into calculations, let us briefly summarize the main idea. In Section \ref{sec:bulk-theory}, we found that the averaged partition function of the symmetric orbifold CFT is partially reproduced by a sum over bulk geometries which include nontrivial vortices. Roughly, these vortices are holographically dual to the twisted-sector states of the orbifold theory on the boundary, see Figure \ref{fig:bulk-boundary-twist-field}. Now, let us consider a correlation function of twist fields\footnote{The exact definition of twist fields is discussed in detail below.}
\begin{equation*}
\Braket{\sigma_{\pi_1}(x_1)\cdots\sigma_{\pi_n}(x_n)}\,,
\end{equation*}
where $\pi_i\in S_N$ are permutations specifying the monodromy of fundamental fields around the point $x_i$. Just as in the case of the partition function of the symmetric orbifold, such correlators are determined by passing to a covering space $\Sigma\to\mathbb{CP}^1$ which is branched over the points $x_1,\ldots,x_n$, such that the branching structure is induced by the monodromies $\pi_1,\ldots,\pi_n$. Upon averaging, one is then instructed to sum over bulk geometries $\widetilde{M}$ filling in $\Sigma$, specified by a choice of Lagrangian sublattice of $H_1(\Sigma,\mathbb{Z})$.

\begin{figure}
\centering
\begin{tikzpicture}[scale = 0.75]
\draw[thick, dashed] (0,0) [partial ellipse = 0:180:4 and 1];
\fill[white] (0,0) circle (2);
\fill[gray, opacity = 0.2] (0,0) circle (2);
\draw[thick] (0,0) circle (2);
\draw[thick, red] (45:4) -- (45:1.8);
\draw[thick, red] (135:4) -- (135:1.8);
\draw[thick, red] (225:4) -- (225:1.8);
\draw[thick, red] (-45:4) -- (-45:1.8);
\draw[thick] (0,0) circle (4);
\draw[thick] (0,0) [partial ellipse = 0:-180:2 and 0.5];
\draw[thick, dashed] (0,0) [partial ellipse = 0:180:2 and 0.5];
\draw[thick] (0,0) [partial ellipse = 0:-180:4 and 1];
\node[above right] at (45:4) {$x_1$};
\node[above left] at (135:4) {$x_2$};
\node[below left] at (225:4) {$x_3$};
\node[below right] at (-45:4) {$x_4$};
\end{tikzpicture}
\caption{The bulk geometry dual to the four-point function of twist fields in the symmetric orbifold. In principle, the vortices can branch and tangle in complicated ways in the bulk.}
\label{fig:generic-four-point-vortex}
\end{figure}
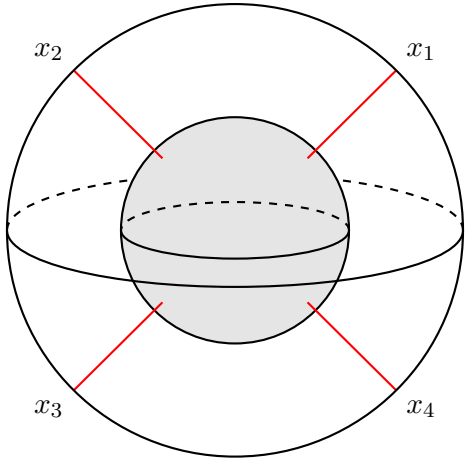

Holographically, one would expect this correlator to be dual to a bulk computation involving vortices $\mathcal{V}_{\pi_i}$ ending at the points $x_i$ on the boundary. Such vortices will need to end somewhere in the bulk, and in principle one should sum over all bulk configurations of the vortices. In terms of the bulk $G\wr S_N$ Chern-Simons theory, the calculation of the path integral in the presence of a nontrivial vortex configuration corresponds to passing to the branched covering of $\mathbb{H}^3$ over the locus of the vortices, with the branching structure determined by the monodromy specified on the components of the branching locus. Given that the vortices end on the boundary at the points $x_1,\ldots,x_n$, a branched cover of $\mathbb{H}^3$ will induce a branched cover of the boundary. However, the precise organization of the vortices in the bulk can drastically effect the topology of the 3-dimensional branched covering. Thus, each topological choice of configuration of vortices in the bulk will specify a 3-manifold $\widetilde{M}$ which is a branched covering of $\mathbb{H}^3$, and whose boundary is a branched covering $\Sigma$ of $\mathbb{CP}^1$ branched at $x_1,\ldots,x_n$ with monodromies $\pi_1,\ldots,\pi_n$.

The above discussion suggests that it is possible to reproduce the Narain average of symmetric orbifold correlators by summing over topologically distinct vortex configurations in $G\wr S_N$ Chern-Simons theory on $\mathbb{H}^3$. However, this is not always possible. For the same reasons discussed in Section \ref{sec:4.3}, while it is always true that a branched cover $\widetilde{M}$ over $\mathbb{H}^3$ always has boundary $\Sigma$ which is a branched cover of $\mathbb{CP}^1$, the converse is much more difficult to satisfy. Given a branched cover $\Sigma\to\mathbb{CP}^1$, it is not always the case that a given 3-manifold $\widetilde{M}$ with boundary $\Sigma$ admits the structure of a branched cover of $\mathbb{H}^3$ (or any other 3-manifold with boundary $\mathbb{CP}^1$ for that matter).

For the examples we explicitly consider in this section, however, this turns out to not be an issue. Specifically, in this section we consider the symmetric orbifold $\mathbb{T}^D\wr S_2$, for which there is a unique twist operator $\sigma_{(1\,2)}$.\footnote{It is possible to consider $\mathbb{T}^D\wr S_N$ theories with $N>2$. We leave this to future work.} We consider the correlation functions of the form
\begin{equation}
\Braket{\sigma_{(1\,2)}(x_1)\cdots\sigma_{(1\,2)}(x_n)}\,.
\end{equation}
Just as in \cite{Benjamin:2021wzr}, we find that the Narain averages of these correlators are indeed reproduced by a sum over bulk vortex configurations in $\text{U}(1)^D\times\text{U}(1)^D\wr S_2$ Chern-Simons theory, for which the vortices are constrained to lie in \textit{rational tangles}.\footnote{Our analysis differs slightly from that of \cite{Benjamin:2021wzr}, in that they consider orbifolds of the form $\mathbb{T}^D/\mathbb{Z}_n$, which are qualitatively different from symmetric orbifolds. As such, their bulk theory is of the form $\text{U}(1)^D\times\text{U}(1)^D\rtimes\mathbb{Z}_n$.} As such, we are able to reproduce the averaged symmetric orbifold correlation function via a bulk Chern-Simons calculation.\footnote{We emphasize that for $N>2$, the analysis does not work out so cleanly, and not every term in the averaged correlation functions will be reproducible by a bulk Chern-Simons calculation on a classical background.}

\subsection{Correlators in permutation orbifolds}

Before considering the ensemble average of correlation functions, let us first review the calculation of correlation functions in permutation orbifolds. For the rest of this section, we only concern ourselves with correlators of fields on the sphere ($\mathbb C\mathbb P ^1$) for simplicity.

As an orbifold theory, a permutation orbifold $X\wr\,\Omega$ contains non-local operators known as twist-fields which implement a monodromy on the fundamental fields of the theory. Denoting by $\boldsymbol{\Phi}=(\Phi^{(1)},\ldots,\Phi^{(N)})$ the collective fundamental fields of the tensor theory $X^{\otimes N}$, a twist operator $\sigma_{\pi}$ associated to a permutation $\pi\in\Omega$ is defined by the monodromy relation
\begin{equation}
\boldsymbol{\Phi}(e^{2\pi i}z+\zeta)\,\sigma_{\pi}(\zeta)=(\pi\cdot\boldsymbol{\Phi})(z+\zeta)\,\sigma_{\pi}(\zeta)\,,
\end{equation}
where
\begin{equation}
\pi\cdot\boldsymbol{\Phi}=(\Phi^{(\pi(1))},\ldots,\Phi^{(\pi(N))})\,.
\end{equation}
On their own, twist fields $\sigma_{\pi}$ are not gauge-invariant.We can see this as follows. Consider the monodromy of an element $\lr{\rho\cdot \boldsymbol{\Phi}}(z)$ around the twist field $\sigma_\pi(\zeta)$. The twist field acts on $\boldsymbol{\Phi}(z)$:
\begin{equation}
	\lr{\rho\cdot \boldsymbol{\Phi}}(e^{2\pi i}z+\zeta) \sigma_{\pi}(\zeta)= \lr{\rho\cdot\pi\cdot \boldsymbol{\Phi}}(z+\zeta)\sigma_{\pi}(\zeta) = \lr{\rho\cdot\pi\cdot\rho^{-1}\cdot\lr{\rho\cdot\boldsymbol{\Phi}}}(z+\zeta)\sigma_{\pi}(\zeta) \ .
\end{equation}
From this we can infer the action of the twist fields on the field  $\lr{\rho\cdot \boldsymbol{\Phi}}(z)$, which lies on the same gauge slice as $\boldsymbol{\Phi}(z)$:
\begin{equation}
	\lr{\rho\cdot \boldsymbol{\Phi}}(e^{2\pi i}z+\zeta)\sigma_{\pi}(\zeta) = \lr{\rho\cdot \boldsymbol{\Phi}}(e^{2\pi i}z+\zeta)\sigma_{\rho\pi\rho^{-1}}(\zeta)\ .
\end{equation}
Where on the RHS the twist field acts on $\lr{\rho\cdot \boldsymbol{\Phi}}$ and on the LHS on $\boldsymbol{\Phi}(z)$. Hence under an overall permutation $\rho\in\Omega$, twist fields transform as
\begin{equation}
\sigma_{\pi}\to\sigma_{\rho\pi\rho^{-1}}\,.
\end{equation}
From a twist field $\sigma_{\pi}$, we can construct a gauge-invariant twist field $\sigma_{[\pi]}$ by
\begin{equation}
\sigma_{[\pi]}=\mathcal{N}_{[\pi]}\sum_{\rho\in[\pi]}\sigma_{\rho}\,,
\end{equation}
where $\mathcal{N}_{[\pi]}=1/\sqrt{|[\pi]|}$ is a normalization factor so that $\sigma_{[\pi]}$ is canonically normalized.\footnote{Here, by `canonically normalized' we mean that the two-point function satisfies $\braket{\sigma_{[\pi]}(x_1)\sigma_{[\pi]}(x_2)}=1/(x_1-x_2)^{2h(\pi))}$, where $h(\pi)$ is the conformal weight of the twist field.} Note that $\sigma_{[\pi]}$ only depends on the conjugacy class of $\pi$ in $\Omega$. A generic correlator of gauge-invariant twist fields thus takes the form
\begin{equation}\label{eq: gauge inv. cor.}
\Braket{\prod_{i=1}^{n}\sigma_{[\pi_i]}(x_i)}=\prod_{i=1}^{n}\left(\mathcal{N}_{[\pi_i]}\right)\sum_{\rho_1,\ldots,\rho_n\in\Omega}\Braket{\prod_{i=1}^{n}\sigma_{\rho_i\pi_i\rho_i^{-1}}(x_i)}\,.
\end{equation}
That is, the correlators of gauge-invariant twist fields can be expressed purely in terms of an appropriate sum over correlators of the `pure' twist fields $\sigma_{\pi}$. From now on, we calculate only the correlators of pure twist fields $\sigma_{\pi}$, keeping in mind that we should sum over conjugacy classes to obtain a gauge-invariant result. Expression \eqref{eq: gauge inv. cor.} will in general have disconnected contributions, meaning contributions for which some terms of the right hand side factorize. Here we focus on the connected part of the correlators. This can be done by looking at correlators  for which the group elements that appear in the twist fields generate a transitive subgroup of the permutation group acting on the elements of $\{1,2,...,N\}$ that appear in the correlator (see e.g. \cite{Dei:2019iym}). 

Given a set of twist fields $\sigma_{\pi_i}$, we can compute the (sphere) correlation function
\begin{equation}
\braket{\sigma_{\pi_1}(x_1)\cdots\sigma_{\pi_n}(x_n)}
\end{equation}
in the following way \cite{Lunin:2000yv,Lunin:2001pw}. Within the path integral, the twist fields define twisted boundary conditions of the fundamental field on the punctured sphere $\mathbb{CP}^1\setminus\{x_1,\ldots,x_n\}$, and so we can pass to a covering space $\Sigma$ of $\mathbb{CP}^1$ which is ramified over the points $x_i$ such that the fundamental fields are single-valued on $\Sigma$.\footnote{Note that, unlike in previous sections, we are now using $\Sigma$ as opposed to $\widetilde{\Sigma}$ to refer to the covering space, since the base space is already specified to be $\mathbb{CP}^1$.} The covering space $\Sigma$ is related to the base sphere via a holomorphic map $\Gamma:\Sigma\to\mathbb{CP}^1$, and so we can exploit the conformal symmetry to pull back the fields of the seed theory $X$ to the covering space $\Sigma$, at the expense of a conformal anomaly in the path integral measure when $X$ has non-zero central charge. The result is that the correlation function of twist fields is given by
\begin{equation}\label{eq:correlationfunctiontwistfieldsgeneral}
\braket{\sigma_{\pi_1}(x_1)\cdots\sigma_{\pi_n}(x_n)}=e^{-S_L[\Phi_{\Gamma}]}Z_X(\Sigma)\,,
\end{equation}
where $Z_X(\Sigma)$ is the partition function of the seed theory on the surface $\Sigma$. The genus of the covering surface $\Sigma$ is fixed by the Riemann -  Hurwitz formula. Specifically, it is given in terms of the genus $g$ of the base surface, the number $M$ of distinct elements of $\{1,2,...,N\}$ that appear in the correlator and the lengths of the cycles $w_j$, $j=1,2,...,n$. Concretely, we have\begin{equation}
		g'-1 = M (g-1) + \frac12 \sum_{j=1}^{n}(w_j-1)\ . 
\end{equation} Since we focus on sphere correlators, $g=0$ and thus \begin{equation}
g'=1-M +  \frac12 \sum_{j=1}^{n}(w_j-1)\, .
\end{equation} 
Here, the Liouville action $S_L$ is given by
\begin{equation}\label{eq:liouville-action}
S_L[\Phi]=\frac{c}{48\pi}\int_{\Sigma}\mathrm{d}^2z\sqrt{g}\left(-\Phi\Delta\Phi+R\Phi\right)\,,
\end{equation}
where $c$ is the central charge of the seed CFT $X$, $R$ is the scalar curvature on $\Sigma$, and $\Delta$ is the Laplacian on $\Sigma$.\footnote{In our conventions, we have $\sqrt{g}\,\Delta\Phi=\partial(\sqrt{g}\,\overline{\partial}\Phi)+\overline{\partial}(\sqrt{g}\,\partial\Phi)$.} The conformal anomaly is found by evaluating this action on the scalar
\begin{equation}
\Phi_{\Gamma}=\log|\partial\Gamma|^2\,.
\end{equation}
This corresponds to a metric on the covering space \begin{equation}
    ds^2=e^{\Phi_{\Gamma}}\mathrm{d}z\,\mathrm{d}\overline{z}
\end{equation}
where $z, \overline{z}$ are local coordinates on the covering space.

\subsection{Narain-averaging correlators}

Let us now specify our CFT to be a Narain theory $X=\mathbb{T}^D$. We  write the correlators of twist fields in $\mathbb{T}^D\wr\,\Omega$ as
\begin{equation}
\braket{\sigma_{\pi_1}(x_1)\ldots\sigma_{\pi_n}(x_n)}=e^{-S_L[\Phi_{\Gamma}]}Z(m,\Sigma)\,,
\end{equation}
where $Z(m,\Sigma)$ is the partition function of the $\mathbb{T}^D$ theory on the surface $\Sigma$. We can now average this result over the moduli $m$ and we find 
\begin{equation}
\int_{\mathcal{M}_D}\mathrm{d}\mu(m)\,\braket{\sigma_{\pi_1}(x_1)\ldots\sigma_{\pi_n}(x_n)}=e^{-S_L[\Phi_{\Gamma}]}\int_{\mathcal{M}_D}\mathrm{d}\mu(m)\,Z(m,\Sigma)\,.
\end{equation}
However, the average of the partition function $Z$ is readily written down in terms of a sum over Lagrangian sublattices of $H_1(\Sigma,\mathbb{Z})$, i.e.
\begin{equation}
\int_{\mathcal{M}_D}\mathrm{d}\mu(m)\,\braket{\sigma_{\pi_1}(x_1)\ldots\sigma_{\pi_n}(x_n)}=e^{-S_L[\Phi_{\Gamma}]}\sum_{\substack{\Gamma_0\subset H_1(\Sigma,\mathbb{Z})}} Z_{\text{CS}}(\Gamma_0)\,,
\end{equation}
where $Z_{\text{CS}}(\Gamma_0)$ is shorthand for the $\text{U}(1)^D\times\text{U}(1)^D$ Chern-Simons partition function on the bulk manifold defined by the Lagrangian sublattice  $\Gamma_0$. In the case that $\Sigma$ is a connected surface, we can instead express the averaged correlation function in terms of a sum over handlebodies bounded by $\Sigma$, namely
\begin{equation}\label{eq:averaged-correlators-connected}
\int_{\mathcal{M}_D}\mathrm{d}\mu(m)\,\braket{\sigma_{\pi_1}(x_1)\ldots\sigma_{\pi_n}(x_n)}=e^{-S_L[\Phi_{\Gamma}]}\sum_{\substack{\text{handlebodies}\\\partial M=\Sigma}} Z_{\text{CS}}(\mathcal{M})\,.
\end{equation}
In the rest of this section, we will consider examples of such correlators and interpret them in terms of correlators of vortices of $\text{U}(1)^D\times\text{U}(1)^D\wr\,\Omega$ Chern-Simons theory on the hyperbolic ball. Completely analogously to \cite{Benjamin:2021wzr}, we find that the sum over geometries on the covering space can be interpreted as a sum over configurations of bulk vortices.

\subsection[Four-point functions]{\boldmath Four-point functions}\label{sec:n=2-four-point}

We begin with the simplest nontrivial example: the four-point function of the twist field $\sigma_{(1\,2)}$ in the orbifold $\mathbb{T}^D\wr S_2$. Since $S_2\cong\mathbb{Z}_2$, this analysis is very similar to that of \cite{Benjamin:2021wzr}. 

Without loss of generality, we can use the $\text{SL}(2,\mathbb{C})$ isometry on the sphere to place three points at $0,1,\infty$ and leave the fourth point arbitrary. That is, we consider the correlator
\begin{equation}
\braket{\sigma_{(1\,2)}(0)\sigma_{(1\,2)}(1)\sigma_{(1\,2)}(u)\sigma_{(1\,2)}(\infty)}\,.
\end{equation}
Since $S_2$ is abelian, this is actually a fully gauge-invariant correlator. This correlator is also connected as the permutation $\lr{12}$ acts in a transitive way on the elements $\{1,2\}$. The covering space on which the fundamental fields are single-valued is given by the torus whose modular parameter is related to the cross-ratio $u$ via the modular $\lambda$ function \cite{Dixon:1986qv}. That is,
\begin{equation} \label{eq:lambda-function}
		u(\tau) = 1- \lambda(\tau) = 1- \frac{\vartheta_{2}(\tau)^4}{\vartheta_3(\tau)^4}\,.
\end{equation}
The explicit form of the covering map $\Gamma$ is
\begin{equation}
\Gamma(z)=\frac{\mathfrak{p}(z;\tau)-\mathfrak{p}(\frac{1}{2};\tau)}{\mathfrak{p}(\frac{\tau}{2};\tau)-\mathfrak{p}(\frac{1}{2};\tau)}\,,
\end{equation}
where $\mathfrak{p}$ is the Weierstrass function. Given the form of the covering map, the conformal anomaly $e^{-S_L[\Phi_{\Gamma}]}$ can be explicitly worked out and, after regularization, takes the form \cite{Dixon:1986qv}
\begin{equation}
e^{-S_L[\Phi_{\Gamma}]}=\frac{1}{2^{2c/3}|u(1-u)|^{c/12}}\,.
\end{equation}
Therefore overall :
\begin{equation}
\braket{\sigma_{(1\,2)}(0)\sigma_{(1\,2)}(1)\sigma_{(1\,2)}(u)\sigma_{(1\,2)}(\infty)}=\left(\frac{1}{2^{2c/3}|u(1-u)|^{c/12}}\right)\left(\frac{\Theta(m,\tau)}{\left|\eta(\tau)\right|^{2D}}\right)\,.
\end{equation}
Putting everything together, the average of the four-point function of $\sigma_{(1\,2)}$ is computed using the Siegel-Weil formula, and we find
\begin{equation}\label{eq:averaged-2-4-point}
\begin{split}
\int_{\mathcal{M}_D}\mathrm{d}\mu(m)\,&\braket{\sigma_{(1\,2)}(0)\sigma_{(1\,2)}(1)\sigma_{(1\,2)}(u)\sigma_{(1\,2)}(\infty)}\\
&=\frac{1}{2^{2c/3}|u(1-u)|^{c/12}}\sum_{\gamma\in\Gamma_{\infty}\setminus\text{SL}(2,\mathbb{Z})}\frac{1}{|\eta(\gamma\cdot\tau)|^{2D}}\,,
\end{split}
\end{equation}
where $\tau$ is obtained from $u$ by inverting \eqref{eq:lambda-function} and the central charge $c$ is just the dimension $D$ of the torus target.\footnote{We keep the central charge $c$ arbitrary since, for example, for Narain theories with supersymmetry, the central charge is instead $c=3D/2$.} Specifically, $\tau$ admits a closed-form expression in terms of hypergeometric functions
\begin{equation}
\tau=\frac{_2F_1(\frac{1}{2},\frac{1}{2},1:u)}{_2F_1(\frac{1}{2},\frac{1}{2},1:1-u)}\,.
\end{equation}

\subsubsection*{The bulk interpretation}

As in \cite{Benjamin:2021wzr}, we can interpret the sum over modular images in \eqref{eq:averaged-2-4-point} in terms of $\text{U}(1)^D\times\text{U}(1)^D\wr\,S_2$ Chern-Simons theory in the following way. Consider the CFT sphere on which we calculate the correlation functions to be the boundary of the ball $\mathbb{H}^3$. We extend each twist field $\sigma_{(1\,2)}$ as a vortex in the bulk which meets the boundary at the point $x_i$. Each vortex in the bulk implements a monodromy $A_{(1)}\to A_{(2)}$ in the bulk gauge field. Since a vortex cannot just end at one point, we need to join pairs of boundary points by vortices. Let us choose for the moment a vortex which joins the point at $x=0$ with the point at $x=1$ and another which joins $x=u$ with $x=\infty$. The two strands are in principle allowed to cross and link in the bulk in an arbitrary fashion, so long as they do not cross. For example, the `trivial' configuration $\mathcal{T}_0$ on the left of Figure \ref{fig:rational-tangles} connects $0$ to $1$ and $u$ to $\infty$ in the simplest way possible -- with no crossings in the bulk. The bulk geometry found by taking the double branched cover of the ball over the vortex $\mathcal{T}_0$ has the property that the cycle generated by a loop encircling $u$ and $\infty$ is contractible in the bulk. For an illustration, see Figure \ref{fig:rational-tangles-bulk}. Indeed, as was noted in \cite{Benjamin:2021wzr}, the branched cover of the geometry in Figure \ref{fig:rational-tangles-bulk} is a handlebody with torus boundary.\footnote{This can be seen by `cutting open' the hyperbolic ball along branch cuts associated to the vortices and gluing a second copy along the same branch cuts, see Figure 2 of \cite{Benjamin:2021wzr}.}
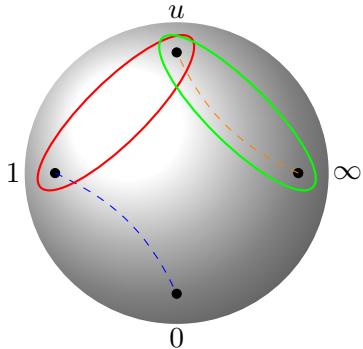
\begin{figure}
\centering
\ifdraft
\else
\begin{tikzpicture}[scale=2]
	\coordinate (O) at (0,0,0);
	\def\r{1}
	\shade[ball color=white] (O) circle (\r);
	\foreach \i in {1,2,...,4} {
		\coordinate (P\i) at ({90+90*(\i-1)}:\r);
		\coordinate (P\i') at ({90+90*(\i-1)}:\r*0.8);
		\filldraw[black] (P\i') circle (0.03);
	}
		\draw[blue,dashed,bend left=20] (P2') to (P3');
	\draw[orange,dashed,bend left=20] (P4') to (P1');
	\coordinate (C) at ($0.5*(P1')+0.5*(P2')$);
	\draw[red,thick,rotate=45] (C) ellipse (0.7cm and 0.2cm);
	\coordinate (C) at ($0.5*(P1')+0.5*(P4')$);
	\draw[green,thick,rotate=-45] (C) ellipse (0.7cm and 0.2cm);
	\coordinate (C) at ($1.2*(P1')$);
	\node[above] at (C) { $u$ };
	\coordinate (C) at ($1.2*(P4')$);
	\node[right] at (C) { $\infty$ };
	\coordinate (C) at ($1.2*(P2')$);
	\node[left] at (C) { $1$ };
	\coordinate (C) at ($1.2*(P3')$);
	\node[below] at (C) { $0$ };
\end{tikzpicture}
\fi
\caption{An illustration of the (non)contractibility of loops encircling points on the sphere. The green loop is contractible when continued in the bulk whereas the red loop is not (it has to cross the orange and blue vortices that live inside the sphere).  }
\label{fig:rational-tangles-bulk}
\end{figure}

\begin{figure}
\centering
\ifdraft
\else
\begin{tikzpicture}[scale = 0.6]
\begin{scope}[xshift = -12cm]
\node[below left] at (-2.15,-2.15) {$0$};
\node[above left] at (-2.15,2.15) {$1$};
\node[above right] at (2.15,2.15) {$u$};
\node[below right] at (2.15,-2.15) {$\infty$};
\begin{knot}[
clip width = 5,
flip crossing = 1,
]
\strand[very thick] (45:3cm) to[out = -135, in = 90] (0.5,0)
to[out = -90, in = 135] (-45:3cm);
\strand[very thick] (-135:3cm) to[out = 45, in = -90] (-0.5,0)
to[out = 90, in = -45] (135:3cm);
\end{knot}
\draw[thick] (0,0) circle (3);
\end{scope}
\begin{scope}[xshift = -4cm, rotate = 90]
\node[below right] at (-2.15,-2.15) {$\infty$};
\node[below left] at (-2.15,2.15) {$0$};
\node[above left] at (2.15,2.15) {$1$};
\node[above right] at (2.15,-2.15) {$u$};
\begin{knot}[
clip width = 5,
flip crossing = 1,
]
\strand[very thick] (45:3cm) .. controls +(-1,0) and +(1,0) .. (0,-0.5)
.. controls +(-1,0) and +(1,0) .. (135:3cm);
\strand[very thick] (-135:3cm) .. controls +(1,0) and +(-1,0) .. (0,0.5)
.. controls +(1,0) and +(-1,0) .. (-45:3cm);
\end{knot}
\draw[thick] (0,0) circle (3);
\end{scope}
\begin{scope}[xshift = 4cm, rotate = 90]
\node[below right] at (-2.15,-2.15) {$\infty$};
\node[below left] at (-2.15,2.15) {$0$};
\node[above left] at (2.15,2.15) {$1$};
\node[above right] at (2.15,-2.15) {$u$};
\begin{knot}[
clip width = 5,
flip crossing = 2,
]
\strand[very thick] (-45:3cm) to[out = 135, in = 0] (0.5,0.5) to[out = 180, in = 0] (-0.5,-0.5) to[out = 180, in = -45]  (135:3cm);
\strand[very thick] (-135:3cm) to[out = 45, in = 180] (-0.5,0.5)
to[out = 0, in = 180] (0.5,-0.5) to[out = 0, in = -135] (45:3cm);
\end{knot}
\draw[thick] (0,0) circle (3);
\end{scope}
\end{tikzpicture}
\fi
\caption{Some rational tangles contributing to the four-point function of twist fields in the averaged $\mathbb{T}^D\wr S_2$ orbifold.}
\label{fig:rational-tangles}
\end{figure}
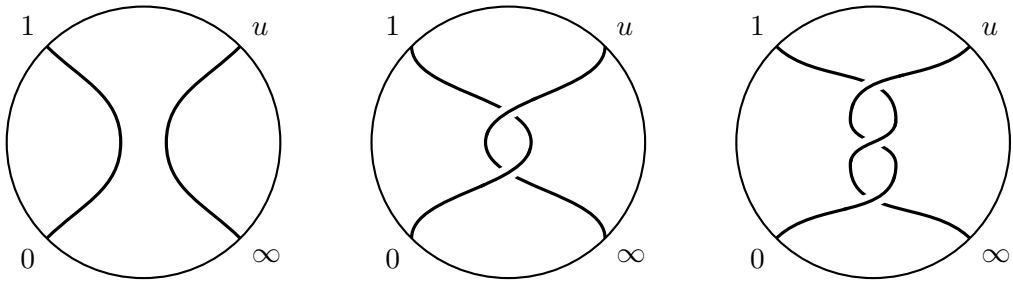

In \cite{Benjamin:2021wzr}, the sum over modular images in equation \eqref{eq:averaged-2-4-point} was argued to arise from a sum over vortex configurations in the bulk which are topologically `rational tangles'. A rational tangle is a vortex configuration which is obtained by applying successive exchanges (braidings) of the points $0,1,u,\infty$ on the trivial tangle $\mathcal{T}_0$. For example, the three tangles shown in Figure \ref{fig:rational-tangles} are the trivial tangle $\mathcal{T}_0$, the tangle obtained by starting with $\mathcal{T}_0$ and braiding the ends at $1,u$ around each other twice, and the tangle obtained from $\mathcal{T}_0$ by braiding the ends at $u,\infty$ three times. Note that the direction of swapping matters.

Mathematically, a rational tangle is obtained from the trivial tangle $\mathcal{T}_0$ in the following way: Let $B_4(\text{S}^2)$ be the braid group of points on the sphere. There is a natural action of $B_4(\text{S}^2)$ on the ends of the vortices at $x=0$, $x=1$, $x=u$, and $x=\infty$. A basis of generators for $B_4$ are the `standard braids' $\sigma_1,\sigma_2,\sigma_3$ which swap the pairs $(0,1)$, $(1,u)$ and $(u,\infty)$ respectively (with specified orientation) as in Figure \ref{fig: braid group}. All rational tangles can be seen as the action of an element of $B_4(\text{S}^2)$ on the trivial tangle $\mathcal{T}_0$.

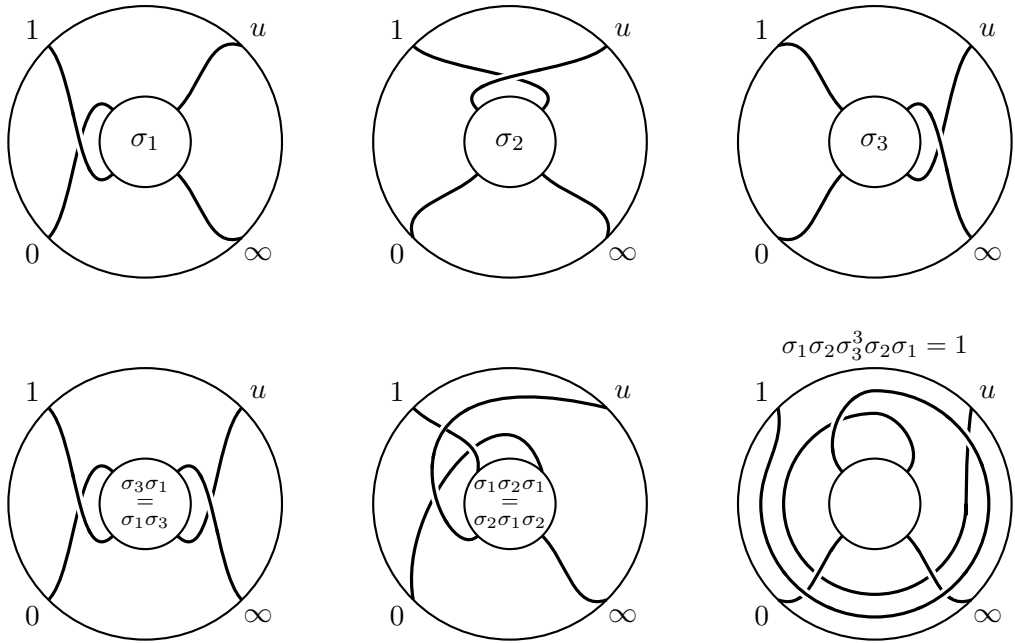
\begin{figure}
\centering
\ifdraft
\else
\begin{tikzpicture}[scale=0.6]

	\begin{scope}[xshift = -12cm, yshift=0cm]
	\node at (0,0) {$\sigma_1$};
	\node at (225:3.5cm) {$0$};
	\node at (135:3.5cm) {$1$};
	\node at (45:3.5cm) {$u$};
	\node at (-45:3.5cm) {$\infty$};
		\begin{knot}[
			clip width = 3,
			flip crossing = 2,
			]
			\strand[very thick] (135:3cm) .. controls (-1.5,1.5) and (-1.5,-1.5) ..  (225:1cm);
			\strand[very thick] (225:3cm)  .. controls (-1.5,-1.5) and (-1.5,1.5) ..  (135:1cm);
			\strand[very thick] (45:3cm) to[out = 160, in = 45] (45:1cm);
			\strand[very thick] (-45:3cm) to[out = -160, in = -45] (-45:1cm);
		\end{knot}
		\draw[thick] (0,0) circle (3);
		\draw[thick] (0,0) circle (1);
	\end{scope}
	
		\begin{scope}[xshift = -4 cm,rotate=-90]
		\node at (0,0) {$\sigma_2$};
		\node at (225:3.5cm) {$1$};
	\node at (135:3.5cm) {$u$};
	\node at (45:3.5cm) {$\infty$};
	\node at (-45:3.5cm) {$0$};
		\begin{knot}[
			clip width = 3,
			flip crossing = 2,
			]
			\strand[very thick] (135:3cm) .. controls (-1.5,1.5) and (-1.5,-1.5) ..  (225:1cm);
			\strand[very thick] (225:3cm)  .. controls (-1.5,-1.5) and (-1.5,1.5) ..  (135:1cm);
			\strand[very thick] (45:3cm) to[out = 160, in = 45] (45:1cm);
			\strand[very thick] (-45:3cm) to[out = -160, in = -45] (-45:1cm);
		\end{knot}
		\draw[thick] (0,0) circle (3);
		\draw[thick] (0,0) circle (1);
	\end{scope}
		\begin{scope}[xshift = 4 cm, yshift=0cm,rotate=180]
		\node at (0,0) {$\sigma_3$};
		\node at (225:3.5cm) {$u$};
		\node at (135:3.5cm) {$\infty$};
		\node at (45:3.5cm) {$0$};
		\node at (-45:3.5cm) {$1$};
		\begin{knot}[
			clip width = 3,
			flip crossing = 2,
			]
			\strand[very thick] (135:3cm) .. controls (-1.5,1.5) and (-1.5,-1.5) ..  (225:1cm);
			\strand[very thick] (225:3cm)  .. controls (-1.5,-1.5) and (-1.5,1.5) ..  (135:1cm);
			\strand[very thick] (45:3cm) to[out = 160, in = 45] (45:1cm);
			\strand[very thick] (-45:3cm) to[out = -160, in = -45] (-45:1cm);
		\end{knot}
		\draw[thick] (0,0) circle (3);
		\draw[thick] (0,0) circle (1);
	\end{scope}
	\begin{scope}[xshift = -12 cm, yshift=-8cm,rotate=0]
	\node at (0,0) {$\substack{\sigma_3\sigma_1\\=\\\sigma_1\sigma_3}$};
	\node at (225:3.5cm) {$0$};
	\node at (135:3.5cm) {$1$};
	\node at (45:3.5cm) {$u$};
	\node at (-45:3.5cm) {$\infty$};
	\begin{knot}[
		clip width = 3,
		flip crossing = 1,
		]
			\strand[very thick] (225:3cm)  .. controls (-1.5,-1.5) and (-1.5,1.5) ..  (135:1cm);
		\strand[very thick] (135:3cm) .. controls (-1.5,1.5) and (-1.5,-1.5) ..  (225:1cm);
		\strand[very thick] (-45:3cm)  .. controls (1.5,-1.5) and (1.5,1.5) ..  (45:1cm);
		\strand[very thick] (45:3cm) .. controls (1.5,1.5) and (1.5,-1.5) ..  (-45:1cm);
	\end{knot}
	\draw[thick] (0,0) circle (3);
	\draw[thick] (0,0) circle (1);
\end{scope}

\begin{scope}[xshift = -4 cm, yshift=-8cm,rotate=0]
	\node at (0,0) {$\substack{\sigma_1\sigma_2\sigma_1\\=\\ \sigma_2\sigma_1\sigma_2}$};
	\node at (225:3.5cm) {$0$};
	\node at (135:3.5cm) {$1$};
	\node at (45:3.5cm) {$u$};
	\node at (-45:3.5cm) {$\infty$};
		\begin{knot}[
			end tolerance=0.3cm
			clip width = 5,
			flip crossing = 4,
			flip crossing = 3,
			]
			\strand[very thick] (-45:3cm) to[out = -160, in = -45] (-45:1cm);
			\strand[very thick] (45:3cm) .. controls (-3.5,3.5) and (-1.5,-1.5) ..  (225:1cm);
			\strand[very thick] (225:3cm)  .. controls (-2.5,-0.) and (0,3.0) ..  (45:1cm);
			\strand[very thick] (135:3cm) to[out = -45, in = 80] (135:1cm);
		\end{knot}
		\draw[thick] (0,0) circle (3);
		\draw[thick] (0,0) circle (1);
\end{scope}
\begin{scope}[yshift=-8cm, xshift=4cm]
	\node at (0,3.5) {$\text{\small{$\sigma_1\sigma_2\sigma_3^3\sigma_2\sigma_1=1$}}$};
	\node at (225:3.5cm) {$0$};
	\node at (135:3.5cm) {$1$};
	\node at (45:3.5cm) {$u$};
	\node at (-45:3.5cm) {$\infty$};
		\begin{knot}[
			end tolerance=0.1cm
			clip width = 5,
			flip crossing = ,
			flip crossing = 2,
						flip crossing = 4,
						flip crossing = 5,
						flip crossing = 6,
						flip crossing = ,
			]
			
\strand[very thick] (-45:3cm) to[out = -160, in = -45] (-45:1cm);
\strand[very thick] (225:3cm) to[out = -20, in = -135] (225:1cm);

\strand[very thick] (45:3cm) to[out = -100, in = -90] (0:2cm)  to[out = -90, in = 0] (-90:2cm)  to[out = 180, in = -90] (180:2cm)  to[out = 90, in = 180] (90:2cm) to[out = 0, in = 45] (45:1cm);
\strand[very thick] (135:3cm) to[out = -80, in = 90] (180:2.5cm)  to[out = -90, in = 180] (-90:2.5cm)  to[out = 0, in = -90] (0:2.5cm)  to[out = 90, in = 0] (90:2.5cm) to[out = 180, in = 135] (135:1cm);

		\end{knot}
		\draw[thick] (0,0) circle (3);
		\draw[thick] (0,0) circle (1);
\end{scope}

\end{tikzpicture}
\fi
\caption{The action of the braid of points on the sphere $B_4(\text{S}^2)$. The two circles depict cross sections of two-spheres. Inside the smaller two-sphere the strands could be arbitrarily tangled. The first row shows the action of the generators, the second depicts some of the relations that these generators satisfy. In the last drawing, the strands can be ``untangled": the one that connects to $1$ from the front of the inner $\text{S}^2$ and the one connected to $u$ behind the inner $\text{S}^2$.  
}
\label{fig: braid group}
\end{figure}

The braid group $B_4(\text{S}^2)$ is generated by $\sigma_1,\sigma_2,\sigma_3$, which satisfy the following relations:
\begin{equation}\label{eq:b4-relations}
\begin{gathered}
\sigma_1\sigma_3=\sigma_3\sigma_1\,,\quad\sigma_1\sigma_2\sigma_1=\sigma_2\sigma_1\sigma_2\,,\quad\sigma_2\sigma_3\sigma_2=\sigma_3\sigma_2\sigma_3\\
\sigma_1\sigma_2\sigma_3\sigma_3\sigma_2\sigma_1=1\,.
\end{gathered}
\end{equation}
The last of these relations is specific to the braid group on the sphere. Note that the braid group does not act on rational tangles faithfully. In particular, the combination $\sigma_1\sigma_3^{-1}$ acts trivially on any rational tangle since it corresponds to a reflection about the East-West axis as in Figure \ref{fig: reflection}, see \cite{Maloney:2016kee}. 
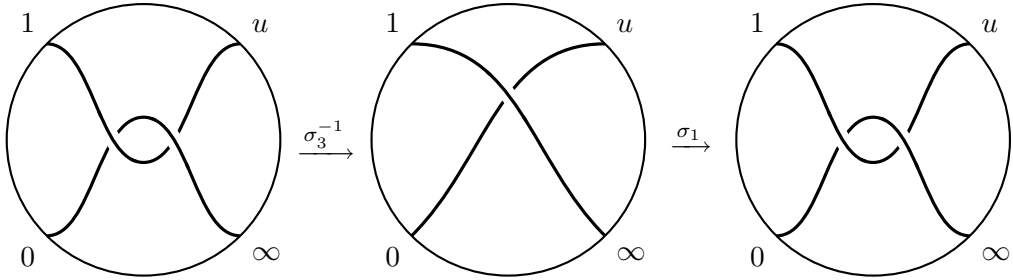
\begin{figure}
\centering
\ifdraft
\else
\begin{tikzpicture}[scale=0.6]

\begin{scope}[xshift = -4cm]
\node[below left] at (-2.15,-2.15) {$0$};
\node[above left] at (-2.15,2.15) {$1$};
\node[above right] at (2.15,2.15) {$u$};
\node[below right] at (2.15,-2.15) {$\infty$};
\begin{knot}[
clip width = 5,
flip crossing = 1,
]
\strand[very thick] (45:3cm) .. controls +(-1,0) and +(1,0) .. (0,-0.5)
.. controls +(-1,0) and +(1,0) .. (135:3cm);
\strand[very thick] (-135:3cm) .. controls +(1,0) and +(-1,0) .. (0,0.5)
.. controls +(1,0) and +(-1,0) .. (-45:3cm);
\end{knot}
\draw[thick] (0,0) circle (3);
\node at (4,0) {$\xrightarrow{\sigma^{-1}_3}$} ;
\end{scope}

\begin{scope}[xshift = 4cm]
\node[below left] at (-2.15,-2.15) {$0$};
\node[above left] at (-2.15,2.15) {$1$};
\node[above right] at (2.15,2.15) {$u$};
\node[below right] at (2.15,-2.15) {$\infty$};
\begin{knot}[
clip width = 5,
flip crossing = 2,
]
\strand[very thick] (-45:3cm) to[out = 135, in = 0] (135:3cm);
\strand[very thick] (-135:3cm) to[out = 45, in = 180]  (45:3cm);
\end{knot}
\draw[thick] (0,0) circle (3);
\node at (4,0) {$\xrightarrow{\sigma_1}$} ;
\end{scope}

\begin{scope}[xshift = 12cm]
\node[below left] at (-2.15,-2.15) {$0$};
\node[above left] at (-2.15,2.15) {$1$};
\node[above right] at (2.15,2.15) {$u$};
\node[below right] at (2.15,-2.15) {$\infty$};
\begin{knot}[
clip width = 5,
flip crossing = 1,
]
\strand[very thick] (45:3cm) .. controls +(-1,0) and +(1,0) .. (0,-0.5)
.. controls +(-1,0) and +(1,0) .. (135:3cm);
\strand[very thick] (-135:3cm) .. controls +(1,0) and +(-1,0) .. (0,0.5)
.. controls +(1,0) and +(-1,0) .. (-45:3cm);
\end{knot}
\draw[thick] (0,0) circle (3);

\end{scope}

\end{tikzpicture}
\fi
\caption{The action of $\sigma_1 \sigma^{-1}_3$ on the tangle obtained by acting with $\sigma_3^2$. We see that this corresponds to the same tangle.}
\label{fig: reflection}
\end{figure}

Let $\mathcal{N}\braket{\sigma_1\sigma_3^{-1}}$ be the normal subgroup of $B_4(\text{S}^2)$ generated by $\sigma_1\sigma_3^{-1}$.\footnote{That is, $\mathcal{N}\braket{\sigma_1\sigma_3^{-1}}$ is the smallest normal subgroup of $B_4(\text{S}^2)$ which contains $\sigma_1\sigma^{-1}_3$.} Since this is a normal subgroup, the quotient $B_4(\text{S}^2)/\mathcal{N}\braket{\sigma_1\sigma_3^{-1}}$ is a group which acts in a well-defined manner on the space of tangles. This quotient effectively imposes $\sigma_1\sim\sigma_3$ in the defining relations \eqref{eq:b4-relations} of the braid group. It turns out that this quotient is nothing more than the modular group, i.e.
\begin{equation}
B_4(\text{S}^2)/\mathcal{N}\braket{\sigma_1\sigma_3^{-1}}\cong\text{PSL}(2,\mathbb{Z})\,.
\end{equation}
The isomorphism is seen by the direct matrix identification
\begin{equation}
\sigma_1,\sigma_3\to
\begin{pmatrix}
1 & 1\\
0 & 1
\end{pmatrix}\,,\quad\sigma_2=
\begin{pmatrix}
1 & 0\\
-1 & 1
\end{pmatrix}\,.
\end{equation}
Finally, note that not even $B_4(\text{S}^2)/\mathcal{N}\braket{\sigma_1\sigma_3^{-1}}$ acts faithfully on the space of rational tangles. This is because $\sigma_1\cdot\mathcal{T}_0=\mathcal{T}_0$. Thus, the set of operations which acts faithfully on the set of rational tangles is
\begin{equation}\label{eq:braid-group-vs-modular-group}
(B_4(\text{S}^2)/\mathcal{N}\braket{\sigma_1\sigma_3^{-1}})/\braket{\sigma_1}\cong\Gamma_{\infty}\backslash\text{SL}(2,\mathbb{Z})\,.
\end{equation}
Thus, the sum over rational tangles produces a sum over the coset $\Gamma_{\infty}\backslash\text{SL}(2,\mathbb{Z})$. This sum also precisely reproduces the sum over Lagrangian sublattices, since each rational tangle admits a combination of the cycles shown in Figure \ref{fig:rational-tangles-bulk} which becomes contractible in the bulk. Furthermore, the branched double cover of each rational 2-tangle is a handlebody with torus boundary.\footnote{This follows either from the fact that the double cover of $\mathcal{T}_0$ is a torus handlebody and the fact that all rational 2-tangles are homeomorphic to $\mathcal{T}_0$. Intuitively, the action of the braid group on $\mathcal{T}_0$ can be thought of as implementing Dehn surgery on the double cover of $\mathcal{T}_0$. In the mathematics literature, this is often referred to as the `Montesinos Trick' \cite{knots-groups-manifolds}.}

\subsection{Higher-point functions}

The above analysis tells us how to interpret the four-point function of twist fields in the $S_2$ permutation orbifolds in terms of rational tangles of vortices in the bulk. Let us now consider the case of higher-point functions. In order to have a non-vanishing correlation function, we require that the number of twist field insertions is even, and so we consider the correlator
\begin{equation}\label{eq:s2-higher-point-function}
\braket{\sigma_{(1\,2)}(x_1)\cdots\sigma_{(1\,2)}(x_{2g+2})}\,,
\end{equation}
where $g$ is some non-negative positive integer. The covering space associated to this correlation function is a genus-$g$ hyperelliptic curve defined by the equation 
\begin{equation}\label{eq:hyperelliptic}
\Sigma_g=\left\{(x,y)\in\mathbb{CP}^1\times\mathbb{CP}^1\,\bigg|\,y^2=\prod_{i=1}^{2g+2}(x-x_i)\right\}\,.
\end{equation}
Indeed, around the point $x=x_i$, the solutions $y(x)$ have a square-root branch cut, as required by the form of the correlator. The correlation function can be expressed as
\begin{equation}\label{eq:higher-point-function}
\braket{\sigma_{(1\,2)}(x_1)\cdots\sigma_{(1\,2)}(x_{2g+2})}=e^{-S_L[\Phi_{\Gamma}]}Z(\Sigma_g)\,,
\end{equation}
where $\Gamma$ is the covering map 
\begin{equation}
\begin{split}
\Gamma:&\Sigma_g\to\mathbb{CP}^1\\
&(x,y)\mapsto x\,.
\end{split}
\end{equation}
This gives a double cover of $\mathbb{CP}^1$ by $\Sigma_g$ since each value of $x$ has two values of $y$ satisfying \eqref{eq:hyperelliptic}. Here and in what follows we will not be explicit about the form of the conformal anomaly.

Upon ensemble averaging, we have
\begin{equation}\label{eq:2-higher-point-sum}
\int_{\mathcal{M}_D}\mathrm{d}\mu(m)\,\braket{\sigma_{(1\,2)}(x_1)\cdots\sigma_{(1\,2)}(x_{2g+2})}=e^{-S_L[\Phi_{\Gamma}]}\sum_{\gamma\in P\backslash\text{Sp}(2g,\mathbb{Z})}Z_{\text{CS}}(\gamma\cdot\Omega)\,,
\end{equation}
where $\Omega$ is the period matrix of $\Sigma_g$, and $Z_{\text{CS}}(\gamma\cdot\Omega)$ is shorthand for the expression
\begin{equation}
Z_{\text{CS}}(\gamma\cdot\Omega)=\frac{(\det\text{Im}(\gamma\cdot\Omega))^{D/2}}{(\det\text{Im}(\Omega))^{D/2}|\det'\bar{\partial}_{\Sigma_g}|^D}\,,
\end{equation}
which is the $\text{U}(1)^D\times\text{U}(1)^D$ Chern-Simons partition function on a handlebody bounded by $\Sigma_g$ \cite{Maloney:2020nni}. The sum over modular images of $\Omega$ under the action of $P\backslash\text{Sp}(2g,\mathbb{Z})$ defines a sum over different inequivalent handlebodies with boundary $\Sigma_g$. For later convenience, we introduce a basis for the homology group of $\Sigma_g$ by choosing the branch cuts in equation \eqref{eq:hyperelliptic} to be between neighboring points $x_{2i-1}$ and $x_{2i}$ for $i=1,\ldots,g+1$, and choosing the $\mathcal{A}$ and $\mathcal{B}$ cycles of $\Sigma_g$ as in Figure \ref{fig:HomologyBasis}. \begin{figure}
    \centering
     \begin{tikzpicture}[scale=1]
    	\node (x1) at (-3,0) {$x_1$};
    	\node (x2) at (-2,0) {$x_2$};
    	\node (x3) at (-1,0) {$x_3$};
    	\node (x4) at (0,0) {$x_4$};
    	\node (x5) at (1,0) {$x_5$};
    	\node (x6) at (2.3,0) {$x_6$};
    	\node (x7) at (3.2,0) {$x_7$};
    	\node (x8) at (4,0) {$x_8$};
    	\node[inner sep=0pt] (hom) at (0,0) {\includegraphics[width=0.9\textwidth]{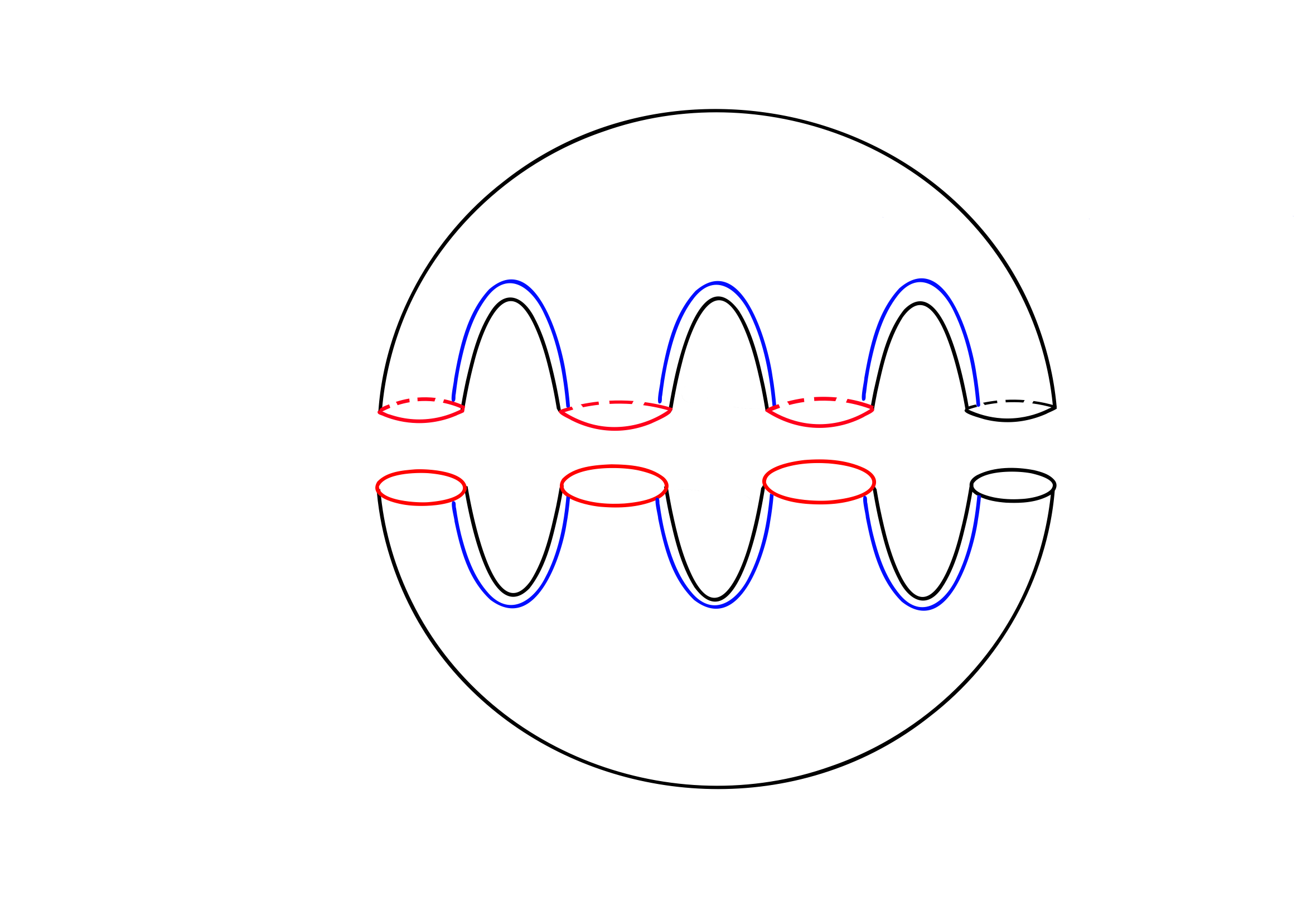}};
    \end{tikzpicture}
    \caption{The homology basis of the genus $g$ surface $\Sigma_g$. The red and blue curves represent the $\mathcal{A}$ and $\mathcal{B}$ cycles respectively. In the preimage, the red cycles are given by the branch cuts between $x_i$ and $x_{i+1}$.}
    \label{fig:HomologyBasis}
\end{figure}
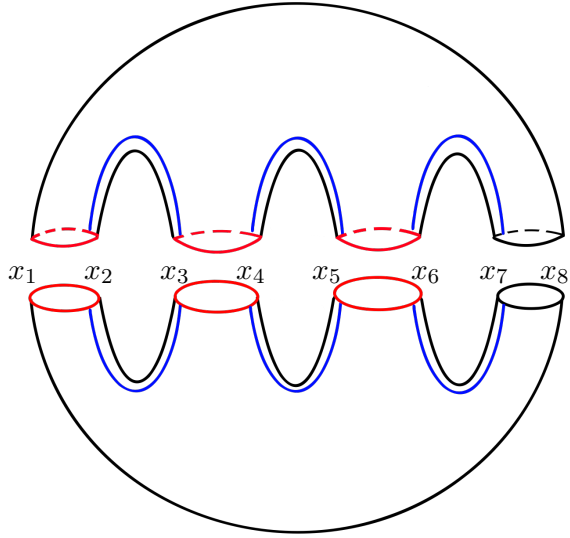
Just as in the case of the four-point function, we can interpret the sum over handlebodies as a sum over rational tangles of vortices in the bulk. We do this in the following way: let $\mathcal{T}_0$ be the `trivial' tangle defined by connecting $x_{2i-1}$ to $x_{2i}$ as shown in Figure 
\ref{fig: trivial tangle many points}. 
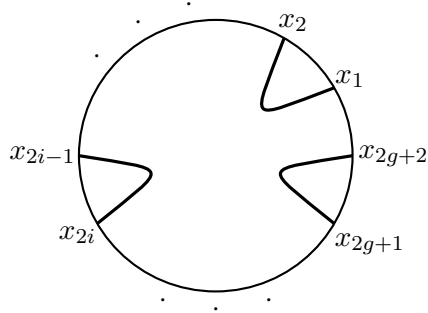
\begin{figure}
	\centering
	\ifdraft
	\else
	\begin{tikzpicture}[scale=0.6]
		
		\begin{scope}[xshift = -4cm]
			\node at (30:3.4cm) {$x_1$};
			\node at (60:3.4cm) {$x_2$};
			\node at (100:3.4cm) {$\cdot$};
			\node at (120:3.4cm) {$\cdot$};
			\node at (140:3.4cm) {$\cdot$};
			\node at (180:3.8cm) {$x_{2i-1}$};
			\node at (210:3.5cm) {$x_{2i}$};
			\node at (250:3.4cm) {$\cdot$};
			\node at (270:3.4cm) {$\cdot$};
			\node at (290:3.4cm) {$\cdot$};
			\node at (330:3.9cm) {$x_{2g+1}$};
			\node at (360:3.9cm) {$x_{2g+2}$};
				\draw[very thick] (30:3cm) .. controls (50:1cm) and (40:1cm) .. (60:3cm);
				\draw[very thick] (330:3cm) .. controls (350:1cm) and (340:1cm) .. (360:3cm);
				\draw[very thick] (180:3cm) .. controls (200:1cm) and (190:1cm) .. (210:3cm);
			\draw[thick] (0,0) circle (3);
		\end{scope}
	\end{tikzpicture}
	\fi
	\caption{The trivial tangle $\mathcal{T}_0$.}
	\label{fig: trivial tangle many points}
\end{figure}
The double cover of the ball branched over this surface will be a genus $g$ handlebody such that the $\mathcal{A}$ cycles of $\Sigma_g$ (those surrounding the branch cuts between $x_{2i-1}$ and $x_{2i}$) are contractible in the bulk. Now, let us pick out the point $x_{2g+2}$ to be special. We can generate all rational tangles connecting the points $x_1,\ldots,x_{2g+2}$ via actions of the braid group $B_{2g+1}$ (where we treat the points $x_1,\ldots,x_{2g+1}$ as individual strands, keeping $x_{2g+2}$ fixed) on the trivial tangle $\mathcal{T}_0$. Via the double cover $\Gamma:\Sigma_g\to\mathbb{CP}^1$ associated to this configuration of points, the action of the braid group on the configuration $X$ can be shown to induce an action on the fundamental group $\pi_1(\Sigma_g)$ or, similarly, its abelianization $H_1(\Sigma_g,\mathbb{Z})$. Specifically, if we let $\sigma_k\in B_{2g+1}$ be the element of the braid group which swaps the points $x_k$ and $x_{k+1}$ (with specified orientation), and we pick a homology basis $\mathcal{A}_i,\mathcal{B}_i$ on $\Sigma_g$, then we can take the action of the braid group on $H_1(\Sigma_g,\mathbb{Z})$ to be (breaking up cases for $k$ even and odd)
\begin{equation}
\begin{split}
\sigma_{2i}\cdot\mathcal{A}_{i}&=\mathcal{A}_{i}+\mathcal{B}_i\,,\\
\sigma_{2i}\cdot\mathcal{B}_i&=\mathcal{B}_i\,,\\
\sigma_{2i}\cdot\mathcal{A}_{i+1}&=\mathcal{A}_{i+1}-\mathcal{B}_i\,,\\
\sigma_{2i-1}\cdot\mathcal{B}_{i-1}&=\mathcal{B}_{i-1}+\mathcal{A}_i\,,\\
\sigma_{2i-1}\cdot\mathcal{A}_i&=\mathcal{A}_i\,,\\
\sigma_{2i-1}\cdot\mathcal{B}_{i}&=\mathcal{B}_{i}-\mathcal{A}_i\,.
\end{split}
\end{equation}
Furthermore, all other actions are trivial. This action defines a representation $\rho:B_{2g+1}\to\text{Aut}(H_1(\Sigma_g,\mathbb{Z}))\simeq\text{GL}(2g,\mathbb{Z})$ of dimension $2g$ of the braid group by integer matrices. Furthermore, it is not hard to see that the above action preserves the intersection form $\braket{\mathcal{A}_i,\mathcal{B}_j}=\delta_{ij}$, and thus the representation of the braid group is actually via symplectic matrices. That is, there is a homomorphism
\begin{equation} 
\rho:B_{2g+1}\to\text{Sp}(2g,\mathbb{Z})\,.
\end{equation}
This is known as the \textit{symplectic} representation of the braid group \cite{ACampo1979}. Given an element $\sigma\in B_{2g+1}$, the symplectic representation $\rho(\sigma)$ acts on the period matrix of $\Sigma_g$ as
\begin{equation}
\Omega\to\sigma\cdot\Omega=(A\Omega+B)(C\Omega+D)^{-1}\,,
\end{equation}
where $A,B,C,D$ are the block entries of $\rho(\sigma)$ as a symplectic matrix.

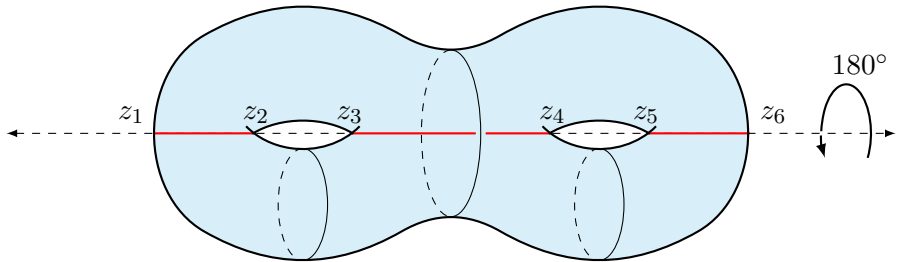
\begin{figure}
\centering
\ifdraft
\else
\begin{tikzpicture}[scale = 1.3]
\fill[CornflowerBlue, opacity = 0.2] (0,1) to[out=30,in=150] (2,1) to[out=-30,in=210] (3,1) to[out=30,in=150] (5,1) to[out=-30,in=30] (5,-1) to[out=210,in=-30] (3,-1) to[out=150,in=30] (2,-1) to[out=210,in=-30] (0,-1) to[out=150,in=-150] (0,1);
\draw[dashed] (2.5,0) [partial ellipse = 90:270:0.3 and 0.85];
\fill[white] (1,-0.37) [partial ellipse = 47:133:0.75 and 0.5];
\fill[white] (1,0.49) [partial ellipse = -47:-133:0.75 and 0.65];
\fill[white] (4,-0.37) [partial ellipse = 47:133:0.75 and 0.5];
\fill[white] (4,0.49) [partial ellipse = -47:-133:0.75 and 0.65];
\draw[dashed, latex-] (-2,0) -- (1.475,0);
\draw[dashed] (3.525,0) -- (5.5,0);
\draw[thick, red] (-0.5,0) -- (0.525,0);
\draw[thick, red] (1.475,0) -- (2.75,0);
\draw[thick, red] (2.85,0) -- (3.525,0);
\draw[thick, red] (4.475,0) -- (5.525,0);
\draw[smooth, thick] (0,1) to[out=30,in=150] (2,1) to[out=-30,in=210] (3,1) to[out=30,in=150] (5,1) to[out=-30,in=30] (5,-1) to[out=210,in=-30] (3,-1) to[out=150,in=30] (2,-1) to[out=210,in=-30] (0,-1) to[out=150,in=-150] (0,1);
\draw[thick] (1,-0.37) [partial ellipse = 47:133:0.75 and 0.5];
\draw[thick] (1,0.49) [partial ellipse = -40:-140:0.75 and 0.65];
\draw[thick] (4,-0.37) [partial ellipse = 47:133:0.75 and 0.5];
\draw[thick] (4,0.49) [partial ellipse = -40:-140:0.75 and 0.65];

\draw[thick, -latex] (6.5,0) [partial ellipse = -30:210:0.25 and 0.5];
\draw[ultra thick, white] (6.1,0) -- (6.3,0);
\draw[dashed, -latex] (5.5,0) -- (7,0);
\node[above right] at (6.25, 0.5) {$180^{\circ}$};
\draw (1,-0.725) [partial ellipse = -90:90:0.25 and 0.57];
\draw[dashed] (1,-0.725) [partial ellipse = 90:270:0.25 and 0.57];
\draw (4,-0.725) [partial ellipse = -90:90:0.25 and 0.57];
\draw[dashed] (4,-0.725) [partial ellipse = 90:270:0.25 and 0.57];
\draw (2.5,0) [partial ellipse = -90:90:0.3 and 0.85];
\node[above left] at (-0.5,0) {$z_1$};
\node[above] at (0.525,0) {$z_2$};
\node[above] at (1.475,0) {$z_3$};
\node[above] at (3.525,0) {$z_4$};
\node[above] at (4.475,0) {$z_5$};
\node[above right] at (5.525,0) {$z_6$};
\end{tikzpicture}
\fi
\caption{A handlebody filling in the hyperelliptic curve $\Sigma_g$ ($g=2$ shown). The hyperelliptic involution induces a 180$^{\circ}$ rotation around the symmetry axis, and the fixed points of this rotation are $g+1$ intervals. The $\mathbb{Z}_2$ quotient of the handlebody under the hyperelliptic involution is homeomorphic to a ball with $g+1$ vortices. The points $z_i$ are the preimages of the points $x_i$ in the correlator \eqref{eq:higher-point-function}. Since a tangle is homeomorphic to the trivial tangle if and only if it is rational, the double-cover of a rational $(g+1)$-tangle is always a handlebody.}
\label{fig:hyperelliptic-handlebody}
\end{figure}

Now, while the action of the braid group on the $2g+2$ endpoints of the $g+1$ vortices induces $\text{Sp}(2g,\mathbb{Z})$ transformations on the period matrix $\Omega$ of the hyperelliptic curve $\Sigma$, it is also possible to show that the resulting rational tangle (obtained by acting on the trivial tangle with the braid group) indeed has a double cover which is a handlebody of genus $g$, whose boundary is $\Sigma_g$.\footnote{This follows again by the Montesinos trick -- braid group actions on the trivial tangle lift to Dehn twists on the branched double cover, and so the double cover of any rational $(g+1)$-tangle is related to a genus $g$ handlebody by a Dehn twist, and is therefore also a genus $g$ handlebody.} Each rational tangle in the $G\wr S_2$ Chern-Simons theory, therefore, reproduces a handlebody in the sum \eqref{eq:2-higher-point-sum}. However, it is not clear that every element of the sum \eqref{eq:2-higher-point-sum} over handlebodies is reproduced by a rational tangle of vortices in the Chern-Simons theory. Indeed, this is not the case. To see this, consider the image of the symplectic representation $\rho$. For $g=1,2$ this image is indeed the full modular group $\text{Sp}(2g,\mathbb{Z})$, but for $g\geq 3$, $\rho(B_{2g+1})$ is only a finite-index subgroup of $\text{Sp}(2g,\mathbb{Z})$ \cite{ACampo1979}. Thus, for $g\geq 3$, i.e. for $n$-point functions of twist fields with $n\geq 8$, the sum over rational tangles cannot reproduce the correct sum over the symplectic group.

The reason why is because many handlebodies in the sum \eqref{eq:2-higher-point-sum} do not respect the covering map structure of the boundary. Specifically, the covering map $\Gamma:\Sigma_g\to\mathbb{CP}^1$ has the structure of a $\mathbb{Z}_2$ quotient by the so-called \textit{hyperelliptic involution} $\iota$, which sends $(x,y)\to(x,-y)$ in equation \eqref{eq:hyperelliptic}. Given a choice of Lagrangian sublattice $\Gamma\subset H_1(\Sigma_g,\mathbb{Z})$ it is not guaranteed that the hyperelliptic involution extends to a symmetry of the bulk (when $\iota$ does extend to a symmetry of the bulk, it is represented by a $180^{\circ}$ rotation around a fixed axis \cite{brendle2011point,Brendle_2014}, see Figure \ref{fig:hyperelliptic-handlebody}).\footnote{For low genera ($g=1,2$), however, the hyperelliptic involution always extends into the bulk.} Those handlebodies $M$ which admit an extension of the hyperelliptic involution are then branched covers $M\to M/\iota$ of 3-balls, for which the branching loci are rational tangles. Those handlebodies $M$ for which the hyperelliptic involution does not extend into the bulk cannot, as far as we know, be thought of as 2-fold branched covers of a 3-manifold with spherical boundary, and thus represent another example of terms in the averaged symmetric orbifold theory which cannot be recovered by a semiclassical Chern-Simons calculation in the bulk.

\subsection{The conformal anomaly}\label{subsec:conformal-anomaly}

In the above discussion, we demonstrated how, in certain special cases, the ensemble average of correlators of twist fields in the symmetric orbifold can be reproduced by an appropriate sum over vortex configurations in a bulk $\text{U}(1)^D\times\text{U}(1)^D\wr S_N$ Chern-Simons theory. In doing so, we only discussed the origin of the Eisenstein series appearing in the averaged correlation function from the Chern-Simons perspective, and we ignored the conformal anomaly which appears in the symmetric orbifold correlators. Let us now briefly comment on the origin of this prefactor from the Chern-Simons side.

First, let us recall the origin of the conformal anomaly in the symmetric orbifold setting. The 2-sphere $\mathbb{CP}^1$ can be locally taken to have fiducial metric $\mathrm{d}x\,\mathrm{d}\overline{x}$. In calculating the correlator of twist-fields on $\mathbb{CP}^1$, each contribution is given by an appropriate covering map $\Gamma:\Sigma\to\mathbb{CP}^1$. The associated contribution to the symmetric orbifold path integral arises from calculating the correlation functions of the pullback of the twist fields under $\Gamma$ on the surface $\Sigma$. If $\Gamma$ has the correct branching properties, then the pullbacks of these fields are untwisted, and the calculation reduces to the computation of a $\mathbb{T}^D$ correlator on $\Sigma$, see \cite{Lunin:2000yv}.

Importantly, in this construction, the path integral on $\Sigma$ is evaluated with respect to the pullback metric
\begin{equation}\label{eq:pullback-metric-cft}
\Gamma^*(\mathrm{d}x\,\mathrm{d}\overline{x})=|\partial\Gamma(z)|^2\,\mathrm{d}z\,\mathrm{d}\overline{z}\,,
\end{equation}
where $z$ is a local coordinate on $\Sigma$. This metric can be locally put back into the form of the fiducial metric $\mathrm{d}z\,\mathrm{d}\overline{z}$ via a Weyl rescaling $e^{-\Phi}$ with
\begin{equation}
\Phi=\log|\partial\Gamma|^2\,.
\end{equation}
This Weyl rescaling comes at the cost of the conformal anomaly
\begin{equation}
\exp\left(-\frac{c}{6}S_L[\Phi]\right)=\exp\left(-\frac{c}{48\pi}\int_{\Sigma}\mathrm{d}^2z\,\sqrt{g}\left(-\frac{1}{2}\Phi\Delta\Phi+R\Phi\right)\right)\,,
\end{equation}
where the metric $g$ on $\Sigma$ is the fudicial metric $\mathrm{d}z\,\mathrm{d}\overline{z}$, or more accuraltely the Weyl-transformed metric $e^{-\Phi}\Gamma^*\left(\mathrm{d}s^2_{\mathbb{CP}^1}\right)$.

On the Chern-Simons theory side, the conformal anomaly arises in a nearly identical way. Let us for the moment consider $G\wr S_N$ Chern-Simons theory for some generic group $G$. Given a bulk manifold $M$ with spherical boundary $\partial M=\mathbb{CP}^1$ (in our case $M=\mathbb{H}^3$ is the 3-ball), we can consider the Chern-Simons path integral with vortex locus $L$. To compute this path integral, we must consider the branched cover $\widetilde{M}$ of $M$ branched along the vortex locus $L$. Let $\gamma:\widetilde{M}\to M$ be the appropriate branched covering. We define $\Gamma=\gamma|_{\partial M}$ to be the induced branched covering $\Gamma:\Sigma\to\mathbb{CP}^1$ on the boundaries, where $\Sigma=\partial\widetilde{M}$. Now, we can gauge fix the Chern-Simons path integral by explicitly choosing a metric $g$ on $M$. Then the $G\wr S_N$ Chern-Simons path integral is given by the $G$ Chern-Simons path integral evaluated on $\widetilde{M}$ with metric $\gamma^*g$.

The Chern-Simons path integral is not completely independent of the choice of metric, and in particular depends on the boundary value of the metric on a specific bulk manifold. It is convenient to locally pick $g|_{\partial M}=\mathrm{d}x\,\mathrm{d}\overline{x}$. Then the induced boundary metric on $\widetilde{M}$ is simply the pullback of the boundary metric of $M$ by $\gamma$, which is exactly the metric \eqref{eq:pullback-metric-cft}. We can perform a Weyl transformation by $e^{-\Phi}$ on the metric on $\widetilde{M}$\footnote{Here, $\Phi$ is understood to be a scalar field on $\widetilde{M}$ which reduces to $\log|\partial\Gamma|^2$ on the boundary.} in order to bring the boundary metric of $\widetilde{M}$ into canonical form. However, the path integral of Chern-Simons theory is also not invariant under such transformations, and picks up a conformal anomaly \cite{Hu:2001ue}
\begin{equation}
\exp\left(-\frac{\text{dim}(G)}{12}S_L[\Phi]\right)=\exp\left(-\frac{\text{dim}(G)}{96\pi}\int_{\Sigma}\mathrm{d}^2z\,\sqrt{g}\left(-\frac{1}{2}\Phi\Delta\Phi+R\Phi\right)\right)\,,
\end{equation}
where the integral is over the boundary $\Sigma=\partial\widetilde{M}$ evaluated with the Weyl-transformed boundary metric. For the case of $\text{U}(1)^D\times\text{U}(1)^D\wr S_N$ Chern-Simons theory, this reproduces the correct conformal anomaly of the symmetric orbifold theory since
\begin{equation}
\text{dim}(G)=2D=2c\,.
\end{equation}
The fact that the Chern-Simons theory picks up the same conformal anomaly as a boundary $\mathbb{T}^D$ CFT is, of course, to be expected. By the CS/WZW correspondence \cite{Witten:1988hf,Cabra_1997}, the partition function of $\text{U}(1)^D\times\text{U}(1)^D$ Chern-Simons theory on a smooth handlebody with boundary conditions specified in Section \ref{sec:preliminaries} should reproduce the vacuum character $|\chi_{\text{vac}}|^2$ of a (non-chiral) $\text{U}(1)^D$ WZW model on the boundary. The vacuum character of a WZW model, however, picks up the appropriate conformal anomaly under Weyl transformations of the boundary metric.

\section{Adding supersymmetry to the Narain ensemble}
\label{sec:6}

\subsection{Narain averaging with supersymmetry}

In this section we consider ensemble averaging the supersymmetric variants of the Narain theories. The supersymmetric $\mathbb{T}^D$ theories are obtained by simply adding $D$ free fermions to the bosonic theory. Each fermion is a partner to one of the $D$ free scalars. As discussed in \ref{sec:preliminaries}, we must make a choice of boundary conditions for the fermions as they traverse nontrivial cycles. For a CFT on a torus we label the spin structure by two half integers $\alpha,\beta$ such that
\begin{equation}
\psi(A\cdot z)=e^{2\pi i\alpha}\psi(z)\,,\quad\psi(B\cdot z)=e^{2\pi i\beta}\psi(z)\,,
\end{equation}
where again we take $A$ to be the $\tau$-cycle while $B$ is the $1$-cycle. The supersymmetric $\mathbb{T}^D$ partition function on the torus with a choice of spin structure is then given by 
\begin{equation}
Z_{\mathbb{T}^D} \begin{bmatrix}\alpha\\\beta\end{bmatrix}(\tau)= Z_{\text{ferm}} \begin{bmatrix}\alpha\\\beta\end{bmatrix}(\tau)^D Z_{\mathbb{T}^D}(\tau)=|\det\overline{\partial}_{\alpha,\beta}|^{D} \frac{\Theta_D(m,\tau)}{|\eta(\tau)|^{2D}} \,,
\end{equation}
where we have split the partition function into the bosonic and fermionic parts. Here, $\overline{\partial}_{\alpha,\beta}$ is the Dirac operator acting on spin $(\tfrac{1}{2},0)$ fermions with spin structure $\alpha,\beta$. The determinant of these operators are given by theta functions. Specifically,
\begin{equation}
\left|\det\overline{\partial}_{\alpha,\beta}\right|=\frac{1}{|\eta(\tau)|} \times 
\begin{cases}
|\vartheta_1(\tau)|\,,\quad \alpha=\beta=0\,,\\
|\vartheta_2(\tau)|\,,\quad\alpha=\frac{1}{2},\quad\beta=0\,,\\
|\vartheta_3(\tau)|\,,\quad\alpha=\beta=\frac{1}{2}\,,\\
|\vartheta_4(\tau)|\,,\quad\alpha=0\,,\quad\beta=\frac{1}{2}\,.
\end{cases}
\end{equation}
The exact form of the theta functions will not matter much in what follows. It suffices to remark on their modular properties. The partition function $Z_{\text{ferm}}$ of a single free fermion, given by the above determinant, satisfies the modularity property
\begin{equation} \label{eqn:FermionModTransformation}
Z_{\text{ferm}}\begin{bmatrix}a\alpha+b\beta\\c\alpha+d\beta\end{bmatrix}\left(\frac{a\tau+b}{c\tau+d}\right)=Z_{\text{ferm}}\begin{bmatrix}\alpha\\\beta\end{bmatrix}(\tau)\,,
\end{equation}
when acted on by an element of $\text{SL}(2,\mathbb{Z})$. The partition function of the supersymmetric Narain theory can now be readily averaged, since only the theta function $\Theta_D$ depends on the Narain moduli. The result is
\begin{equation}
\int_{\mathcal{M}_D}\mathrm{d}\mu\,Z_{\mathbb{T}^D}\begin{bmatrix}\alpha\\\beta\end{bmatrix}(\tau)=Z_{\text{ferm}}\begin{bmatrix}\alpha\\\beta\end{bmatrix}(\tau)^D \sum_{\gamma\in\Gamma_{\infty}\backslash\text{SL}(2,\mathbb{Z})}\frac{1}{|\eta(\gamma\cdot\tau)|^{2D}}\,.
\end{equation}
The above partition function is recovered straightforwardly from the partition function of $\mathcal{N}=(1,1)$ Chern-Simons theory, summed over geometries. The boundary fermions in the Chern-Simons theory have specified spin structure on the boundary, namely the same spin structure specified for the Narain CFT.

As explained around equation \eqref{eqn:TorusAverage}, each element of the sum over geometries is specified by a choice $(c,d)$ of coprime integers such that the cycle $cA+dB$ becomes contractible in the bulk. Because the fermion picks up a sign of $(-1)^{2\alpha}$ around the $A$-cycle and $(-1)^{2\beta}$ around the $B$-cycle, the monodromy around the contractible cycle is given by $(-1)^{2(c\alpha+d\beta)}$. This flies in the face of the usual intuition that the boundary fermions must satisfy antiperiodic boundary conditions around a contractible bulk cycle. This only occurs if $c\alpha+d\beta$ is half-integer.

To be concrete, let us take the boundary theory to be in the $\text{NS}$ sector, i.e. $\alpha=\beta=1/2$. Then in the bulk geometry associated to $(c,d)$, the boundary fermion is periodic when $c+d$ is even, and antiperiodic when $c+d$ is odd. We can break up the sum over geometries into two sums:
\begin{equation} \underbrace{Z_{\text{ferm}}\begin{bmatrix}\tfrac{1}{2}\\\tfrac{1}{2}\end{bmatrix}(\tau)^D\sum_{\substack{(c,d)=1\\c+d\text{ odd}}}\frac{1}{|\eta(\gamma\cdot\tau)|^{2D}}}_{\text{`good' boundary conditions}}+\underbrace{Z_{\text{ferm}}\begin{bmatrix}\tfrac{1}{2}\\\tfrac{1}{2}\end{bmatrix}(\tau)^D\sum_{\substack{(c,d)=1\\c+d\text{ even}}}\frac{1}{|\eta(\gamma\cdot\tau)|^{2D}}}_{\text{`bad' boundary conditions}}\,.
\end{equation}
The `good' boundary conditions correspond to the fermions being anti-periodic around the contractible cycle, while the bad correspond to the opposite. We give an explanation of this from the bulk perspective in the next subsection.

We can also consider the symmetric product orbifold of this supersymmetric theory, the only modification being the inclusion of the fermion partition function. It is easier to directly consider the grand canonical partition function in equation \eqref{eqn:supersymmetricgrandZ} which we showed can be written in terms of supersymmetric Hecke operators $\mathcal{T}_k$
\be
\mathfrak{Z} \begin{bmatrix}\alpha \\ \beta\end{bmatrix} (p,\tau)=\exp\left(\sum_{k=1}^{\infty}p^k\mathcal{T}_k Z_{\mathbb{T}^D}  \begin{bmatrix}\alpha \\ \beta\end{bmatrix} (\tau)\right)\,.
\ee
We can average the above partition function in the same way we carried out the bosonic average. Since the fermions do not depend on the moduli they factor out of the average. It is convenient to consider the average of the connected part of the partition function
\begin{equation} \label{eqn:susyboundaryGrandCanonicalZ}
\begin{split}
&\int_{\mathcal{M}_D}\mathrm{d}\mu\,\log\mathfrak{Z} \begin{bmatrix}\alpha \\ \beta\end{bmatrix}(p, \tau)=\sum_{k=1}^{\infty}p^k\Braket{\mathcal{T}_k Z \begin{bmatrix}\alpha \\ \beta\end{bmatrix} (\tau)}\\
&=\sum_{k=1}^\infty \frac{p^k}{k} \sum_{a d = k} \sum_{b=0}^{d-1} Z_{\text{ferm}}\begin{bmatrix} a\alpha + b \beta \\ d\beta\end{bmatrix} \lr{\frac{a\tau+b}{d}}^D \left\langle Z_{\mathbb{T}^D} \lr{\frac{a \tau + b}{d}} \right\rangle,
\end{split}
\end{equation}
where we have split it up into the bosonic part which the average acts on, and the fermionic part. We will again reproduce the above average by performing a summation over bundles and vortices, now including fermions. 

\subsection{The bulk dual}

The natural candidate bulk theory is a supersymmetric version of Chern-Simons. We will now briefly explain this supersymmetric theory, leaving our conventions and additional details to appendix \ref{sec:susyappendix}. In \cite{Belyaev:2008xk} supersymmetric Chern-Simons in flat Minkowski space was considered in the presence of a boundary. The boundary broke half the supersymmetry down to $\mathcal{N}=(1,0)$. In the conventions of \cite{Belyaev:2008xk} the flat space action is given by 
\be \label{eqn:SCSN10}
S^{\mathcal{N}=(1,0)}_{\text{CS}} = \int_M d^3 x (\epsilon^{\mu \nu \rho} A_\mu \pd_\nu A_\rho + \overline{\lambda} \lambda) - \frac{1}{2} \int_{\partial M} d^2 x \sqrt{h} (h^{m n} A_m A_n + \ol{\chi}_- \g^m \pd_m \chi_-).
\ee
In the above $\lambda, \chi$ are Majorana fermions and the notation $\chi_{\pm}$ means we project the fermion onto it's top/bottom component respectively. The boundary is located at $x^3=0$ and the coordinates on the boundary are given by $(x^1, x^2)$ indexed by the label $m$, the boundary metric is $h$ and $\g$ are the gamma matrices.

Due to the presence of the boundary, the action is not invariant under the most general supersymmetry transformation. However, it is invariant under half of the supersymmetry transformations $\mathcal{N}=(1,0)$.\footnote{The action is invariant under these transformations without having to impose any boundary conditions on the fields. This approach was advocated as ``supersymmetry without boundary conditions'' in \cite{Belyaev:2008xk}.}  These transformations are given by
\begin{align}
\delta A_\mu &= (\ol\lambda \g_\mu \epsilon_{+}) + (\ol\epsilon_{+} \pd_\mu \chi_{-}), \nonumber\\
\delta \lambda_a &= -\epsilon^{\mu \nu \rho} (\g_\rho \epsilon_+)_a \pd_\mu A_\nu,\\
\delta \chi_{-} &=(\g^\mu \epsilon_{+}) A_\mu= (\g^m \epsilon_{+}) A_m. \nonumber
\end{align}
Where $\epsilon_+$ is a two component spinor projected onto only it's top component. In the path integral we must choose boundary conditions that satisfy the variational principle and that are left invariant under the above supersymmetry transformations. It was found in \cite{Belyaev:2008xk,Berman:2009kj} that one such choice of boundary conditions is given by fixing given by $A_- = 0$ and $2 \g^2 \lambda_+ + \pd_- \chi_-=0$, where we have defined $A_{\pm} = A_1 \pm A_2$ and $\pd_{\pm} = \pd_1 \pm \pd_2$. The boundary action in terms of these fields is given by\footnote{For this one uses the identity  $\g^1\chi_-=-\g^2\chi_-$.}
\be \label{eqn:susybdyaction1}
S_{\text{bdy}}^{\mathcal{N}=(1,0)} = \frac{1}{2} \int_{\pd M} d^2 x (A_+ A_- + \ol{\chi}_- \g^2 \pd_- \chi_-).
\ee
The second boundary condition we fixed, relating $\chi_-$ and $\lambda_+$, guarantees that our boundary condition for the gauge field is invariant under the $\epsilon_+$ supersymmetry transformation $\delta A_-=0$. Similarly, the other boundary condition $2 \g^2 \lambda_+ + \pd_- \chi_-=0$ is also invariant under $\epsilon_+$ transformations. We don't need to set any additional boundary conditions because the second term in the boundary action \eqref{eqn:susybdyaction1} varies into an equation of motion for $\chi_-$ on the boundary. One of the interesting features of this theory is that there is a dynamical boundary fermion $\chi_-$, decoupled from the other fields, without any corresponding bulk action.

Similarly, there is a $\mathcal{N}=(0,1)$ supersymmetric Chern-Simons theory invariant under $\epsilon_-$ transformations, given by the action 
\be 
S^{\mathcal{N}=(0,1)}_{\text{CS}} = \int_M d^3 x (\epsilon^{\mu \nu \rho} A_\mu \pd_\nu A_\rho + \overline{\lambda} \lambda) + \frac{1}{2} \int_{\partial M} d^2 x \sqrt{g} (g^{m n} A_m A_n + \ol{\chi}_+ \g^m \pd_m \chi_+).
\ee
In the above $\lambda, \chi_+$ are again Majorana fermions, except now $\chi$ has been projected onto the top component. The boundary action can be re-written as
\be \label{eqn:susybdyaction2}
S_{\text{bdy}}^{\mathcal{N}=(0,1)} = \frac{1}{2} \int_{\pd M} d^2 x (-A_+ A_- + \ol{\chi}_+ \g^2 \pd_+ \chi_+).
\ee
A consistent choice of boundary conditions is given by $A_+=0$ and $\left(-2 \g^2 \lambda_- + \pd_+ \chi_+\right)=0$. We can consider the combined action
\be
S = S^{\mathcal{N}=(1,0)}_{\text{CS}} - S^{\mathcal{N}=(0,1)}_{\text{CS}},
\ee
where in total the bulk theory has $\mathcal{N}=(1,1)$ supersymmetry with each half realized independently by one of two theories. The total action will depend on two independent Chern-Simons fields $A,B$, two independent auxiliary fields $\lambda_1, \lambda_2$ and two, fermions $\chi_{\pm}$ which have been projected onto the top/bottom component and effectively function as single component fermions. When considering the partition function of the total theory it will factorize into a contribution from the independent Chern-Simons fields, the two copies of the auxiliary fields $\lambda_i$, and the boundary fermions $\chi_-, \chi_+$. Since we are integrating over the auxiliary fields $\lambda_i$ there is no restriction on the field configurations the boundary fermions take. 

After analytically continuing to Euclidean signature and defining the theory on a bulk handlebody with an asymptotic boundary torus the full partition function is given by the product of contributions of: $U(1) \times U(1)$ Chern-Simons, a holomorphic and anti-holomorphic 2d free fermion, and an overall normalization given by integrating over the auxiliary $\lambda_i$. Dropping the normalization given by the auxiliary fields gives the partition function
\be \label{eqn:ZSCS1} Z_{\text{ferm}}\begin{bmatrix}\alpha\\\beta\end{bmatrix}(\tau) Z_{\text{CS}}(\tau) = \left|\det\overline{\partial}_{\alpha,\beta}\right| \frac{1}{|\eta(\tau)|^2} \,.
\ee
where the first factor comes from the Chern-Simons contribution and the second comes from the free fermions.

We take $D$ copies of the above theory and perform a summation over all bulk handlebodies. The choice of asymptotic boundary conditions fixes the spin structures $\alpha, \beta$ around two particular boundary cycles $A,B$. When summing over bulk handlebodies the standard prescription is to only include handlebodies which can inherit the spin structure specified at the asymptotic boundary \cite{Maloney:2007ud}. That is, if the cycle $c A + d B$ is contractible then it must be true that the fermions are anti-periodic around that cycle. However, in our case the fermions reduce to a boundary term, and so we do not have such a constraint since the spin structure does not need to be extended to the entire handlebody. Summing over all handlebodies, taking into account that the fermions give identical contributions due to  \eqref{eqn:FermionModTransformation}, we find
\begin{equation} \label{eqn:BulkZSCS}
Z_{\text{Bulk}}(\tau)=Z_{\text{ferm}}\begin{bmatrix}\alpha\\\beta\end{bmatrix}(\tau)^D \sum_{\gamma\in \Gamma_{\infty}\setminus\text{SL}(2,\mathbb{Z})}\left|\frac{1}{\eta(\gamma\cdot\tau)^2}\right|^D\,.
\end{equation}
We see that the bulk supersymmetric Chern-Simons theory precisely reproduce the boundary ensemble average.

Let us now explain how the above is modified when we consider the symmetric product orbifold of the supersymmetric Narain theories. We are again interested in implementing twisted boundary conditions along the contractible and non-contractible cycles for both the gauge fields and the fermions. 

Let's first consider the non-contractible cycle. In the case of the gauge fields we found that our gauge group was given by $(\text{U}(1)^D \times \text{U}(1)^D) \wr S_N$, and the summation over non-trivial $S_N$ bundles gave us twisted boundary conditions for the gauge fields along the non-contractible cycle. The story for the fermions is similar. Fermions are sections of a spinor bundle with fiber denoted by $S$, and before quotienting the $S_N$ symmetry the $N$ fermions takes values in this spinor bundle. After quotienting we take the fiber of the spinor bundle to be given by $S \otimes \{1,\ldots, N\}$, and we must sum over all non-trivial bundles. The group $S_N$ acts on the fiber by permutations acting on the the set $\{1,\ldots, N\}$. This implements a sum over all possible twisted boundary conditions for the fermions $\psi_I \to e^{2\pi i \alpha} \psi_{\pi(I)}$ when the $S_N$ portion of the fiber is non-trivial, identical to the case of the gauge fields in Section \ref{sec:4.1}.

For the contractible cycle, we can again use a vortex operator $\mathcal{V}$ to implement twisted boundary conditions for the fermions. We implement the action of the operator by specifying the monodromies it implements on the fermions, namely $\psi_I \to e^{2\pi i \beta} \psi_{\pi(I)}$ as we travel around the contractible cycle. Summing over all possible choices of vortices, similar to the gauge theory case in section \ref{sec:4.1}, gives a summation over all twisted boundary conditions along the contractible cycle. 

Combining these two ingredients together we find that a summation over bundles and vortices again implements the twisted boundary conditions necessary for the symmetric product orbifold. Since the fermions and gauge fields both transform in the adjoint of the discrete gauge group they acquire the same monodromy $A_I \to A_{\pi(I)}$ and $\psi_I \to e^{2\pi i \alpha} \psi_{\pi(I)}$ as they travel around a vortex. Since the fields acquire the same monodromies it immediately follows that that, identical to the bosonic case, performing a summation over bundles and vortices implements a summation over all degree $N$ covering spaces of the base torus.  

If we consider the grand canonical partition function of the supersymmetric Chern-Simons theory on handlebody $M$, since we are summing over covering spaces, we find that it is again given by the exponential of connected covers
\begin{equation}
\mathfrak{Z}_{\text{SCS}}(M,p)=\exp\left(\sum_{k=1}^{\infty}p^k \mathcal{T}_k Z \begin{bmatrix}\alpha\\\beta\end{bmatrix}\lr{\tau}\right)\,.
\end{equation}
Comparing to the boundary answer in equation \eqref{eqn:susyboundaryGrandCanonicalZ}, we immediately have that the connected covering space contributions match between the bulk and the boundary theories after summing over handlebodies $M$
\begin{equation}
\int_{\mathcal{M}_D}\mathrm{d}\mu\,\log\mathfrak{Z} \begin{bmatrix}\alpha\\\beta\end{bmatrix}(\tau,p)=\sum_{M}\log\mathfrak{Z}_{\text{SCS}}(M,p)\,. 
\end{equation}

\section{Conclusions and discussion}\label{sec:7}

In this work our goal was to provide a bulk dual for the ensemble average over $\text{Sym}^N(\mathbb{T}^D)$ CFTs. In the case of a torus boundary, we showed that a Chern-Simons theory with gauge group $\lr{\text{U}(1)^D \times \text{U}(1)^D}\wr S_N$, with the inclusion of bulk vortices, reproduces what we denoted as the ``semiclassical'' portion of the boundary average. However, there remain many terms that do not have a clear semiclassical interpretation. 

We argued that some of these terms can be included by appropriately modifying the rules of the gravity path integral. For instance, disconnected handlebodies that break replica symmetry can be reproduced by letting independent gauge fields $A_I$ live on ``independent'' bulk manifolds $M_I$ with distinct contractible cycles but the ``same'' asymptotic boundary, see the discussion around equation \eqref{eqn:sublatticebreaksreplicasym}. Furthermore, we argued that the simplest wormhole geometry in equation \eqref{eqn:wormholesublattice} can be reproduced by having two vortices in the bulk. It's possible that other wormholes can also be reproduced by more complicated vortex configurations. 

We will briefly contrast our results with the expectations for pure AdS$_3$ gravity \cite{Witten:2007kt,Maloney:2007ud}. With a boundary torus, the only classical saddle-points for the Einstein-Hilbert action are handlebody geometries. However, in \cite{Benjamin:2021wzr} it was argued that another reasonable class of geometries to include are conical defect geometries which are singular orbifolds. We note this is quite similar to what we have found. We also sum over handlebody geometries and the gauge theory analogue of conical defect geometries, since the bulk vortices make the field strength singular at the vortex. In addition, in AdS$_3$ gravity it is expected that additional off-shell geometries should be included \cite{Maxfield:2020ale,Cotler:2020ugk} in the path integral. While off-shell geometries seem to differ from the non-semiclassical geometries we have found, it is interesting that we also find contributions beyond simple handlebodies. 

\subsubsection*{Ensemble Averaging Strings} \label{sec:string-theory}

As mentioned in the introduction, one motivation to consider the ensemble averaged symmetric orbifold comes from string theory. It has recently been shown that the tensionless string theory on $\text{AdS}_3\times\text{S}^3\times\mathbb{T}^4$ is precisely dual to the (super) symmetric orbifold $\text{Sym}^N\mathbb{T}^4$ CFT at large $N$ \cite{Eberhardt:2018ouy,Eberhardt:2019ywk,Eberhardt:2020akk,Eberhardt:2020bgq,Eberhardt:2021jvj}. The tensionless string has string length equal to the AdS length, and so the theory is far from any sensible semiclassical limit. While all observables of the theory can be calculated through worldsheet path integrals, there is no known sense in which the string partition function can be approximated by a gravitational path integral with a sum over smooth bulk geometries. One hope is that a suitable sum over geometries may emerge if we average over the moduli of the tensionless string. We now explain how this is partially realized.

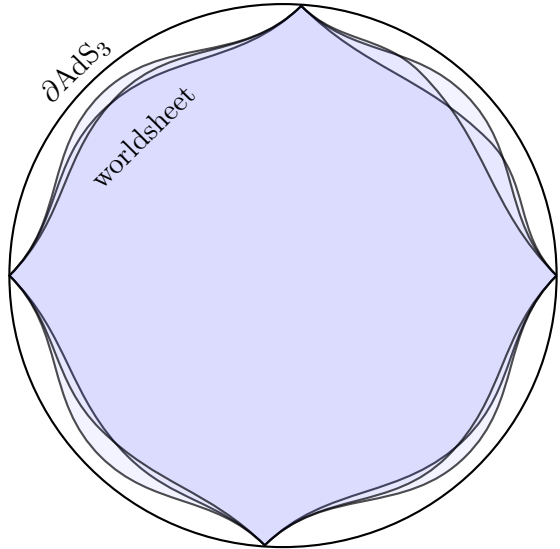
\begin{figure}
\centering
\ifdraft
\else
\begin{tikzpicture}[scale = 1.2]
\draw[thick] (0,0) circle (3);
\path[draw, closed=true, thick, black, opacity = 0.7] (3,0) to[out = 135, in = -45] (2.25,1.5) to[out = 135, in = -45] (0.2,2.98) to[out = -135, in = 45] (-2,1.8) to[out = -135, in = 45] (-3,0) to[out = -45, in = 135] (-1.7,-2) to[out = -45, in = 135]  (-0.2,-2.98) to[out = 45, in = -135] (2,-1.8) to[out = 45, in = -135] (3,0);
\path[fill, closed=true, thick, blue, opacity = 0.05] (3,0) to[out = 135, in = -45] (2.25,1.5) to[out = 135, in = -45] (0.2,2.98) to[out = -135, in = 45] (-2,1.8) to[out = -135, in = 45] (-3,0) to[out = -45, in = 135] (-1.7,-2) to[out = -45, in = 135]  (-0.2,-2.98) to[out = 45, in = -135] (2,-1.8) to[out = 45, in = -135] (3,0);
\path[draw, closed=true, thick, black, opacity = 0.7] (3,0) to[out = 135, in = -45] (2,2) to[out = 135, in = -45] (0.2,2.98) to[out = -135, in = 45] (-2,2) to[out = -135, in = 45] (-3,0) to[out = -45, in = 135] (-2,-2) to[out = -45, in = 135]  (-0.2,-2.98) to[out = 45, in = -135] (2,-2) to[out = 45, in = -135] (3,0);
\path[fill, closed=true, thick, blue, opacity = 0.05] (3,0) to[out = 135, in = -45] (2,2) to[out = 135, in = -45] (0.2,2.98) to[out = -135, in = 45] (-2,2) to[out = -135, in = 45] (-3,0) to[out = -45, in = 135] (-2,-2) to[out = -45, in = 135]  (-0.2,-2.98) to[out = 45, in = -135] (2,-2) to[out = 45, in = -135] (3,0);
\path[draw, closed=true, thick, black, opacity = 0.7] (3,0) to[out = 135, in = -45] (1.5,2.25) to[out = 135, in = -45] (0.2,2.98) to[out = -135, in = 45] (-1.8,2) to[out = -135, in = 45] (-3,0) to[out = -45, in = 135] (-2,-1.7) to[out = -45, in = 135]  (-0.2,-2.98) to[out = 45, in = -135] (1.8,-2) to[out = 45, in = -135] (3,0);
\path[fill, closed=true, thick, blue, opacity = 0.05] (3,0) to[out = 135, in = -45] (1.5,2.25) to[out = 135, in = -45] (0.2,2.98) to[out = -135, in = 45] (-1.8,2) to[out = -135, in = 45] (-3,0) to[out = -45, in = 135] (-2,-1.7) to[out = -45, in = 135]  (-0.2,-2.98) to[out = 45, in = -135] (1.8,-2) to[out = 45, in = -135] (3,0);

\node[below, rotate = 45] at (-1.7,1.7) {worldsheet};
\node[above, rotate = 45] at (-2.1,2.1) {$\partial\text{AdS}_3$};
\end{tikzpicture}
\fi
\caption{A cartoon representation of a worldsheet propagating within the bulk in the tensionless limit. The dynamics of the string are constrained to the boundary of spacetime.}
\label{fig:tensionless-worldsheet}
\end{figure}

The natural partition function for the tensionless string is given by the grand canonical ensemble of $\text{Sym}(\mathbb{T}^4)$ \cite{Eberhardt:2021jvj}. The genus-one partition function of the grand canonical ensemble is given by\footnote{Here and in what follows we ignore spin structures.}
\begin{equation}\label{eq:grand-canonical-string}
\mathfrak{Z}(p,\tau)=\exp\left(\sum_{k=1}^{\infty} \underbrace{p^kT_kZ(\tau)}_{\text{worldsheets}}\right)\,.
\end{equation}
The Hecke operators $T_k$ enumerate over connected covering spaces that cover the boundary of AdS$_3$. Holographically, the connected covering spaces are to be interpreted as worldsheets of strings propagating in the target AdS$_3$ which wind the boundary $k$ times. The argument of the exponential only includes contributions with a single string on the AdS$_3$ background. The grand canonical partition function $\mathfrak{Z}$, after expanding the exponential, then includes any number of disconnected strings propagating on the AdS$_3$ background. The free energy $F=\log\mathfrak{Z}$ is then, in this sense, a `first quantized' partition function of the string theory, while the partition function $\mathfrak{Z}$ can be thought of as a `second quantized' quantity.

Within the main text, we considered averaging both the connected and disconnected contributions of the symmetric orbifold, or equivalently we considered the averages of both the first and second quantized partition functions. From the point of string theory, these averages have the following interpretation:
\begin{itemize}

    \item \textbf{First quantized average:} Averaging the string theory at the level of the worldsheet sigma model on $\text{AdS}_3\times\text{S}^3\times\mathbb{T}^4$. This results in considering only an average of the free energy $F=\log\mathfrak{Z}$, which in turn is given by
    \begin{equation} \label{eqn:firstquantizedstring}
    \Braket{F(p,\tau)}=\sum_{k=1}^{\infty}p^k\Braket{T_kZ(\tau)}\,.
    \end{equation}
    As discussed in Section \ref{sec:bulk-theory}, this average is reproduced by a sum over torus handlebodies of the free energy of the grand-canonical $\text{U}(1)^D\times\text{U}(1)^D\wr S_N$ Chern-Simons theory with vortices in the bulk.
    
    \item \textbf{Second quantized average:} Averaging the string theory at the level of the second quantized theory of strings, i.e. at the level of string field theory on $\text{AdS}_3\times\text{S}^3\times\mathbb{T}^4$. The resulting partition function is
    \begin{equation} \label{eqn:secondquantizedstring}
    \Braket{\mathfrak{Z}(p,\tau)}=\Braket{\exp\left(\sum_{k=1}^{\infty}p^kT_kZ(\tau)\right)}\,.
    \end{equation}
    As we have seen throughout this paper, the resulting expansion can be reproduced partially by bulk $\text{U}(1)^D\times\text{U}(1)^D\wr S_N$ Chern-Simons quantities on semiclassical bulk geometries, but many of the terms in the average cannot be recovered this way, and are thought of as arising from `generalized' geometries.
    
\end{itemize}
Of course, from the point of view of string theory, these are both completely valid course-grainings of the $\mathbb{T}^4$ compactification. It's interesting to note that a single string \eqref{eqn:firstquantizedstring} propagating on AdS$_3$ has a simple ensemble averaged interpretation in terms of a sum over handlebodies with vortices. When we average over multiple strings \eqref{eqn:secondquantizedstring} on one background we get additional ``non-semiclassical'' geometries connecting the strings, for which we have no standard holographic interpretation. 

One subtlety with the above is the issue of convergence. As explained in Section \ref{sec:narain-review}, the average of $\mathbb{T}^D$ and therefore of the $\text{Sym}(\mathbb{T}^D)$ partition functions only converges when $D > g + 1$. For the specific case of the tensionless string, $D=4$, this means that most contributions to the grand canonical average diverge. However, the first quantized average \eqref{eqn:firstquantizedstring} converges. Nevertheless it is still tempting to consider each individual term in the average as arising from a bulk geometry. A bulk interpretation of the divergence is that the sum over wormhole geometries diverges when we have too many asymptotic boundaries. 

\subsubsection*{Semiclassical Limit}

It would be interesting if there was an appropriate semiclassical limit of $\langle Z_{\mathbb{T}^D \wr S_N} (\tau)\rangle $ where the dominant geometry is given by the $\lr{U(1)^D \times U(1)^D}\wr S_N$ Chern-Simons theory, as opposed to one of the more mysterious contributions. We do not have a coupling constant we can tune, but there does appear to be a natural notion of a semiclassical limit for the Narain average which is given by sending $D \to \infty$ \cite{Maloney:2020nni,Collier:2021rsn}.

For $N=2$ this can be argued very explicitly. The proper limit is to take $D \to \infty$ while holding the temperature $\tau = i \beta$ sufficiently large or small. In this limit, the contribution of wormholes to $\langle Z^2 \rangle$ is exponentially suppressed relative to the handlebodies \cite{Collier:2021rsn}. Furthermore, it's straightforward to see that the disconnected handlebodies that dominate $\langle Z^2 \rangle$ will be replica symmetric in the large $D$ limit.\footnote{This is because non-replica symmetric disconnected covers are subleading due to the Hawking-Page transition at large/small temperatures.} The symmetric disconnected covers are captured by the Chern-Simons calculation, and therefore we find in this limit
\be
\lim_{D \to \infty} \lim_{\beta \to \infty } \langle Z_{\mathbb{T}^D \wr S_2} (i \beta )\rangle = \sum_{\substack{\text{handlebodies $M$,} \\ \text{vortices}}} Z_{G \wr S_2} (M) + \mathcal{O}(\beta^{-c D}),
\ee
where the subleading terms are exponentially suppressed in $D$, with $c$ some constant \cite{Collier:2021rsn}, relative to the dominant term in the semiclassical contribution. An obvious generalization of this equation is true when $\beta \to 0$. Thus, at least in this limit a semiclassical geometry dominates the average. It would be interesting to know if this holds for general $N$, which would require understanding whether disconnected symmetric handlebodies dominate $\langle Z^N \rangle$ in the large $D$ limit. 

\subsection{Open Questions and future directions}
We will end with a discussion of some open questions and possible future directions. \\

\noindent\textbf{OPE Randomness Hypothesis:} In \cite{Belin:2020hea} the OPE randomness hypothesis was proposed. This amounts to a generalization of ETH to an ansatz for the OPE coefficients of chaotic CFTs. The proposal conjectures that OPE coefficients involving heavy operators are random variables obeying an approximately Gaussian distribution. Both the $\mathbb{T}^D$ and $\text{Sym}(\mathbb{T}^D$) theories should evade these proposals since the theories are not chaotic. The symmetry currents furnish selection rules which should lead to deviations away from the aforementioned subleading non-Gaussianities. It would be interesting to determine the statistics of the OPE coefficients in the Narain ensemble.\\

\noindent\textbf{Deforming Towards Supergravity:} There is a general expectation that the tensionless regime can be deformed to the semiclassical supergravity regime \cite{Seiberg:1999xz} by turning on an exactly marginal operator in a twisted sector of the orbifold theory, see for example ~\cite{David:2002wn,Burrington:2012yq,Gaberdiel:2015uca,Guo:2020gxm,Apolo:2022fya,Benjamin:2022jin}. It would be interesting to understand if it is possible to ensemble average over the Narain moduli space with this marginal deformation turned on. One potential approach would be to use conformal perturbation theory, and our discussion of correlators in Section \ref{sec:correlators} might be useful for such an approach. It would be very interesting to see how the sum over bulk geometries is modified as we move away from the tensionless point. \\

\noindent\textbf{Averaging the D1/D5 moduli space:} As mentioned above, symmetric product orbifold theories $\text{Sym}^N(\mathbb{T}^D)$ live in a larger moduli space than that of their Narain seed theory, and specifically for seed theories of central charge $c\leq 6$, symmetric product orbifolds admit marginal deformations by twist fields, which break the orbifold structure. One special such example is the D1/D5 system compactified on $\mathbb{T}^4$ \cite{Seiberg:1999xz}, whose moduli space contains the `orbifold point' $\text{Sym}^N(\mathbb{T}^4)$ along with its Narain moduli, as well as four exactly marginal operators in the twist-2 sector.\footnote{These operators roughly correspond to introducing a non-zero `t Hooft coupling $\lambda$.} It would be interesting to consider an average over the full 20-dimensional moduli space of the D1/D5 CFT. Such a computation, however, would require knowledge of unprotected quantities on the full D1/D5 moduli space, which are currently poorly understood beyond low-orders in conformal perturbation theory.
\\

\noindent \textbf{Large $N$ and Phase Transitions:} The grand canonical ensemble of $\text{Sym}(\mathbb{T}^4$) is dual to the tensionless string with any number of strings propagating on the background geometry. It was argued in \cite{Eberhardt:2020bgq} that at large $N$ the grand canonical partition function is dominated by a configuration of strings that wind $\sim N$ times around the AdS$_3$ boundary. This corresponds to the average of the grand canonical ensemble being dominated by Chern-Simons with gauge group $G \wr S_N$ with large $N$. In the case of a single winding string this would correspond to a geometry with a single vortex, whereas multiple winding strings would partially be reproduced by multiple disconnected covering spaces with multiple vortices. It was shown that there are phase transitions between different stringy configurations as the chemical potential was tuned \cite{Eberhardt:2020akk}, it would be interesting to better understand these transitions from the perspective of the Chern-Simons geometries considered in this work.\\

\noindent\textbf{Stringy Ensemble-Averaging:} 
In this work we have attempted to construct an effective theory of a course-grained string theory.
In view of the universal complexity of full-fledged UV compactifications there may be some logic in considering some form of course-graining as a starting point. It would be highly interesting to seek out other examples of string compactifications amenable to some form of ensemble averaging. See \cite{Benjamin:2021wzr} for work in this direction.

\subsection*{Acknowledgments}
We thank Alex Belin, Ivano Basile, Lorenz Eberhardt, Matthias Gaberdiel, Francesco Galvagno, Victor Gorbenko, Hans Jockers, Christoph Keller, Alex Maloney, Hirosi Ooguri, Geoff Penington, Julian Sonner and Jakub Vo\v{s}mera for useful discussions. Special thanks to Alex Belin and Ida Zadeh for comments on a draft of the manuscript. A.K. is supported by the “Onassis Foundation” as an “Onassis Scholar” (Scholarship ID: F ZO 030/2 (2020-2021)). The work of B.K. is supported by the Swiss National Science Foundation through a personal grant and via the NCCR SwissMAP. MU is supported in part by the NSF Graduate Research Fellowship Program under grant DGE1752814, by the Berkeley Center for Theoretical Physics, by the DOE under award DE-SC0019380 and under the contract DE-AC02-05CH11231, by NSF grant PHY1820912, by the Heising-Simons Foundation, the Simons Foundation, and National Science Foundation Grant No. NSF PHY-1748958.

\newpage

\appendix

\section{Supersymmetric Chern-Simons}
\label{sec:susyappendix}
In this section we include additional details on our conventions and on $\mathcal{N}=1$ supersymmetric $U(1)$ Chern Simons in the presence of a boundary, closely following \cite{Belyaev:2008xk,Berman:2009kj}. We first discuss our conventions for Lorentzian signature supersymmetry. The gamma matrices satisfy the standard algebra
\be \label{eqn:lorentziangamma}
\{\g^\mu, \g^\nu\} = 2 \eta^{\mu \nu}, \qquad \g^\mu \g^\nu = \eta^{\mu\nu}+\g^{\mu\nu}=\eta^{\mu \nu} - \epsilon^{\mu \nu \rho} \g_{\rho},
\ee
(the first of the right equations is a definition for $\g^{\mu\nu}$ and they are explicitly given by
\be
\gamma^1=\left(\begin{array}{cc}
0 & -1 \\
1 & 0
\end{array}\right), \quad\gamma^2=\left(\begin{array}{cc}
0 & 1 \\
1 & 0
\end{array}\right), \quad \gamma^3=\left(\begin{array}{cc}
1 & 0 \\
0 & -1
\end{array}\right).
\ee

The metric takes the form $\eta_{\mu \nu} = \operatorname{diag}(-1,1,1)$, and the spacetime coordinates are  $x^\mu = (x^1,x^2,x^3)$ where $x^1$ is the time component, while the boundary is located at ${x^3=0}$. We will use indices $m,n$ to indicate components restricted to the boundary ${x^m = (x^1, x^2)}$. All spinors considered are Majorana $\ol{\lambda}^a \equiv C^{a b}\lambda_b$ where $C=-C^\intercal$ is the charge conjugation matrix. Spinor indices are contracted top right to bottom left: $\ol\lambda \chi = \lambda^a \chi_a$, $\ol\lambda \g^\mu \chi=\lambda^a (\g^\mu)_a^{~b} \chi_b$, where the gamma matrices implicitly have the index structure $(\g^\mu)_a^{~b}$. Spinor indices $a,b$ are raised and lowered with the antisymmetric charge conjugation matrix $C$: $\lambda^a = C^{a b} \lambda_b$, which is defined through $C \g^\mu C^{-1}=-(\g^\mu)^{\intercal}$. We introduce projection operators $P_{\pm} = \frac{1}{2}(1\pm \g^3)$ and define fields $\chi_{\pm}$ with $P_{\pm} \chi = \chi_{\pm}$, which projects onto the top or bottom component of the spinor respectively.

\subsection*{$\mathcal{N}=(1,0)$ Lorentzian Chern Simons}
We first discuss the case of $\mathcal{N}=(1,0)$ supersymmetric $U(1)$ Chern Simons in the presence of a boundary in Minkowski space. The action can be constructed through the use of spinor superfields,\footnote{The spinor superfield formalism includes an additional complex scalar that is the superpartner of $\chi$, but it does not end up contributing to the action. We exclude it below, but see \cite{Belyaev:2008xk}.} including appropriate boundary terms it is given by \cite{Belyaev:2008xk,Berman:2009kj} 
\be \label{eqn:MinkSCS}
S^{\mathcal{N}=(1,0)}_{\text{CS}} = \int_M d^3 x (\epsilon^{\mu \nu \rho} A_\mu \pd_\nu A_\rho + \overline{\lambda} \lambda) - \frac{1}{2} \int_{\partial M} d^2 x \sqrt{h} (h^{m n} A_m A_n + \ol{\chi}_- \g^m \pd_m \chi_-).
\ee
In the above $\lambda, \chi$ are Majorana fermions, with $\chi_{-} = P_{-} \chi$ being a purely boundary fermion. The dynamical boundary fermion $\chi_{-}$ is unusual, and is required to cancel the boundary terms produced by the susy variation. We also have the boundary metric $h_{m n}$. The full susy variations under $\epsilon$ are given by \cite{Belyaev:2008xk}
\begin{align}\label{eq: full susy var}
    \delta A_\mu &= -(\ol\epsilon \gamma_\mu \lambda) + (\ol\epsilon \pd_\mu \chi), \ \nonumber\\
    \delta \lambda_a &= -\epsilon^{\mu \nu \rho} (\g_\rho \epsilon)_a \pd_\mu A_\nu, \\
    \delta \chi_a &= (\gamma^\mu \epsilon)_a A_\mu.\nonumber
\end{align}
However, the action \eqref{eqn:MinkSCS} is only invariant up to a boundary term under a general variation (see \eqref{eq: total variation plus}). For the full action to be invariant we must restrict to variations $\epsilon_{+}=P_{+}\epsilon$, which means the supersymmetry is broken down to $\mathcal{N}=(1,0)$ in the presence of the boundary. This is true even without imposing any boundary conditions on the fields. The explicit susy variations under $\epsilon_{+}$ are
\begin{align}
\delta A_\mu &= -(\ol\epsilon_{+} \g_\mu \lambda) + (\ol\epsilon_{+} \pd_\mu \chi_{-}), \nonumber\\
\delta \lambda_a &= -\epsilon^{\mu \nu \rho} (\g_\rho \epsilon_+)_a \pd_\mu A_\nu,\\
\delta \chi_{-} &= (\g^m \epsilon_{+}) A_m, \nonumber
\end{align}

and the action is invariant under this subset of transformations. Using the above, we can obtain the following variations
\begin{align}
 &\delta A_- = \ol\epsilon_+ \left(2 \g^2 \lambda_+ + \pd_- \chi_- \right),\\
   & \delta \left(2 \g^2 \lambda_+ + \pd_- \chi_- \right) = \g^2 \epsilon_+ \pd_+ A_-.
\end{align}  
where we have defined the notation $A_\pm = A_1 \pm A_2$, $\pd_\pm = \pd_1 \pm \pd_2$ which is distinct from the spinor $\pm$ projection notation and used $\gamma^1\lambda_+=\gamma^2\lambda_+$. For a good variational principle, the boundary conditions we impose are $A_- = 0$ and $2\g^2 \lambda_+ + \pd_- \chi_-=0$ \cite{Belyaev:2008xk,Berman:2009kj}. We see that the $\epsilon_+$ transformations leave these boundary conditions invariant. The boundary kinetic term for $\chi_-$ does not need to be cancelled for a good variational principle since it gives the equations of motion. In the path integral with these boundary conditions the integral over $\chi_-$ is effectively not constrained since we also integrate over $\lambda_+$, thus $\chi_-$ is essentially a free dynamical one-component boundary fermion. The integral over $\lambda$ will give some overall normalization constant since it is non-dynamical.

Similarly, we can construct an action with $\mathcal{N}=(0,1)$ invariant under $\epsilon_-$ transformations by modifying the boundary term
\be \label{eqn:otherbdyaction}
S^{\mathcal{N}=(0,1)}_{\text{CS}} = \int_M d^3 x (\epsilon^{\mu \nu \rho} A_\mu \pd_\nu A_\rho + \overline{\lambda} \lambda) + \frac{1}{2} \int_{\partial M} d^2 x \sqrt{g} (g^{m n} A_m A_n + \ol{\chi}_+ \g^m \pd_m \chi_+),
\ee
where $\lambda, \chi_+$ are again Majorana fermions, and $\chi_+$ has been projected onto it's top component. In this case the variations under $\epsilon_-$ are given by
\begin{align}
\delta A_\mu &= -(\ol\epsilon_{-} \g_\mu \lambda) + (\ol\epsilon_{-} \pd_\mu \chi_{+}), \nonumber\\
\delta \lambda_a &= -\epsilon^{\mu \nu \rho} (\g_\rho \epsilon_-)_a \pd_\mu A_\nu,\\
\delta \chi_{+} &= (\g^m \epsilon_{-}) A_m. \nonumber
\end{align}
Which immediately gives us the variations
\begin{align}
 &\delta A_+ =\ol \epsilon_- \left(-2 \g^2 \lambda_- + \pd_+ \chi_+\right),\\
   & \delta \left(-2 \g^2 \lambda_- + \pd_+ \chi_+\right) = - \g^2 \epsilon_- \pd_- A_+.
\end{align}  
Proper boundary conditions in this case correspond to $A_+ = 0$ and $-2\g^2 \lambda_- + \pd_+ \chi_+ = 0$ which are both preserved under $\epsilon_-$ transformations. In the main text we defined a total theory given by the difference of the above actions
\be
S = S^{\mathcal{N}=(1,0)}_{\text{CS}} - S^{\mathcal{N}=(0,1)}_{\text{CS}},
\ee
so that the bulk theory has the full $\mathcal{N}=(1,1)$ supersymmetry realized by different sectors. Since $\ol\chi_- \gamma^m \partial_m \chi_+=\ol\chi_+ \gamma^m \partial_m \chi_-=0$ the boundary fermion term for the above action can be rewritten in a simple form
\begin{equation}
    \int_{\pd M}\mathrm{d}^2x \ol\chi \g^m \pd_m \chi,
\end{equation}
which makes it clear that we have a dynamical 2d free fermion on the boundary. The gauge fields do not interact with the fermions so the full path integral will simply be a product of the Chern-Simons contribution and the fermion contribution. In each case, the boundary fermions $\chi_\pm$ are projected onto the top/bottom component and function as single component spinors, so individually their partition functions will contribute a determinant of either a holomorphic or anti-holomorphic derivative after analytic continuation to Euclidean signature. 

There is one additional subtlety to address, the above theories were defined on a flat background. However, for our purposes we are interested in supersymmetric Chern-Simons on a manifold with an asymptotic boundary torus. In general such manifolds do not admit flat metrics, so we need to consider the theory on a curved background \cite{Closset:2012ru}. While Chern-Simons is itself background independent, the supersymmetric version depends on a choice of background metric $g$ through the quadratic bulk fermion term. Ignoring the boundary term, the action is of the form
\be 
S = \int_M d^3 x (\epsilon^{\mu \nu \rho} A_\mu \pd_\nu A_\rho + \sqrt{g} \hspace{.1cm}\overline{\lambda} \lambda).
\ee
However, the metric dependence is quite mild since the fermion is non-dynamical, and it produces an overall normalization for the partition function. Nevertheless, to preserve supersymmetry we need to choose a bulk metric $g$ that satisfies the killing spinor equations. The end result is that the supersymmetry transformations will mildly depend on the background metric \cite{Closset:2012ru}. In the case of bulk handlebodies we can choose $g$ to be given by the corresponding AdS$_3$ metric \cite{Assel:2016pgi} and supersymmetry will be preserved, but for more general three-manifolds that appear when considering wormhole geometries little is known. 

\subsection{Details of the variations.}
In this subsection we include additional details details\footnote{Some useful identities include $\ol{\lambda}\chi = \ol{\chi}\lambda$,  $\overline{\lambda}\gamma_\mu \epsilon=-\overline{\epsilon}\gamma_\mu \lambda$, and since the spinors are Majorana we have $\ol{\lambda}^a\chi_a = C^{a b}\lambda_b \chi_a$.} on the variation of the supersymmetic Chern-Simons action in equation \eqref{eqn:MinkSCS}. For convenience we split the action into a bulk and boundary piece
\begin{align}
	S &= \int_M \mathrm d^3 x \left( \epsilon^{\mu \nu \rho} A_\mu \pd_\nu A_\rho +  \overline{\lambda} \lambda \right), \\
	S_\pd &=  - \frac{1}{2} \int_{\partial M} \mathrm d^2 x \left( A^m A_m  + \ol{\chi}_- \g^m \pd_m \chi_- \right) .
\end{align}
Where we used $h_{m n}=\operatorname{diag}\lr{-1,1}$, where Latin indices $m, n$ are again boundary indices and take values in $\{1,2\}$. Varying with the full supersymmetric variations \eqref{eq: full susy var} we find 
\begin{align}
\delta S &= 2\int_M \mathrm d^3 x \epsilon^{\mu \nu \rho} \partial_\nu A_\rho \lr{-\ol\epsilon\gamma_\mu\lambda+\ol\epsilon\pd_\mu \chi}+ \int_{\partial M} \mathrm d^2 x\epsilon^{3mn} A_m \lr{- \ol\epsilon \gamma_n \lambda +\ol\epsilon \pd_n \chi}\ \\
&+ 2\int_M\mathrm d^3x \epsilon^{\mu \nu \rho} \pd_\nu A_\rho \left(\ol \epsilon \gamma_\mu \lambda \right).
\end{align}
The second line cancels the first term, and we can rewrite the remaining bulk term as a boundary term to obtain
\be
	\delta S = -\int_{\partial M}\mathrm d^2x \epsilon^{3nm} A_n \lr{\ol\epsilon \gamma_m\lambda + \ol \epsilon \pd _m\chi} \ .
\ee
The variation of the boundary term gives
\begin{align}
		\delta S_\pd &= \int_{\partial M}\mathrm d^2x \left( A^m \lr{\ol\epsilon \gamma_m\lambda-\ol\epsilon\pd_m\chi} + \lr{\pd_m \ol\chi_- \epsilon A^m - \pd_m \ol\chi_-\epsilon^{m\mu\rho}\gamma_\rho \epsilon A_\mu} \right) .
\end{align}
The variation of the full action \eqref{eqn:MinkSCS} is thus a total boundary term
\begin{align}\label{eq: total variation plus}
\delta S^{\mathcal{N}=(1,0)}_{\text{CS}}= &\int_{\partial M}\mathrm d^2x\left(-\epsilon^{3nm}A_n \lr{\ol \epsilon \pd_m \chi + \ol \epsilon \gamma_m \lambda} +A^m \lr{\pd_m \ol \chi_-\epsilon} - \epsilon^{m\mu\nu} A_\mu \lr{\pd_m \ol\chi_- \gamma_\nu \epsilon}\right)  \nonumber \\ + &\int_{\partial M}\mathrm d^2 x A^m \left( \ol\epsilon \gamma_m\lambda  - \ol\epsilon \pd _m \chi \right) \ . 
\end{align}
Now we specialize to $P_+ \epsilon=\epsilon_+$ variations. We immediately have the following cancellations. The third term cancels the last term since $\ol \chi_- \epsilon_+ = \ol \epsilon_+ \chi$, while the first term is cancelled by fourth term since in the fourth term only $\ol \chi_- \g_3 \epsilon_+ = \ol \chi_- \epsilon_+$ survives. Finally, the second term cancels the fifth term  since we have $\epsilon^{3 n m} \g_m = \g^3 \g^n$ and $\ol \epsilon_\pm \g^3 = \mp \ol \epsilon_\pm$ due to the factor of the charge conjugation matrix $C$ in $\ol \epsilon$. Note again that no special choice of boundary conditions was needed to make the action invariant under $\epsilon_+$ variations.  

The variation of the $\mathcal{N}=(0,1)$ supersymmetric Chern-Simons action under $\epsilon_-$ works similarly, and we obtain
\begin{align}\label{eq: total variation minus}
\delta S^{\mathcal{N}=(0,1)}_{\text{CS}}= &\int_{\partial M}\mathrm d^2x\left(-\epsilon^{3nm} A_n\lr{\ol \epsilon \pd_m \chi + \ol \epsilon \gamma_m \lambda}-A^m \lr{\pd_m \ol \chi_+\epsilon}  + \epsilon^{m\mu\nu}  A_\mu \lr{\pd_m \ol\chi_+ \gamma_\nu \epsilon} \right) \nonumber\\
	+&\int_{\partial M}\mathrm d^2 x A^m \lr{-\ol\epsilon \gamma_m\lambda+ \ol\epsilon \pd_m \chi}. 
\end{align}
Specializing to $P_- \epsilon = \epsilon_-$ variations we again find that all the boundary terms cancel in a similar way as in \eqref{eq: total variation plus}.

\bibliography{main}
\bibliographystyle{utphys.bst}

\end{document}